\documentclass[twocolumn]{aastex63}

\usepackage{graphicx}	
\usepackage{amsmath}	
\usepackage{amssymb}	
\usepackage{xcolor}
\usepackage{booktabs}   

\usepackage{graphicx}
\usepackage{amssymb}
\usepackage{amsmath}
\usepackage{natbib}
\usepackage{wrapfig}
\usepackage[utf8]{inputenc}
\usepackage[T1]{fontenc}
\usepackage{url}
\PassOptionsToPackage{hyphens}{url}\usepackage{hyperref}
\hypersetup{breaklinks=true}
\usepackage{hyperref}
\usepackage[letterspace=0]{microtype}

\tabletypesize{\footnotesize}

\newcommand{\G}{G_{mag}}
\newcommand{\BR}{G_{BPmag}-G_{BRmag}}

\newcommand{\masyr}{ \ {\rm{mas \ yr^{-1}}}\>}
\newcommand{\kms}{ \ {\rm{km \ s^{-1}}}\>}

\newcommand{\foo}[1]{}

\newcommand{\kpc}{\>{\rm kpc}}

\newcommand{\degree}{\degr}

\newcommand{\pmra}{\mu_{\alpha\star}}
\newcommand{\pmdec}{\mu_{\delta}}
\newcommand{\parallax}{\varpi}
\newcommand{\vlos}{v_{\mathrm{los}}}

\newcommand{\hyperfootnote}[1][]{\def\ArgI\hyperfootnoteRelay}
\newcommand\hyperfootnoteRelay[2][]{\href{#1#2}{\ArgI}\footnote{\href{#1#2}{#2}}}

\newcommand{\pmx}{\mu_x}
\newcommand{\pmy}{\mu_y}

\def\secrevision#1{#1}
\def\revision#1{#1}

\newcommand{\Gaia} {Gaia }
\newcommand{\Beta}{B} 

\defcitealias{Lokas2010}{\L10}
\defcitealias{Law-Majewski2010}{LM10}
\defcitealias{Fardal2019}{F19}
\defcitealias{Ferguson-Strigari2020}{FS20}
\defcitealias{Vasiliev-Belokurov2020}{VB20}

\accepted{December 22, 2020}
\shorttitle{Sagittarius in 6D}
\shortauthors{del Pino et al.}
\graphicspath{{./}{figures/}}

\begin{document}

\title{\lsstyle
\Large{Revealing the Structure and Internal Rotation of the Sagittarius Dwarf Spheroidal Galaxy with \Gaia and Machine Learning}}

\correspondingauthor{Andres del Pino}
\email{adelpinomolina@stsci.edu}

\author[0000-0003-4922-5131]{Andr\'es del Pino}
\affiliation{Space Telescope Science Institute, 3700 San Martin Drive, Baltimore, MD 21218, USA}

\author[0000-0003-4207-3788]{Mark A. Fardal}
\affiliation{Space Telescope Science Institute, 3700 San Martin Drive, Baltimore, MD 21218, USA}

\author[0000-0001-7827-7825]{Roeland P. van der Marel}
\affiliation{Space Telescope Science Institute, 3700 San Martin Drive, Baltimore, MD 21218, USA}
\affiliation{Center for Astrophysical Sciences, Department of Physics \& Astronomy, Johns Hopkins University, Baltimore, MD 21218, USA}

\author[0000-0001-7138-8899]{Ewa L. {\L}okas}
\affiliation{Nicolaus Copernicus Astronomical Center, Polish Academy of Sciences, Bartycka 18, 00-716 Warsaw, Poland}

\author[0000-0002-6330-2394]{Cecilia Mateu}
\affiliation{Departamento de Astronom\'ia, Instituto de F\'isica, Universidad de la Rep\'ublica, Igu\'a 4225, CP 11400 Montevideo, Uruguay}

\author[0000-0001-8368-0221]{Sangmo Tony Sohn}
\affiliation{Space Telescope Science Institute, 3700 San Martin Drive, Baltimore, MD 21218, USA}

\begin{abstract}

We present a detailed study of the internal structure and kinematics of the core of the Sagittarius dwarf spheroidal galaxy (Sgr). Using machine-learning techniques, we have combined the information provided by 3300 RR Lyrae stars, more than 2000 spectroscopically observed stars, and the \Gaia second data release to derive the full phase space, i.e.\ 3D positions and kinematics, of more than $1.2\times10^5$ member stars in the core of the galaxy. Our results show that Sgr has a bar structure $\sim 2.5$ kpc long, and that tidal tails emerge from its tips to form what it is known as the Sgr stream. The main body of the galaxy, strongly sheared by tidal forces, is a triaxial (almost prolate) ellipsoid with its longest principal axis of inertia inclined $43\degree\pm6\degree$ with respect to the plane of the sky and axis ratios of 1:0.67:0.60. Its external regions are expanding mainly along its longest principal axis, yet the galaxy conserves an inner core of about $500\times330\times300$ pc that shows no net expansion and is rotating at $v_{\rm rot} = 4.13 \pm 0.16 \kms$. The internal angular momentum of Sgr forms an angle $\theta = 18\degree\pm6\degree$ with respect to its orbital angular momentum, meaning that the galaxy is in an inclined prograde orbit around the Milky Way. We compared our results with predictions from $N$-body models with spherical, pressure-supported progenitors and a model whose progenitor is a flattened rotating disk. Only the rotating model, based on preexisting simulations aimed at reproducing the line-of-sight velocity gradients observed in Sgr, was able to reproduce the observed properties in the core of the galaxy.

\end{abstract}




\section{Introduction}\label{sec:intro}

The dwarf spheroidal galaxy of Sagittarius (Sgr) is the nearest and most prominent example of ongoing galactic cannibalism in the entire sky \citep{Ibata1994}. The majority of its stars have been stripped from the main body due to tidal forces from the Milky Way (MW; \citealt{Niederste2010}), and span a $> \, 360^{\circ}$ great circle on the sky forming what is known as the Sgr stream \citep{Majewski2003,Law2016}. This structure is one of the best tracers of the MW's potential well that we have. In turn, Sgr is thought to shake the MW, inducing vertical displacements, kinematic distortions, and star formation in the MW disk \citep{Purcell2011, Antoja2018, Laporte2018, Ruiz-Lara2020}. The stars of the Sgr system show a wide dispersion in age and composition despite the current lack of gas, indicating a gas-rich progenitor galaxy heavily modified by the process of infall \citep[as reviewed in][]{Law2016}. Tidal forces also put their imprint on the elongated structure of the galaxy at the present day \citep{Ibata1997}, but the dwarf's structure still offers encoded clues to the primeval condition of the its progenitor.

While the low Galactic latitude, large spatial extension \citep[$r_h = 342 \pm 12$ arcmin][]{McConnachie2012}, and significant extinction towards the Sgr galaxy place obstacles in the way of its study, each new survey covering the Sgr footprint provides us with a clearer view of its stellar content and kinematic properties. Even so, numerous features of the Sgr system remain to be understood. There have been several attempts at modeling and reproducing the stream \citep{Law-Majewski2010, Gibbons2014, Dierickx-Loeb2017a, Fardal2019, Cunningham2020}. While all are generally consistent with the shape of the stream, many struggle to reproduce its kinematics and the bifurcation of the stream now observed in both the leading and the trailing arms \citep{Belokurov2006, Koposov2012, Ramos2020}. This feature, as well as other kinematic characteristics, could be naturally explained if the Sgr progenitor was an inclined rotating disk in infall into the MW potential well \citep{Penarrubia2010}. Sgr's morphology and kinematics would then had been eroded after several close passages through the MW halo resulting in its current spheroidal shape and stirred kinematics, although some residual rotation signal is expected to remain in the core of the galaxy \citep[][and references therein]{Mayer2010}.

Line-of-sight velocities ($\vlos$) have been derived for several thousand of Sgr's giant stars \citep{Ibata1997, Giuffrida2010, Penarrubia2011, Frinchaboy2012, McDonald2012}, a significant effort that has allowed comparisons with $N$-body model predictions. Sgr's core is being disrupted by MW's tidal forces, but both rotating and pressure-supported $N$-body models have reproduced well the observed $\vlos$ profiles \citep{Lokas2010, Frinchaboy2012}, leaving the rotation as an open question. Sgr's shape is also under discussion. Rotating models normally result in a prolate spheroid with tidally induced bars while pressure-supported models result in more spherical shapes. Different observation techniques, on the other hand, have led to mixed results between prolate \citep{Ibata1997} and triaxial \citep{Ferguson-Strigari2020} spheroids.

Recently the \Gaia second data release, (DR2; \citealt{GaiaDR2}), has given us access to the tangential component of the stellar velocities of its stars, or proper motions (PMs). This has increased current interest in Sgr, as shown in particular by two papers that have appeared while we were working on this particular one \citep{Ferguson-Strigari2020, Vasiliev-Belokurov2020}. The former uses RR Lyrae variable stars to analyze the spheroidal shape of the core of the galaxy, while the latter combines DR2 with spectroscopic $\vlos$ available in the literature to try to find the best $N$-body reproduction of the observed features of the Sgr core (though not the stream).

In this paper, we revisit the galaxy from a different perspective. We did not try to fit an $N$-body model to the data and learn from it. Instead, we used machine-learning (ML) techniques to combine three different data sets and predict the full phase space (3D positions and 3D velocities) of each individual star in the core of the galaxy. This allowed us to construct the full phase space of the core of the galaxy without using any physical prior or constraint, to study its internal dynamics in detail, and to compare the results with physically motivated $N$-body models.

The paper is organized as follows. In Section~\ref{sec:DATA} we present the data and preliminary membership selection. Section~\ref{sec:6D} explains the ML models, their predictions for distance and $\vlos$, and the final selection of member stars. Section~\ref{CMD} shows the color-magnitude diagram (CMD) of the selected member stars. Section~\ref{sec:Centerofmass} shows how the selected member stars are arranged in the PM-parallax space and the bulk properties of the galaxy as it orbits the MW. Section~\ref{sec:Comoving_coo} introduces a comoving coordinate system centered in Sgr's center of mass (COM). Section~\ref{sec:Results_6D} analyzes the internal dynamics of Sgr in 6D using different projections and coordinate systems. In Section~\ref{sec:Nbody_counterparts}, a comparison with three $N$-body models is shown. Our results are discussed in Section~\ref{sec:Discussion}. Finally, a summary and the main conclusions of the paper are presented in Section~\ref{sec:Conclusions}.

\section{First Data Processing}\label{sec:DATA}

We downloaded all of the stars from the \Gaia {\tt gaia\_source} table in a circular region of $5.7 \degree$ radius around Sgr central coordinates (1 half-light radius). Only stars with available colors and astrometric solutions were considered. We imposed parallax $\parallax$ and PM cuts in order to select stars compatible with the bulk and the internal dynamical properties of Sgr. In particular, we selected stars compatible within $3\sigma$ with the Sgr bulk dynamic properties reported by \citet{Helmi2018} and their respective $3\sigma$ errors and with the dispersion expected in the PMs resulting from the projected real intrinsic velocity dispersion observed in $\vlos$ of its stars \citep[11.4 $\kms$][]{McConnachie2012}. This selection was made taking into account \Gaia systematic errors calculated from equations 16 and 18 from \citet{Lindegren2018}. We also accounted for errors being 1.1 times larger than the ones listed in the {\tt gaia\_source} table, following DR2 verification papers' findings that the uncertainties may be underestimated by this factor \citep{Lindegren2018, Arenou2018}. Once the data were downloaded, we multiplied all statistical astrometric errors by this factor prior to any further analysis.

\begin{figure}
\begin{center}
\includegraphics[width=\linewidth]{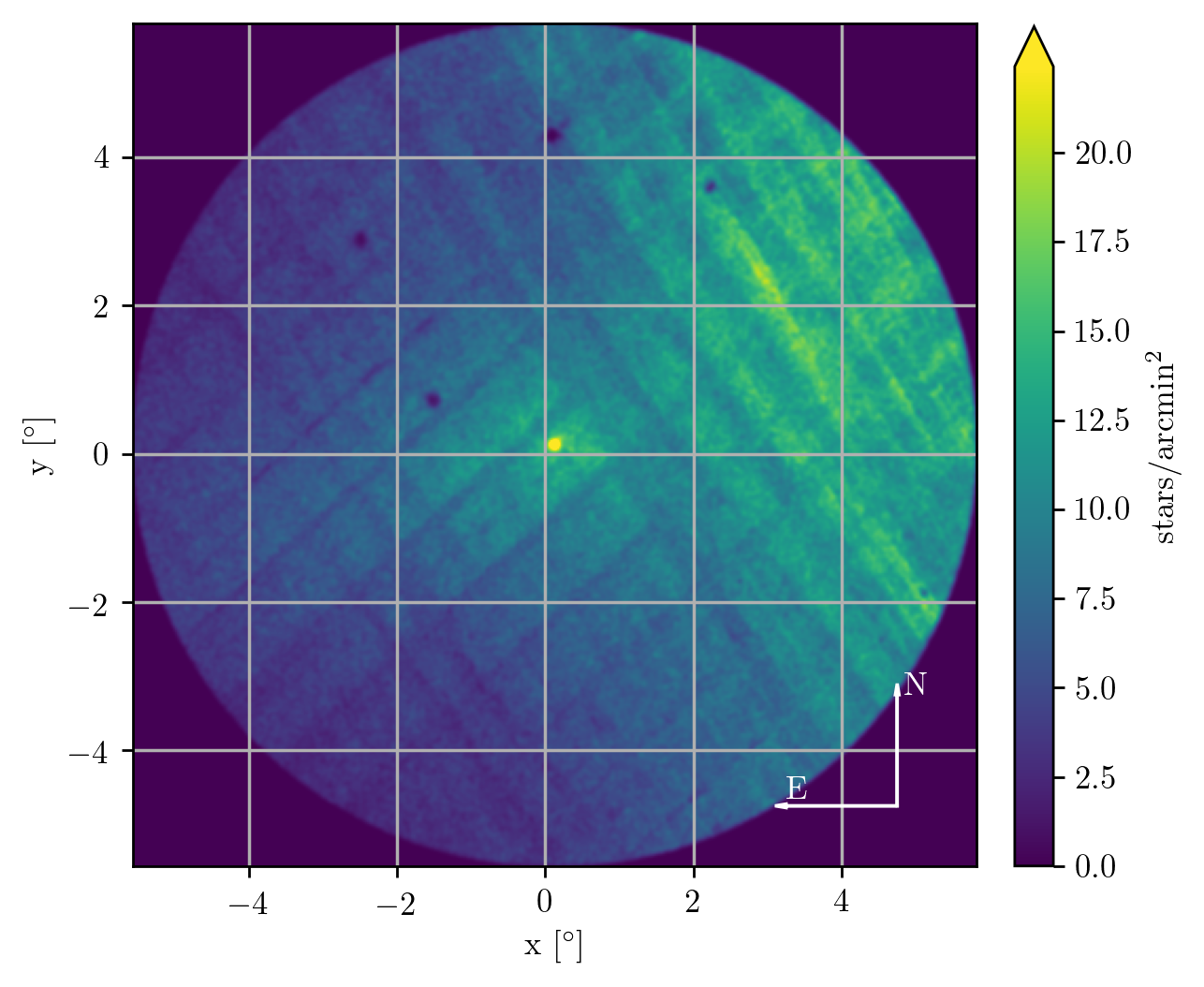}
\caption{Stellar surface density for all the stars passing our first quality and astrometric cuts. Coordinates are in degrees in the celestial sphere. Stripes are related to the \Gaia scanning pattern. At the center of the image, M54 is clearly visible,  whereas MW stars populate mostly the upper right regions.}
\label{fig:raw_data_sky}
\end{center}
\end{figure}

We proceeded to impose various quality cuts intended to screen out bad measurements. First, we removed sources with bad astrometric fits following Equation C.1 in \citet{Lindegren2018} and the GAIA-C3-TN-LU-LL-124\footnote{\url{http://www.rssd.esa.int/doc_fetch.php?id=3757412}} technical note: defining the renormalized unit weight error $\text{RUWE} \equiv (\mbox{\tt astrometric\_chi2\_al} /$  $\mbox{\tt astrometric\_n\_good\_obs\_al} - 5)^{1/2}/$ $u_{0}(\G,$ $\BR)$, we require $\text{RUWE} < 1.2 \times \max(1, \exp(-0.2(G-19.5)))$ and $\text{RUWE} < 1.4$. We used the $u_{0}(\G,\BR)$ reference value available on the ESA \Gaia DR2 Known issues page\footnote{\url{https://www.cosmos.esa.int/documents/29201/1769576/DR2_RUWE_V1.zip/d90f37a8-37c9-81ba-bf59-dd29d9b1438f}}. \secrevision{This selection removed approximately 2\% of the stars from the original list, leaving a sample of 2877172 stars.} Figure~\ref{fig:raw_data_sky} shows the distribution in the sky of the data fulfilling our quality cuts.

We corrected the photometry for distance modulus and reddening. A distance modulus of $(m-M)_{0} = 17.10\pm 0.15$ ($26\pm 2 \kpc$) was adopted for Sgr \citep{McConnachie2012}. The reddening correction was applied individually over each star by cross-correlating their positions with the dust maps of \citet{Schlafly-Finkbeiner2011} and using coefficients from \citet{Evans2018} for the conversion to the \Gaia bands.

We next consider a set of nested criteria for our membership selection based on PM, parallax and  position in the color-magnitude diagram (CMD). We start by selecting stars compatible within $3\sigma$ with a set of {\sc PARSEC} isochrones \citep{Bressan2012}, with ages from 5 to 13.5 Gyr and metallicities spanning [Fe/H] = $-0.4$ \citep{McConnachie2012} to [Fe/H] = $-2.5$. 

The selected stars are then observed in the PM-parallax space, where member stars are expected to cluster around the bulk PMs and parallax values of Sgr. A Gaussian model is fitted to the data in this space. Stars are scored based on their logarithmic-likelihood of belonging to the Gaussian distribution. Stars whose score, $\ell$, is lower than the median of all scores distribution minus $n$ times the sigma of that distribution are rejected. This is the logical expression: 

\begin{equation}\label{eq:log_cond}
\begin{aligned}
\ell \geq \tilde{\ell} - n \sigma(\ell)
\end{aligned}
\end{equation}

that stars must meet. The value that maximized the selection of Sgr stars while keeping MW's pollution low was $n=2$ (see Appendix~\ref{Apx:Choosing_best_sigmas}).

The fitting of the model is performed iteratively until self-convergence, refining the centroid of the distribution while rejecting stars not compatible with it. As a first guess for the parameters of the Gaussian during the first iteration we used the PMs and parallax provided in Table C.2. of \citet{Helmi2018}.

The aforementioned procedure guarantees that stars not compatible with the bulk PMs or parallax of Sgr are rejected, yet some MW stars may lie inside the distribution of selected stars in the PM-parallax space and thus be selected as possible members. Most of these contaminants, however, will show PMs and parallaxes not consistent with those from its surrounding neighbor stars, as they follow different trends in such quantities across the sky. In order to screen out most of these contaminants, we divided the sky into a grid of square cells of $1 \degree$ on a side and applied the procedure described in the paragraph above to each one of the cells individually, rejecting stars not fulfilling our logarithmic-likelihood condition. After convergence is reached in all cells, we introduce a shift in the grid position and repeat the procedure in the displaced cells until convergence. The shifts are introduced randomly in steps of $0.5 \degree$ in the west and north directions independently $\left[\Delta(x), \Delta(y)\right]$. After all shifts have been covered (four in total), the procedure starts again from $\left[\Delta(x), \Delta(y)\right] = (0, 0)$. The whole procedure converges when no further stars are rejected from any cell on any shift after three consecutive iterations. 

\section{Full Phase Space: ML Predictions}\label{sec:6D}

The full phase space for all stars is required in order to fully understand the internal structure and kinematics of the galaxy. Therefore, distances, $D$, and line-of-sight velocities, $\vlos$, must be derived and combined with PMs and sky coordinates. However, and despite its high astrometric performance, \Gaia parallax precision does not allow to derive distances for Sgr stars. Nor can the $\vlos$ be studied through \Gaia in detail; the tip of the Sgr red giant branch (RGB) is located at $G\sim14$, one magnitude below the nominal limiting magnitude with \Gaia $\vlos$ measurements, which results in only a handful of Sgr stars with $\vlos$ scattered along the sky.

To overcome this, we have made use of ML models to derive such quantities for each individual star in our sample. Generally speaking, a model, $f(\mathbf{x})$, is trained over a sample on the dependence of a variable $y_\mathrm{true}$ to a set of independent variables, $\mathbf{x}$, to predict such variable, $y_\mathrm{pred}$, from a different sample of $\mathbf{x}$.

\subsection{Training Sets}\label{EC}

We assembled two different catalogs that will be used later to train prediction models to determine $D$ and $\vlos$ for all stars from the {\tt gaia\_source} table. The $\vlos$ training table consists of 2427 stars with observed line-of-sight velocities, $v_{\mathrm{los, true}}$, either from {\tt gaia\_source}, the APOGEE survey, or one of the following publicly available catalogs: \citet{Ibata1997}, \citet{Giuffrida2010}, \citet{Frinchaboy2012}, and \citet{McDonald2012}. Common stars between the catalogs show consistent measurements in most cases. For stars with less than $ 10 \kms$ of standard deviation between catalogs, we adopted the error-weighted average of their $v_{\mathrm{los, true}}$ as the final value. Stars with larger standard deviation values were rejected.

We composed the distance training table from more than 4000 RR Lyraes stars around Sgr center from the SOS and VariClassifier catalogs published by DPAC making use of \Gaia DR2 \citep{Clementini2019, Holl2019, Rimoldini2019}. Contaminants were removed following the method described in \citet{Rimoldini2019}. Their observed distances, $D_{\mathrm{true}}$, are derived from their pulsation modes, therefore making them independent from the astrometric properties of the star. We estimated an average error of 1.16 kpc in $D_{\mathrm{true}}$ from M54 member stars that accounts for both observational and calibration effects (see Appendix~\ref{Apx:Error_distance}). Details on how the joint SOS and VariClassifier catalog was assembled and the computation of the RR Lyrae distances are available in \citet{Mateu2020}.

\secrevision{Undesired systematic bias effects can occur when the training sample does not fully cover the entire space defined by $\mathbf{x}$ from which later predictions will be made. To avoid such problems, we initially considered a wider PM and $\parallax$ range, as well as a wider region in the sky for the training lists than for the main catalog. Specifically, we selected all stars within $7.7 \degree$ of the central coordinates of Sgr with PMs and $\parallax$ compatible with Sgr by $\pm1\masyr$ and $\pm0.5$ plus 3$\sigma$ mas, respectively.} We then cleaned the two catalogs of contaminants and poorly measured stars following the exact same criteria as for the {\tt gaia\_source} data, ensuring that the three catalogs are subsamples representative of the same set of stars. Up to 3422 RR Lyrae and 2310 stars remained in the distance and line-of-sight velocity training sets, respectively, after the \secrevision{first} membership selection. \secrevision{As a last step before fitting the models to the training lists, we standardized all variables in $\mathbf{x}$ by subtracting their averages and dividing them by their standard deviation.}

\subsection{Predicting Velocities and Distances}\label{ML}

The two training sets were used to create models that predict distances and line-of-sight velocities on the {\tt gaia\_source} table.

To obtain $D$ we used a $k$-neighbors regressor (KNR) model. The KNR models predict $y_\mathrm{pred}$ by local interpolation of the $y_\mathrm{true}$ of the $k$ nearest neighbors in the parameter space defined by $\mathbf{x}$ in the training set. This interpolation is, in our case, weighted by the inverse of the euclidean distance to the target star in the parameter space defined by $\mathbf{x}$. The KNR models are highly scalable and perform fast on large and well-distributed samples. The procedure is as follows.

The KNR model is fitted to the training distances, $y_\mathrm{true}$, using sky coordinates and a set of astrometric magnitudes as $\mathbf{x}$. The number of neighbors, $k$, and best combination of independent variables in $\mathbf{x}$ \revision{are determined through a grid search within a nested cross-validation (NCV)} of the training set minimizing a) the variance in the differences between the predicted and the true values for the distance (generalization error) and b) possible artificial gradients along the sky introduced by the model \secrevision{(see Appendix \ref{Apx:Models_and_Caveats})}. The \secrevision{validated} model is then used to predict distances on the {\tt gaia\_source} based on the same independent variables.

Choosing a KNR over other models has the advantage that the model regularizes (flattens) $y_\mathrm{true}$ \secrevision{when $k > 1$ is used}, avoiding overfitting the data. The relatively large errors observed in our distances ($1.16 \kpc$) and the large full width at half maximum (FWHM) of their distribution ($3.69 \kpc$) compared to previous works \citep{Hamanowicz2016} indicates that regularization and averaging of the training sample are required prior to deriving distances for the main table (for more information about the errors in the training list and the KNR performance, see Appendix~\ref{Apx:Distances}).

The procedure to derive $\vlos$ is similar, but due to the sparse nature of the data (their distribution in the sky is far from uniform) \secrevision{and to their nonlinearity}, we decided to use a stacked regressor (SR) \citep{Wolpert1992, Breiman1996}. The SR models take advantage of different learners by stacking their output through a final regressor that computes the final prediction. They perform much slower than each of the estimators separately, but their results are at least as good as the best of the stacked estimators while improving the results in most cases. We combined six learners\revision{, three nonlinear (extra trees, support vector machine with a radial basis function, and Gaussian process) and three linear (lasso, support vector machine with a linear kernel, and Elastic Net)}, and used a neural network to combine their outputs. \revision{The SR was trained through a out-of-fold scheme: in $k=5$ successive iterations, $k$-1 folds are used to fit the first layer of regressors to make predictions on the remaining subset. The predictions are then stacked and provided as input to the final metaregressor. The optimal metaparameters of the six learners inside our SR model, as well as the best combination of independent variables in $\mathbf{x}$, are optimized through a grid search over the metaparameters space within an NCV.} The resulting model was able to reproduce nonlinear features while showing stability against high-order oscillations on areas where no information is available, such as the outer regions of the galaxy, where the data density is much lower. (For more information about the SR architecture and performance, see Appendix~\ref{Apx:Line-of-sight_velocities}).

The procedure to derive $y_\mathrm{pred}$ \secrevision{and its associated error} is done in a Monte Carlo fashion, considering and propagating the errors from all independent and dependent variables. A total of $10^6$ realizations were performed, randomly sampling each variable from a normal distribution centered on its corresponding nominal value. The predicted values using the nominal values of the independent variables are assumed to be the nominal values of the final result. The quadrature addition of the standard deviation of all realizations plus the \revision{square root of the generalization error derived from the NCV test} is adopted as the total error.

Stars in the main catalog and the two training sets are finally selected based on their full velocity space, i.e.\ 3D velocities, through a 3D Gaussian using the same rejection criteria adopted in Section \ref{sec:DATA}. No rejection was imposed on the spatial coordinates. The whole procedure is repeated until no more stars are rejected and the solution converges. \secrevision{The final lists contain 3305, 1992, and 120168 stars for the RR Lyrae, the $\vlos$, and the final clean sample, respectively.}

Values for $k$ and the best combination of independent variables, $\mathbf{x}$, vary a bit from one iteration to another for the KNR. A larger $k$ produces more stable results but at the cost of flattening the results too much, artificially removing gradients from the data. Our criteria seem to be a good compromise, minimizing such changes in distance gradients and keeping low deviations from the true values. The best results were obtained using $k$ between 10 and 30, and $\mathbf{x} = (x, y, \pmx, \pmy, \parallax)$, where $x, y, \pmx$ and $\pmy$ are the orthographic projections of the usual celestial coordinates and PMs (see Appendix~\ref{Apx:Distance_models}). The optimal metaparameters of the SR model, as well as the best $\mathbf{x}$, also vary between iterations yet producing fully compatible results. See Appendix~\ref{Apx:Models_and_Caveats} for more information.

\subsection{Final Distances and Line-of-sight Velocities}\label{sec:final_D_v}

\revision{After convergence, we evaluated the performance of both models by comparing $y_\mathrm{true}$ and $y_\mathrm{pred}$ from the NCV tests on each corresponding final training list. The NCV procedure uses a series of train, validation, and test set splits to effectively optimize and test the model over different data sets, thus allowing an independent assessment of the models' ability to predict on new data.}

\begin{figure}
\begin{center}
\includegraphics[width=\linewidth]{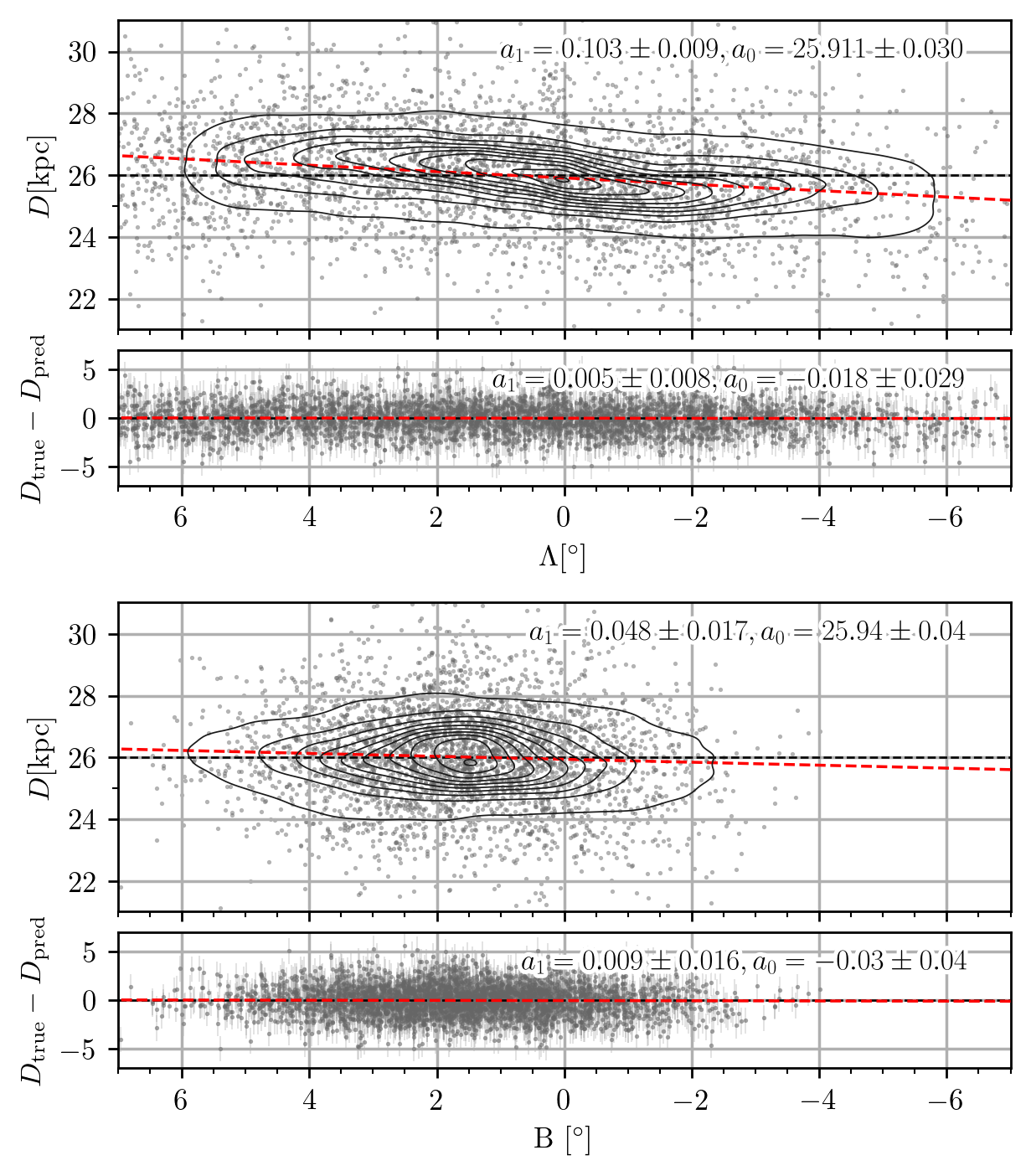}
\caption{Predicted distances for our final sample of stars on the stream coordinate system, $(\Lambda, \Beta)$, shown in the top and bottom panel, respectively \citep{Law-Majewski2010}. Contours show stellar density for the predicted distances, $y_\mathrm{pred}$ \revision{on the clean sample}. Points show the location of RR Lyrae stars selected as Sgr members in our training set. The average distance measured on the training sample is marked by a black horizontal line. \revision{Residuals measured from the NCV on the training list are shown as $D_\mathrm{true} - D_\mathrm{pred}$ along both coordinates. Red dashed lines show a linear fit to the training data and to the residuals. The parameters of the fits are shown in the top right part of each plot.} For clarity, only the error bars of the residuals are shown. The typical error for the RR Lyrae stars distances is $\sim 1.16 \kpc$ (See Appendix~\ref{Apx:Error_distance}).}
\label{fig:distance_fitting}
\end{center}
\end{figure}

\revision{Figure~\ref{fig:distance_fitting} shows the predicted distances, $D$, for our final clean sample and the residuals from the NCV test, $D_\mathrm{true}$ - $D_\mathrm{pred}$, on the coordinate system $(\Lambda, \Beta)$, defined along the stream \citep{Law-Majewski2010}.} Here $\Lambda$ almost coincides with the optical major axis of the system (as projected on the sky) and $\Beta$ with the minor one (slightly offset with respect to the center of the galaxy). A linear fit to the training set, $y_\mathrm{true}$, has been added to shown the general tendency: the western regions of the galaxy (negative $\Lambda$) are, on average, closer to the Sun than the eastern parts (positive $\Lambda$). A less obvious trend can also be seen along $\Beta$, with negative $\Beta$ closer to us.

The agreement between the predicted and true distances is very good, with both distributions showing the same average behavior \secrevision{and a constant variance between $D_\mathrm{true}$ and $D_\mathrm{pred}$} along both axes. Some departures from the linear model (red dashed line) can be easily spotted in \revision{our predicted distances for the clean sample (contours)}, especially along $\Lambda$. These departures are also present on the training sample and point to a real twist in the distance along the line of sight. The KNR does a good job regularizing the distances, with a line-of-sight thickness FWHM $\sim 2.2 \kpc$ for horizontal-branch (HB) stars \secrevision{around the instability strip}. This is closer to the 2.42 kpc found by the OGLE team \citep{Hamanowicz2016} and to our real FWHM estimation of 2.48 kpc than the FWHM measured directly over our RR Lyrae sample ($3.69 \kpc$). These are not worrisome differences to our purposes of analyzing the general distance trends in Sgr, further taking into account that a number of factors could be affecting the measured thickness along the line of sight, such as the number of stars used, metallicity calibration, etc. See Appendix~\ref{Apx:Error_distance} for more information.

\begin{figure}
\begin{center}
\includegraphics[width=\linewidth]{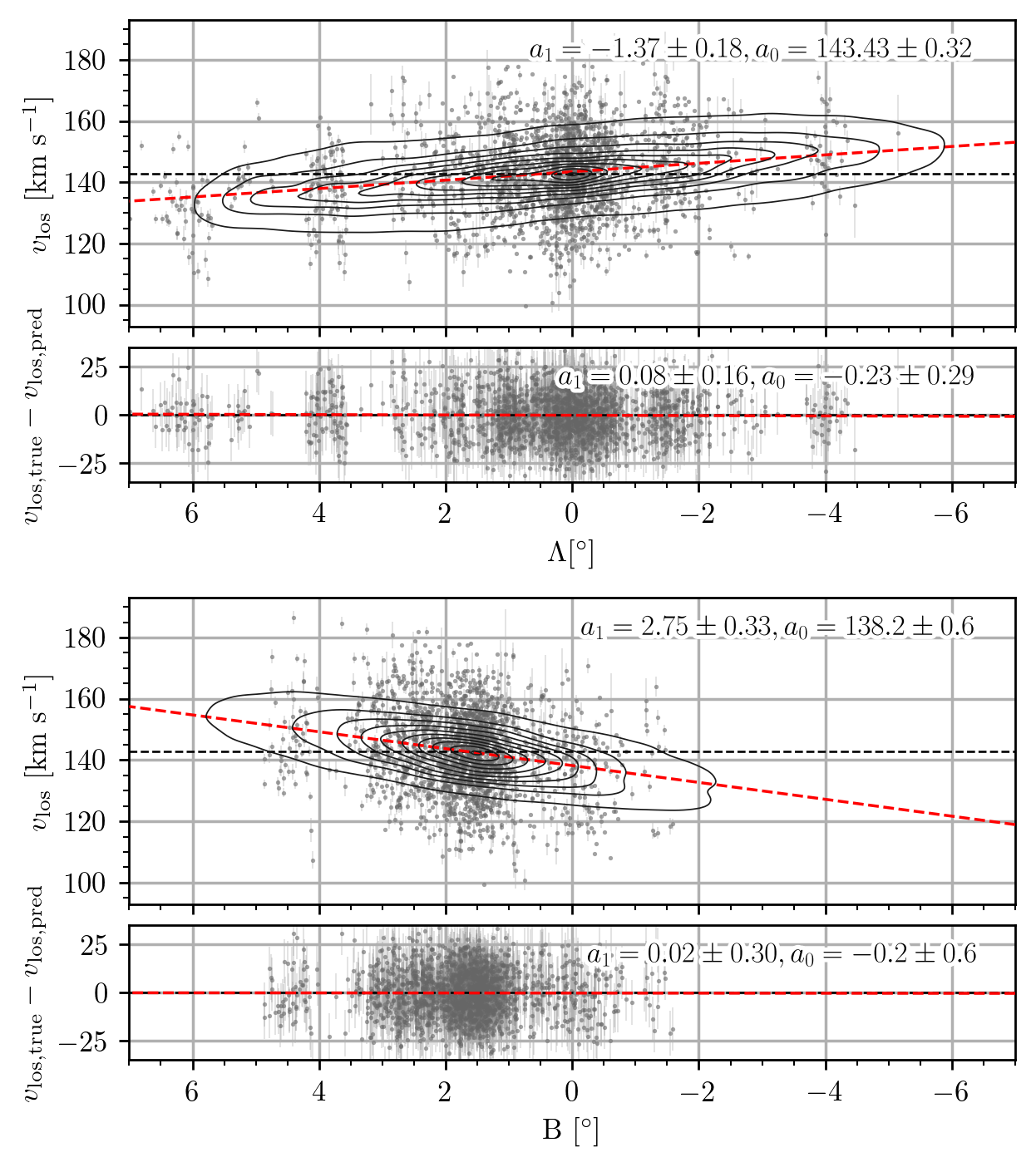}
\caption{Same as Figure~\ref{fig:distance_fitting} for $\vlos$. Error bars show the error in the measured $\vlos$.}
\label{fig:vlos_fitting}
\end{center}
\end{figure}

Figure~\ref{fig:vlos_fitting} shows the predicted and true $\vlos$ for the clean final sample and the training sample, respectively. Again, some departure from the linear fit can be observed in both samples, especially in the very central regions ($|\Lambda| < 1 \degree$), and the most external ones ($|\Lambda| > 4 \degree$). These changes roughly coincide with the points at which changes in distance are also present, indicating the presence of tidal tails with a slightly different kinematics than the core of the galaxy.

Interestingly, the standard deviation of the differences between the predicted $\vlos$ and the real one was found to be $\sigma(\vlos) = 11.7 \kms$, just above the internal velocity dispersion of the galaxy. This, combined with the small observational errors in the training sample and the fact that the spatial distribution of this quantity along the sky is very similar to the one from intrinsic dispersion, suggests that the model fails to reproduce the internal $\vlos$ dispersion of Sgr but reproduces well the general trends in the galaxy. This comes as no surprise, since the model has been regularized to avoid overfitting, which also limits its capacity to fit random velocities. We take that into account during our subsequent Monte Carlo experiments by allowing the stars to have a $\vlos$ in a range that includes this dispersion. For distances, differences were of the order of the error in distance for individual RR Lyraes. In this case, the largest differences were observed for the most extreme cases, i.e.\ stars with the largest or smallest distances in the training sample, indicating that it is a systematic effect caused by the regularization of the predicted distances, rather than random errors. \revision{More information can be found in Appendix~\ref{Apx:Line-of-sight_velocities}.}

\subsection{Possible MW Contaminants}\label{Contaminants}

\begin{figure*}
\includegraphics[width=\textwidth]{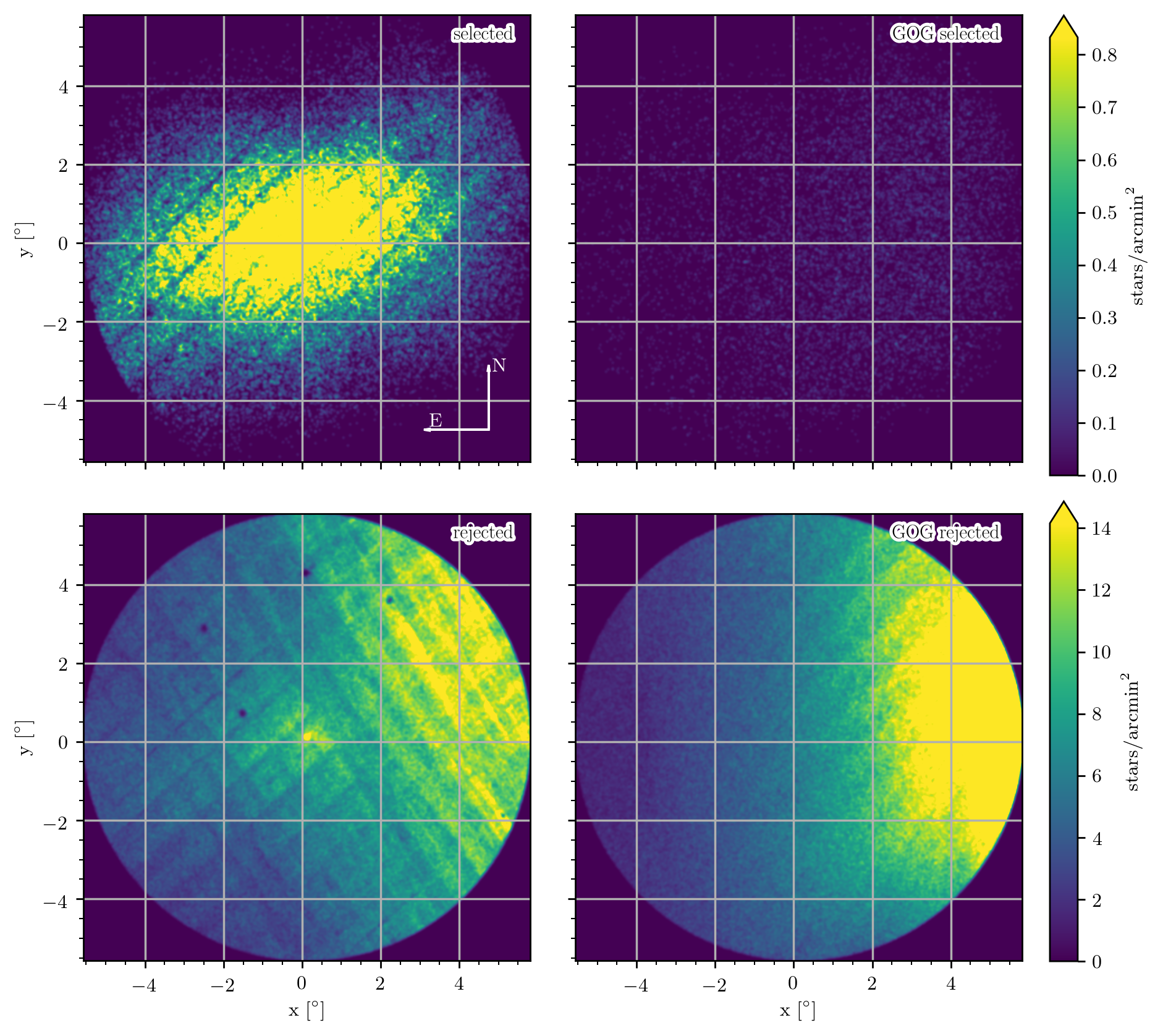}
\caption{Stellar surface density of Sgr selected members (top left), possible MW stars polluting the sample inferred from the GOG model (top right), rejected contaminants \secrevision{(bottom left)}, rejected contaminants from the GOG model \secrevision{(bottom right)}. Each row of panels shares the same color shading. The color shading for the rejected stars has been divided by 20 with respect to the top panels. Stripes crossing the data on a diagonal are due to \Gaia systematic errors.}
\label{fig:members_Sky}
\end{figure*}

Despite all of our efforts to clean up our sample from nonmember stars, some MW contaminants may have passed through our screening process. To assess this contamination rate, we compared our member stars list with an equivalent sample from the Besancon 2016 model of the Galaxy using the \Gaia Object Generator (GOG) model \citep{Luri2014}. We simulated DR2 errors in the GOG model and performed the exact same selections as with real data (see Appendix~\ref{Apx:Contaminants} for further information). This resulted in $\sim7500$ GOG stars passing through our membership selection method, which indicates a $\sim6\%$ of possible contamination from MW stars in our final sample. Figure~\ref{fig:members_Sky} shows the final distribution in the sky of the stars selected as members and stars rejected for the {\tt gaia\_source} table and for the GOG model. 

The stars from the GOG model that have passed our selection criteria are stars whose phase space, color, and magnitude are indistinguishable from Sgr member stars, not only in bulk, but by sectors along the sky. Their contribution to our final measurements is well below the $1\sigma$ level of uncertainty of these; thus, we do not expect any critical impact from their possible presence in our results. It is worth noting that some probable Sgr members are rejected (overdensity in the center area of the bottom left panel in Figure~\ref{fig:members_Sky}). These are mostly stars with large astrometric and photometric errors. Their distribution in the sky follows the stellar density profile of Sgr, and their PMs randomly scatter in the vicinity of the cluster of members in Figure~\ref{fig:members_pmparallax}. These stars are expected to have a null contribution to the bulk dynamical properties of the galaxy \secrevision{ given their small number and the fact that their astrometric properties are indistinguishable from those of Sgr in bulk and their surrounding neighbor stars within $1\degree$ (see Section~\ref{sec:DATA}).}

\section{Color-Magnitude Diagram}\label{CMD}

\begin{figure}
\begin{center}
\includegraphics[width=\linewidth]{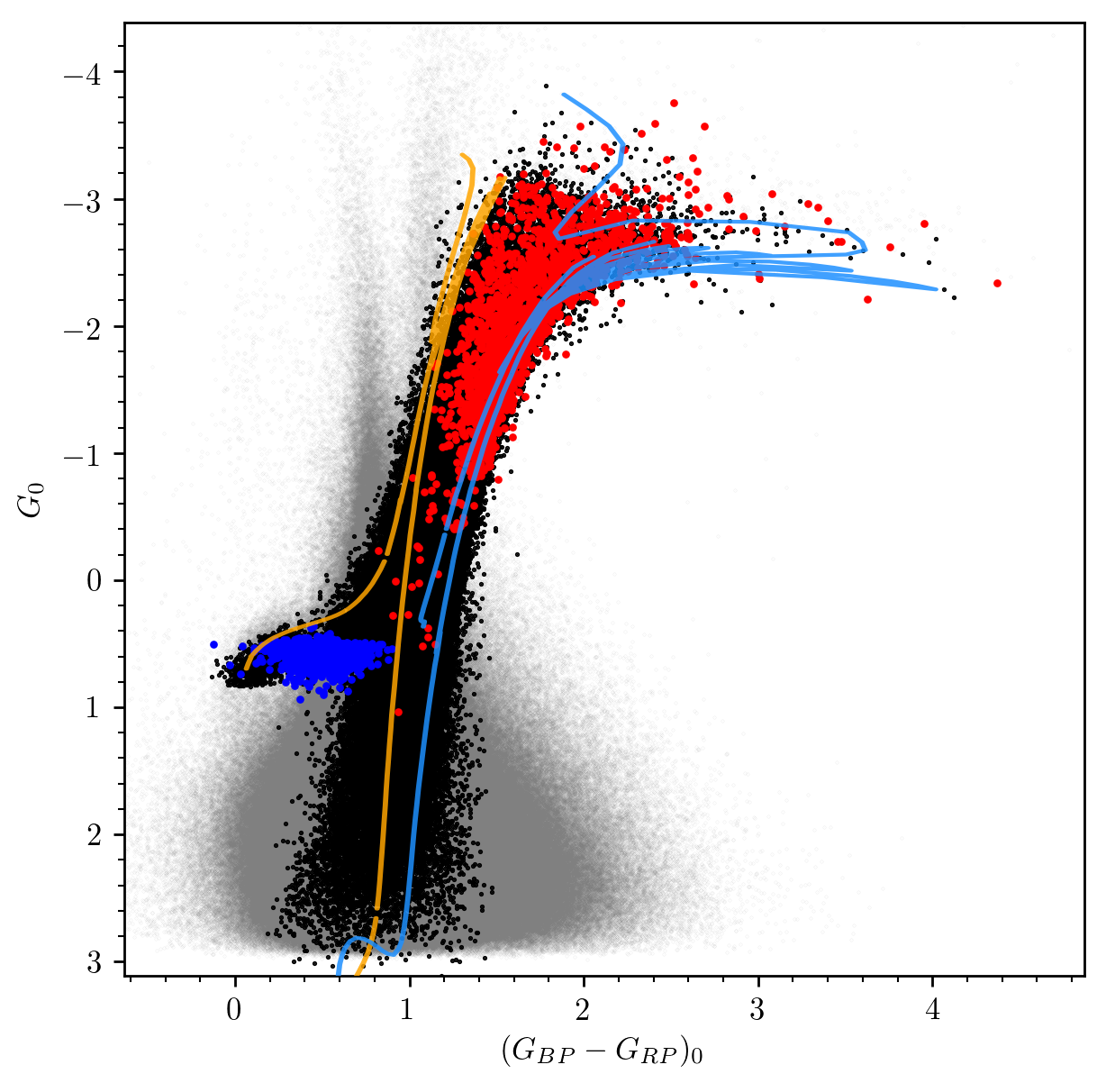}
\caption{Calibrated CMD of Sgr. Black dots are Sgr selected members. Stars screened out as contaminants are shown in gray. The RR Lyrae and spectroscopically observed stars selected as Sgr members are shown in blue and red, respectively. Two isochrones from the PARSEC stellar evolution library have been drawn over the CMD: [Fe/H] = $-$3.3, 13.5 Gyr (orange) and [Fe/H] = $-$0.4, 5 Gyr (light blue).}
\label{fig:members_CMD}
\end{center}
\end{figure}

The distance-calibrated and reddening-corrected CMD of Sgr is shown in Figure~\ref{fig:members_CMD}. A prominent RGB climbs the CMD with its tip at $G_0 \sim -3.2$. Together with a conspicuous HB, visible at $G_0 \sim 0.5$, this shows the presence of mixed stellar populations with ages spanning intermediate to old and metallicities spanning [Fe/H] = $-0.5$ to $<-2$, consistent with our understanding of the Sgr populations. The MW stars are distributed mostly in the range $0 < (G_{BP} - G_{RP})_0 < 2$, forming two vertical plumes.

Stars from both training samples are also included in the CMD. Member RR Lyrae stars from the distance training sample (blue dots) were calibrated using their true distance ($y_\mathrm{true}$). Stars from the $\vlos$ training sample (red dots), on the other hand, were calibrated using their predicted distances ($y_\mathrm{pred}$). The agreement between the three samples is very good, e.g., the location of tip of the RGB coincides for the main sample and for the $\vlos$ training sample, and the RR Lyrae stars lie in the HB, as expected. The dispersion observed for the RR Lyrae stars in $G_0$ is due to their brightness variability and the different reddening maps used to calibrate their distance. These errors are averaged out by the KNR model in the main sample.

\section{Sagittarius Bulk Dynamics}\label{sec:Centerofmass}

\subsection{Sagittarius in the Sky}\label{sec:Sag_in_Sky}

\begin{figure}
\begin{center}
\includegraphics[width=\linewidth]{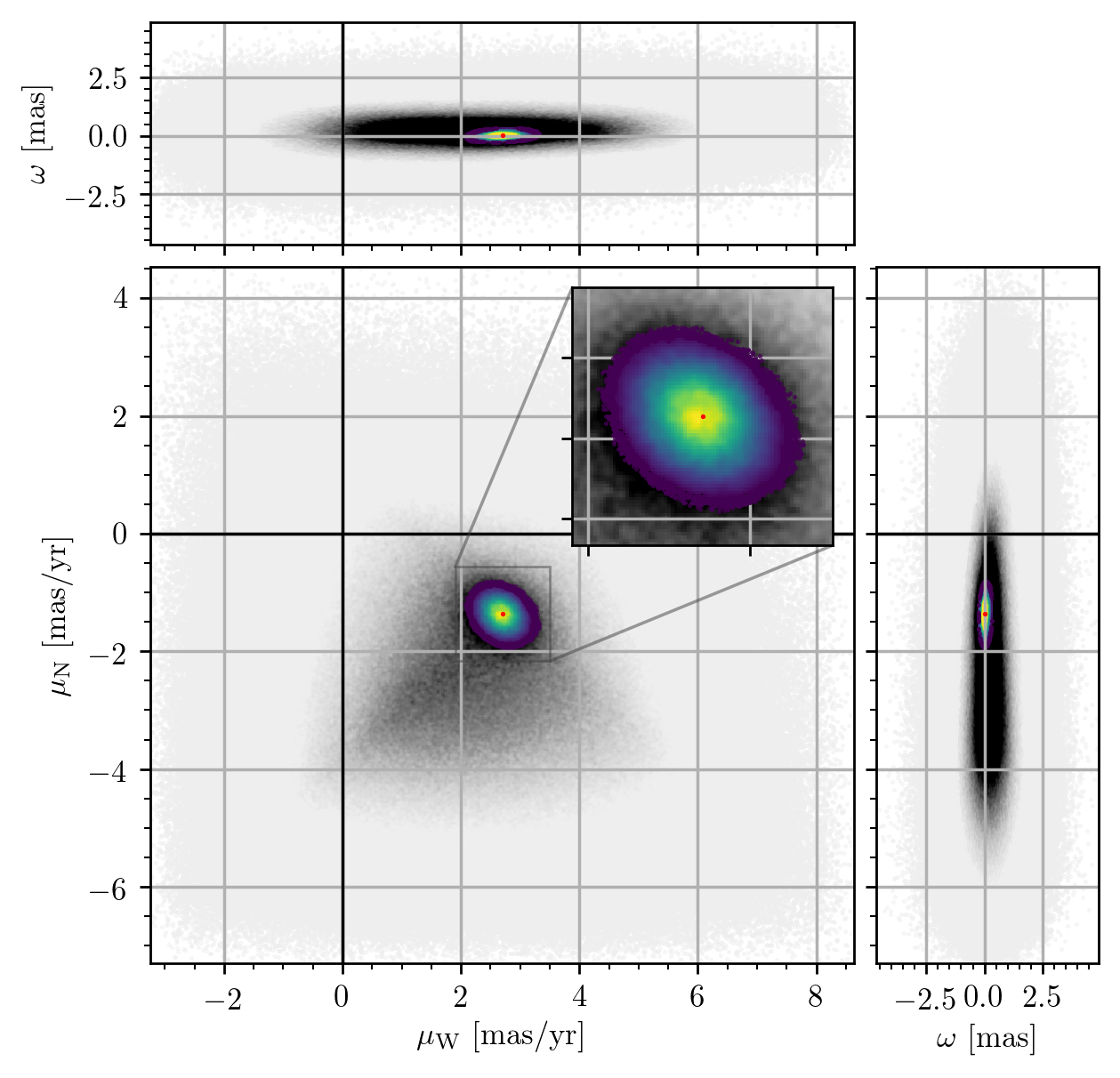}
\caption{Distribution of stars in the PM-parallax space. Member stars ($1.2\times10^5$) are shown by black dots. Gray dots shown stars ruled out as members ($2.74\times10^6$). A red point with error bars (not visible due to their small size) shows the location of the COM.}
\label{fig:members_pmparallax}
\end{center}
\end{figure}

The final selection of member stars is shown in the PM-parallax space in Figure~\ref{fig:members_pmparallax}. The set of stars has its center at $(\pmra, \pmdec, \parallax) = (-2.6924 \pm 0.0006, -1.3713 \pm 0.0005, 0.0176 \pm 0.0003)$, with some correlation between parallax and PMs, which makes the simultaneous fitting of the three variables necessary in order to maximize the members/nonmembers ratio. Our algorithm does a good job rejecting MW stars ($\sim 2.8\times10^6$), while keeping a high fraction of Sgr stars in the sample ($\sim 1.2 \times10^5$).

\begin{deluxetable*}{ccccccc}\label{tab:finalpops_sky}
\tablenum{1}
\tablecaption{Sgr's COM astrometric properties in sky coordinates.}
\tablewidth{0pt}
\tablehead{
    & $\alpha$ & $\delta$ & D & $\pmra$ & $\pmdec$ & $\vlos$ \\
    & (deg) & (deg) & (kpc) & ($\masyr$) & ($\masyr$) & ($\kms$)
}
\startdata
Sag with M54 & $283.909 \pm 0.016$ & $-30.599 \pm 0.007$ & $25.933 \pm 0.006$ & $-2.7049 \pm 0.0027$ & $-1.3648 \pm 0.0025$ & $142.71 \pm 0.08$ \\
 & & & & $(-2.73 \pm 0.05)$ & $(-1.45 \pm 0.05)$ & \\
Sag without M54 & $283.945 \pm 0.019$ & $-30.647 \pm 0.009$ & $25.951 \pm 0.006$ & $-2.6808 \pm 0.0026$ & $-1.3344 \pm 0.0025$ & $141.61 \pm 0.09$ \\
 & & & & $(-2.71 \pm 0.05)$ & $(-1.42 \pm 0.05)$ & \\
M54 & $283.781 \pm 0.009$ & $-30.468 \pm 0.005$ & $25.90 \pm 0.06$ & $-2.722 \pm 0.035$ & $-1.367 \pm 0.033$ & $142.7 \pm 0.4$ \\
 & & & & $(-2.75 \pm 0.06)$ & $(-1.46 \pm 0.06)$ & \\
\enddata
\tablecomments{The first row shows values for the galaxy including M54 stars. The second row shows values after removing M54 stars. Last row shows values for M54. Values within parenthesis show values corrected from large-scale systematic errors (See Appendix~\ref{Apx:Systematic_Errors}).}
\end{deluxetable*}

Our membership selection method iteratively finds the centroid of all stars within $2\sigma$ of a 6D Gaussian fitted to their galactocentric Cartesian phase space $(X, Y, Z, v_X, v_Y, v_Z)$ as the error-weighted average of such coordinates. The resulting centroid is assumed to be the COM of Sgr and its coordinates are then transformed to sky coordinates (R.A., decl., $D$, $\pmra$, $\pmdec$, $\vlos$). Such a transformation is done assuming the distance of the Sun from the Galactic center and the circular velocity of the local standard of rest (LSR) to be $R_0 = 8.29 \pm 0.16$ kpc and $V_0 = 239 \pm 5 \kms$, respectively \citep{McMillan2011}. The solar peculiar velocity with respect to the LSR was taken from the estimates of \citet{Schonrich2010}: $(U_{\rm pec},V_{\rm pec},W_{\rm pec}) = (11.10, 12.24, 7.25) \kms$ with uncertainties of $(1.23, 2.05, 0.62) \kms$.

The M54 stars could be shifting the centroid of Sgr toward the center of the globular cluster. In order to study this influence, we recalculated the COM for our table upon removal of all the stars around M54 up to its tidal radius. We assumed $r_t = r_c 10^c$, where $c = 2.04$ and $r_c = 0^\prime.09$ are the concentration and the core radius of the cluster \citep[][2010 edition]{Harris1996}. This provided a list of 1439 stars, from which we selected 352 very likely members of M54 based on their position on the CMD.

In Table~\ref{tab:finalpops_sky} we list the sky coordinates and velocities for the COM of all member stars from the last iteration of our selection method. The first row shows the COM for the whole sample, including M54 stars. The second row shows the COM after removing M54 stars within its tidal radius. The third row shows the COM for M54 based on our 352 stars selected. The uncertainties listed here were obtained from a Monte Carlo scheme that propagates all observational uncertainties and their correlations, including those for the Sun and possible systematic errors (see Appendix~\ref{Apx:Errors}).

The presence of M54 stars seems to have a negligible influence on the derived COM properties of Sgr, except for its coordinates in the sky, $(\alpha, \delta)$. Such a difference is not surprising; M54 stars are highly concentrated on the sky compared to the rest of the Sgr stars and therefore have a high weight when deriving the Sgr centroid in the sky.

The large difference in coordinates between Sgr and M54 ($\Delta \alpha = 16.9 \pm 2.2 ^\prime, \Delta \delta = 10.8\pm0.6 ^\prime$) is also expected given the different areas considered. The systematics known to be affecting our sample produce a nonuniform surface stellar density along the observed region, thus affecting the centroid. If we focus on the rest of the properties, Sgr and M54 seem to be indistinguishable within the errors. Furthermore, all M54 properties are compatible with those derived for Sgr after removing M54 stars, indicating that both stellar populations share position and dynamics. We will, therefore, consider M54 as part of Sgr regarding astrometry in any respect.

We also studied the effect of large-scale systematics in the PMs of Sgr using quasars (see Appendix~\ref{Apx:Systematic_Errors}). A total of 217 quasars were used to derive the median zero-points in PMs in the field of Sgr. We obtained $\tilde{\mu_{\alpha\star}} = 0.03\pm0.05 \masyr$ and $\tilde{\mu_{\delta}} = 0.09\pm0.05 \masyr$, which indicate that Sgr's motion may be slower toward the north direction. The corrected PMs of Sgr and M54 are listed in Table~\ref{tab:finalpops_sky} in parentheses.

\subsection{Sagittarius around the MW}

\begin{deluxetable*}{ccccccccc}\label{tab:finalpops_gc}
\tablenum{2}
\tablecaption{Same as Table~\ref{tab:finalpops_sky}, but for galactocentric coordinates.}
\tablewidth{0pt}
\tablehead{
    & $X$ & $Y$ & $Z$ & $v_X$ & $v_Y$ & $v_Z$ & $v_R$ & $v_T$\\
    & (kpc) & (kpc) & (kpc) & ($\kms$) & ($\kms$) & ($\kms$) & ($\kms$) & ($\kms$)
}
\startdata
Sgr with M54 & $16.73 \pm 0.16$ & $2.4289 \pm 0.0028$ & $-6.383 \pm 0.007$ & $234.9 \pm 1.3$ & $-14.2 \pm 0.7$ & $203.6 \pm 0.7$ & $143.6 \pm 1.7$ & $276.0 \pm 1.2$\\
             &       &      &      & $(236.3 \pm 2.1)$ & $(-45 \pm 8)$ & $(202 \pm 6)$ & $(140.6 \pm 2.3)$ & $(280 \pm 6)$\\
Sgr without M54 &$16.74 \pm 0.16$ & $2.4157 \pm 0.0035$ & $-6.408 \pm 0.008$ & $233.2 \pm 1.2$ & $-10.0 \pm 0.7$ & $202.8 \pm 0.7$ & $142.6 \pm 1.6$ & $274.3 \pm 1.1$\\
 & & & & $(234.6 \pm 2.1)$ & $(-40 \pm 8)$ & $(201 \pm 6)$ & $(139.7 \pm 2.3)$ & $(278 \pm 6)$\\
M54 & $16.71 \pm 0.17$ & $2.462 \pm 0.006$ & $-6.308 \pm 0.015$ & $235.1 \pm 1.7$ & $-15 \pm 4$ & $205 \pm 4$ & $144.1 \pm 1.8$ & $277 \pm 4$\\
 & & & & $(236.5 \pm 2.4)$ & $(-45 \pm 9)$ & $(203 \pm 7)$ & $(141.0 \pm 2.4)$ & $(282 \pm 7)$\\
\enddata
\end{deluxetable*}

Table~\ref{tab:finalpops_gc} lists the galactocentric positions and velocities for Sgr. The core of Sgr is at a distance of $18.11 \pm 0.15 \kpc$ from the Galactic center, almost at the opposite side from the Sun location and $6.383\pm0.007 \kpc$ below the galactic plane.

Currently, Sgr is on its way to cross the Galactic plane with a net velocity of $311.1 \pm 1.1 \kms$ ($313 \pm 4$ if we correct from large-scale systematics) with respect to the Galactic center. Most of its velocity is tangential, $V_T = 276.0\pm1.2 \kms$ ($280 \pm 6 \kms$), while Sgr currently moves away from the Galactic center with a radial velocity of $143.6 \pm 1.7 \kms$ ($140 \pm 2$). In general, its motion is well aligned with the Sgr stream, which lies close to the $X$--$Z$ plane of the MW (misaligned by $\sim 15^\circ$). The velocity and position of Sgr indicate that it must have passed through its orbit pericenter very recently, and it is now receding with respect to the MW center. 

\section{Comoving Reference Frame}\label{sec:Comoving_coo}

\subsection{Polar Coordinates}\label{sec:Polar_coo}

The position of any point in Sgr can be uniquely determined by its R.A. and decl. 
on the sky, $(\alpha, \delta)$, and its distance, $D$. To simplify our analysis and have a better interpretation of our results in terms of internal kinematics, we introduced a new reference system centered on the galaxy COM with coordinates $(\alpha_0, \delta_0, D_0)$. Angular coordinates $(\rho, \phi)$ are defined on the celestial sphere, where $\rho$ is the is the angular distance between the points $(\alpha, \delta)$ and $(\alpha_0, \delta_0)$ and $\phi$ is the position angle (PA) of the point $(\alpha, \delta)$ with respect to $(\alpha_0, \delta_0)$. The angle $\phi$ is measured counterclockwise starting from the axis that runs in the direction of decreasing R.A. at constant decl. $\delta_0$ \citep[see Fig 1 from][]{Marel2002}. Equations (1)-(3) from \citet{Marel-Cioni2001} allow $(\rho, \phi)$ to be calculated from any $(\alpha, \delta)$:

\begin{equation}\label{eq:xy}
\begin{aligned}
\rho & = \arccos(\cos{\delta}\cos{\delta_0}\cos{(\alpha - \alpha_0)} +\sin{\delta}\sin{\delta_0}) \\
\phi & = \arctan2\left(\frac{\sin\delta\cos\delta_{0}-\cos\delta\sin\delta_{0}\cos(\alpha-\alpha_{0})}{-\cos\delta\sin(\alpha-\alpha_{0})}\right).
\end{aligned}
\end{equation}  

The velocity vector at any position $(\rho, \phi, D)$ can be decomposed into three orthogonal components:

\begin{equation}\label{eq:v1v2v3}
\begin{aligned}
v_1 \equiv \frac{dD}{dt}, v_2 \equiv D\frac{d\rho}{dt}, v_3 \equiv D\sin{\rho}\frac{d\phi}{dt}
\end{aligned}
\end{equation}  

where $v_1$ is the $\vlos$ $(v_{\rm los})$ and $v_2$ and $v_3$ are the radial and the tangential components of the velocity with respect to the COM in the plane of the sky. These components can be derived in kilometers per second from any $(\mu_{\alpha*}, \mu_\delta)$ as

\begin{equation}
\begin{aligned}
v_1 &= v_{\rm los}\\
v_2 &= 4.74 D [\mu_{\alpha*} \sin{\Gamma} + \mu_\delta \cos{\Gamma}]\\
v_3 &= 4.74 D [\mu_{\alpha*} \cos{\Gamma} - \mu_\delta \sin{\Gamma}],
\end{aligned}
\end{equation}  

where $\Gamma$ is the rotation angle of the $(v_2, v_3)$ to the frame on the sky, which is determined by

\begin{equation}\label{eq:gamma}
\begin{aligned}
\sin{\Gamma} & = [ \cos{\delta_0} \sin{(\alpha - \alpha_0)} ] / \sin{\rho} \\
\cos{\Gamma} & = [ \sin{\delta} \cos{\delta_0} \cos{(\alpha - \alpha_0)} - \cos{\delta} \sin{\delta_0}] / \sin{\rho}.
\end{aligned}
.\end{equation}  

\subsection{COM motion} \label{sec:cm}

The COM motion must be subtracted prior to the analysis of internal kinematics. Its contribution to the velocity field can be derived as:

\begin{equation}
\begin{aligned}
v_1 &=  v_t \sin\rho \cos(\phi - \theta_t) + v_{\rm los,0} \cos\rho\\
v_2 &=  v_t \cos\rho \cos(\phi - \theta_t) - v_{\rm los,0} \sin\rho\\
v_3 &= -v_t \sin(\phi - \theta_t)
\end{aligned}
\end{equation}  

where $v_{\rm los,0}$ is the systemic $\vlos$ of the galaxy, $v_t$ is the velocity of the COM projected on the sky, and $\theta_t$ is the angle that defines the direction of $v_t$ defined using the same criterion as $\phi$:

\begin{equation}\label{eq:vt_theta}
\begin{aligned}
v_t &= 4.74 D_0 \sqrt{\mu_{\alpha*0}^2 + \mu_{\delta0}^2}\\
\theta_t &= \arctan2\left(\frac{\mu_{\delta0}} {-\mu_{\alpha*0}}\right)
\end{aligned}
\end{equation}  

We adopted all values listed in Table~\ref{tab:finalpops_sky} in order to derive the COM contribution to $v_1, v_2,$ and $v_3$.

\subsection{Transformation to Cartesian coordinates}\label{sec:Coordinates}

The Cartesian coordinates $(x, y, z)$ can be derived from $(\rho, \phi, D)$ as

\begin{equation}\label{eq:xyz}
\begin{aligned}
x &= \sin{\rho}\cos{\phi} D\\
y &= \sin{\rho}\sin{\phi} D\\
z &= D_0 - \cos{\rho} D
\end{aligned}
,\end{equation}  

while their respective differentials $(v_x, v_y, v_z)$ can be obtained from $(v_1, v_2, v_3)$ as

\begin{equation}\label{eq:vxvyvz}
\begin{aligned}
v_x &=  v_1 \sin{\rho}\cos{\phi} + v_2 \cos{\rho}\cos{\phi} - v_3 \sin{\phi} \\
v_y &=  v_1 \sin{\rho}\sin{\phi} + v_2 \cos{\rho}\sin{\phi} + v_3 \cos{\phi} \\
v_z &= -v_1 \cos{\rho} + v_2 \sin{\rho}
\end{aligned}
.\end{equation}

\section{Sagittarius Internal Dynamics}\label{sec:Results_6D}

The Sgr is moving outward from the MW center and away from the Sun. In order to study its internal dynamical properties, we subtracted the COM motion from the main sample. This is equivalent to observing the galaxy at rest, i.e., removing the solar motion, as well as the Sgr motion around the MW. 

\subsection{Sky Projections}

\begin{figure}
\begin{center}
\includegraphics[width=\linewidth]{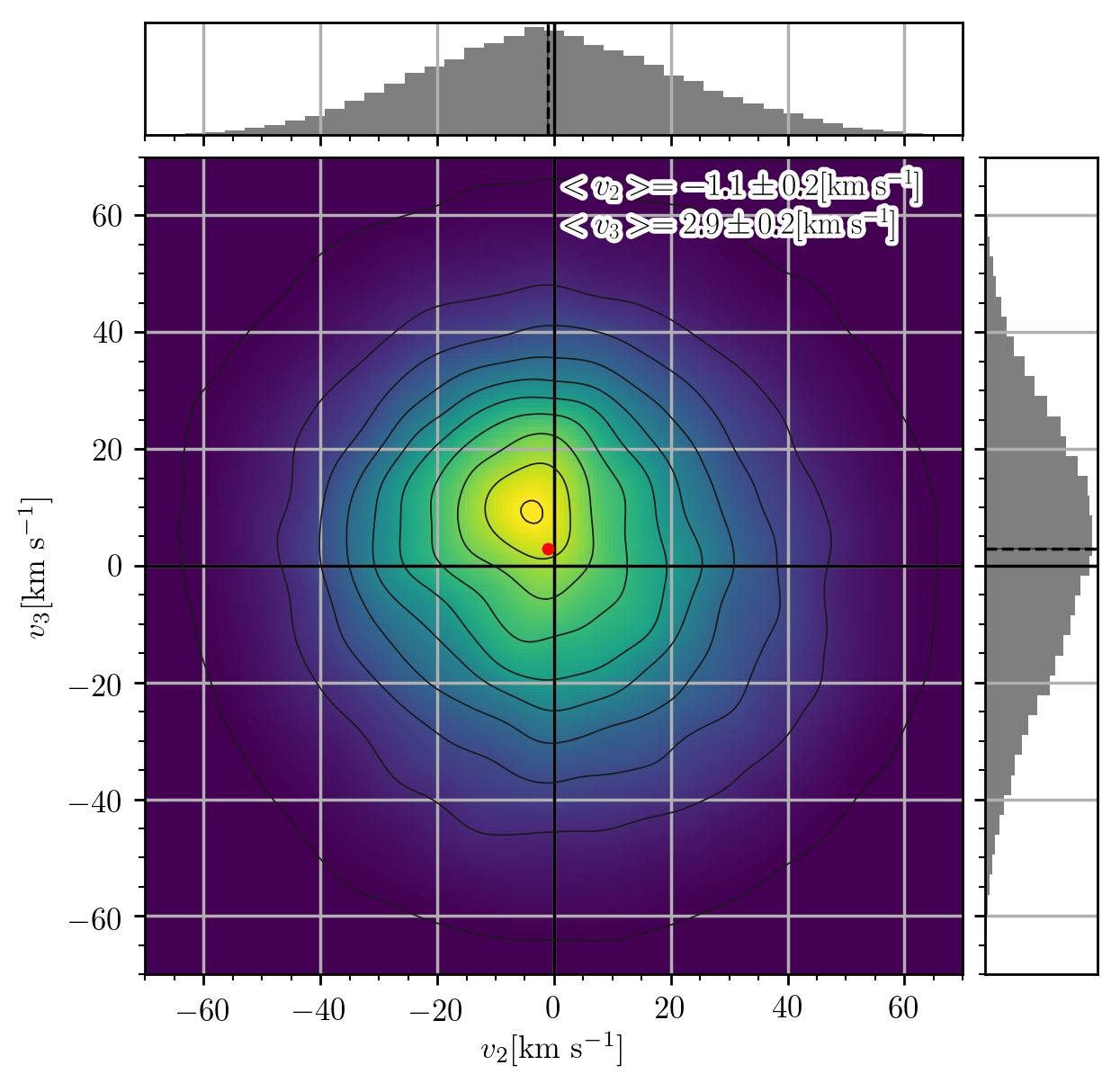}
\caption{Error-weighted distribution of the radial, $v_2$, and tangential, $v_3$, components of the velocity projected in the sky. Contours and color shade indicate concentration. The median of the sample is indicated by a red dot. \secrevision{The error bars of the median are not visible due to their small size.}}
\label{fig:v2v3}
\end{center}
\end{figure}

We derived $(v_1, v_2, v_3)$ for all of the stars in our sample. As expected after the subtraction of the COM motion, the median of the line-of-sight component, $\tilde{v_1} = 0.1 \pm 0.1 \kms$\footnote{\secrevision{The error of the median of the nominal values is determined by bootstrapping the sample.}}, is compatible with zero. Figure~\ref{fig:v2v3} shows the distribution of $v_2$ and $v_3$ for all the stars in the sample. The median radial component of the stellar velocities in the sky, $\tilde{v_2} = -1.1 \pm 0.2 \kms$, indicates that the galaxy is contracting, while its tangential component, $\tilde{v_3} = 2.9 \pm 0.2 \kms$, indicates that Sgr is rotating counterclockwise as projected on the celestial sphere. The medians of the Monte Carlo realizations of the same quantities were closer to zero, but still significant in rotation\footnote{\secrevision{The error of the median of the Monte Carlo experiments is computed as the standard deviation of all the resulting median velocities from each Monte Carlo realization.}}:  $0.1\pm 0.1, -0.15\pm0.17 $, and $1.71\pm 0.16 \kms$ for $\tilde{v_1}$, $\tilde{v_2}$, and $\tilde{v_3}$, respectively. The distribution of $v_2$ and $v_3$ extend towards positive and negative values of $v_2$ and $v_3$, respectively. This is likely to be produced by the influence of tidal tails in the kinematics of the galaxy. Indeed, stars with higher values of $v_2$ and lower of $v_3$ are located at larger distances along the stream. Stars with $v_2 > 0 \kms$ and $v_2 < -50 \kms$ were located at a median absolute $\tilde{\Lambda} = 2.65^\circ$, whereas the rest are closer to the Sgr center: median absolute stream longitude $\tilde{\Lambda} = 1.96^\circ$.

\begin{figure*}
\begin{center}
\includegraphics[width=\linewidth]{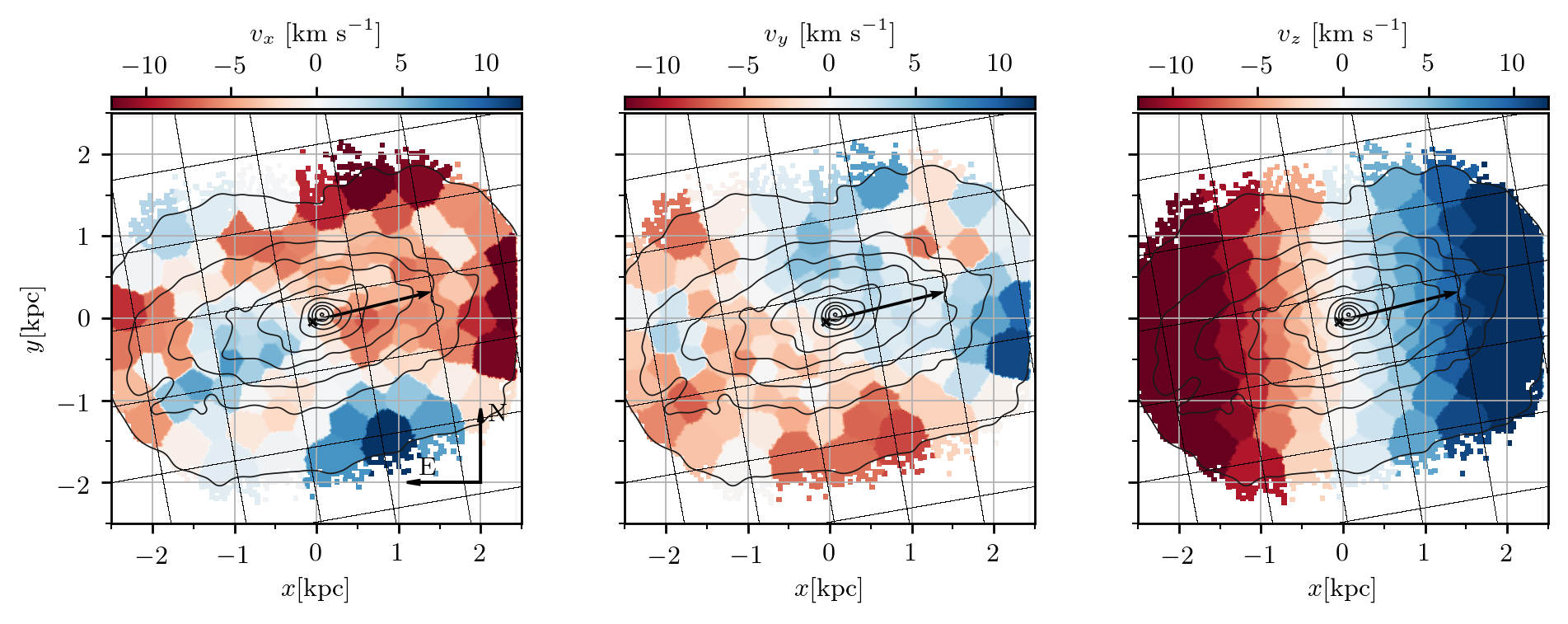}
\caption{Stellar density and kinematics of Sgr projected on the sky plane upon COM subtraction. Stellar density is shown by black contours. From left to right, the color maps indicate the error-weighted average velocity of $v_x$, $v_y$, and $v_z$ inside Voronoi cells of $\sim1000$ stars. \secrevision{The Voronoi cells are the same in the three panels}. A black arrow indicates the sky-projected bulk velocity of the COM. Coordinates were derived from Equation~\ref{eq:xyz} and represent physical distances in kiloparsecs. Velocities were obtained from Equation~\ref{eq:vxvyvz} and are the Cartesian components of the velocities projected over the axes. A grid aligned with the stream coordinates system \citep{Law-Majewski2010} is also plotted.}
\label{fig:voronoi_vxvyvz}
\end{center}
\end{figure*}

Velocity maps provide an immediate qualitative view of the dynamics of Sgr. Figure~\ref{fig:voronoi_vxvyvz} shows the stellar density, as well as the error-weighted average of each velocity component, $(v_x, v_y, v_z)$, for stars within Voronoi cells of approximately 1000 stars defined on their sky positions. Voronoi tessellation ensures a constant number of stars per unit of resolution (``pixel''), thus providing a constant signal-to-noise ratio \citep{Cappellari2003}.

The stellar density projected in the sky has an elliptical profile whose concentration peaks at the position of M54. Results also show a clear velocity gradient ($>10 \kms$) in $v_z$, to first order comparable with the line-of-sight velocity, indicating rotation about the projected optical minor axis of Sgr. This gradient is almost parallel to the bulk tangential velocity of Sgr and to the direction of the stream (along $\Lambda$).

Counterclockwise rotation in the plane of the sky is also clearly visible in $v_x$ and $v_y$, where velocity gradients are inclined with respect to the axes due to the PA of Sgr in the sky ($102\degree.4$). Southern regions of the galaxy are moving west (right), while northern regions are moving east (left). Here $v_y$ shows the same pattern, with eastern regions moving south and western ones moving north. Interesting features can be noticed in the $v_x$ and $v_y$ maps, such as the apparent contraction of the core, the two negative bands ($v_x <0$) coming out towards positive values of $\Lambda$ (east direction), or the region with negative $v_y$ around $(x,y) = (1.5, 1) \kpc$. These are the consequences of the rotation of the galaxy, the inclination of its main body with respect to the plane of the sky, and the presence of tidal tails moving outward from its core. It is also worth noticing an area at $(x,y) \sim (-2,0) \kpc$ with unexpected positive values of $v_y$ and negative $v_x$. This is a region showing unusual small values of $\pmdec$ compared to the average caused by \Gaia systematic errors, resulting in clockwise rotation (see Appendix~\ref{Apx:Systematic_Errors}). Stripes with the same inclination can be seen in Figures~\ref{fig:raw_data_sky} and ~\ref{fig:members_Sky}, and are related to the scanning pattern of the \Gaia satellite. This band would probably show inverted velocities if it were not affected by errors. Systematics also affect the stellar density profiles, although they tend to average out when considering large areas, not changing the general view of the galaxy.

\subsection{3D Projections}

\subsubsection{Sky Projections}

\begin{figure*}
\begin{center}
\includegraphics[width=\linewidth]{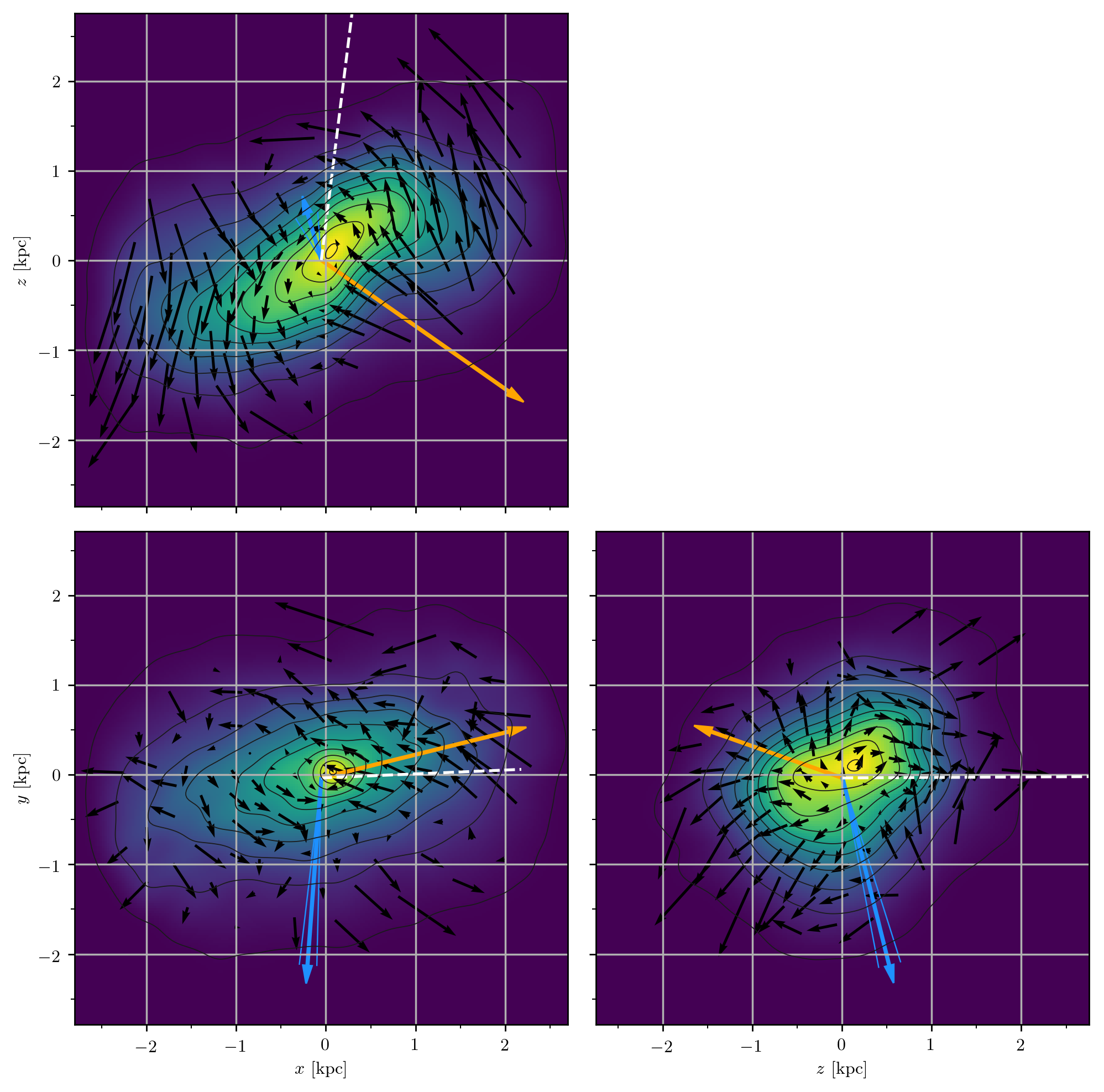}
\caption{Projected stellar density and kinematics of Sgr on the sky on the $z-x$, $x-y$, and $z-y$ planes derived from equations~\ref{eq:xyz} and \ref{eq:vxvyvz}. Stellar density is shown by black contours and color shade. The orange arrow indicates the projected bulk velocity of the COM. The blue arrow shows the projection of the internal angular momentum, $L$. Errors of these quantities are shown by thin lines at both sides of of the arrows. The projected tangential velocities upon subtraction of the COM velocity are shown by the black arrows. The direction towards the MW center is indicated by a dashed white line. Velocities measured in the Voronoi cells are all in the same scale. The scale of these, $L$, and the bulk motion of the galaxy is the same for all panels but differ from one quantity to another in order to accommodate all the arrows in the same plot.}
\label{fig:Sky_distro}
\end{center}
\end{figure*}

The full phase space allows us to represent Sgr in all dimensions (6D), providing a better understanding of the structure and dynamics of the galaxy. Figure~\ref{fig:Sky_distro} shows the projected stellar density and velocities for all the stars in our sample in the coordinate system defined by equations~\ref{eq:xyz} and \ref{eq:vxvyvz}. In this representation, the galaxy is located inside a cube whose sides are the planes $x-y$, $x-z$, and $z-y$. The $x-y$ plane is aligned with the sky with the x axis pointing to the west and the y axis to the north (same as in Figure~\ref{fig:voronoi_vxvyvz}). In planes $x-z$, and $z-y$ the galaxy is projected along the line of sight, with positive values of $z$ pointing to the Sun. The observer is therefore located at $(0, 0, \infty)$.

Agglomerative Voronoi tessellation was applied to the data, creating cells of approximately 1000 stars for which the median values of the star velocities were calculated. These median velocities are represented by black arrows and qualitatively show, how different parts of the galaxy move with respect to the COM. We also have calculated the 3D internal angular momentum of Sgr, $L = I\omega$, where $I$ is the inertia tensor and $\omega$ is the angular speed, for stars in the center of Sgr within 1 kpc radius. It is shown by a blue arrow.

The main body of Sgr is inclined with respect to the plane of the sky, being closer to the Sun for positive values of $x$ and $y$. The data also suggest the presence of a bar-like structure more than $2$ kpc long, from the tips of which depart the tidal tails that form the stream. This bar-and-tails configuration confers Sgr the ``S'' shape observed in the $x-z$ plane, which also projects on the $x-y$ plane. We should note that the tails depart the main body at distances of $1.5 \kpc$ ($\sim 3\degree$ in the sky), which is much closer to the center than the half-light radius found in the literature for Sgr \citep[$5\degree.7$;][]{McConnachie2012}. Studies analyzing the properties of the core of the galaxy should be aware of the effects of such tails.

Rotation is most obvious on the $x-z$ plane (counterclockwise), with some rotation signal also visible on the $x-y$ plane (counter clockwise). The stripe showing unusually high $\pmdec$ is clearly visible in the $x-y$ plane as arrows pointing in the northeast direction (opposite from their neighbors). Least obvious rotation is seen in the $z-y$ plane, with most of the stars located in at positive values of $y$ moving toward positive values of $z$ (right) and stars with negative $y$ moving toward negative values of $z$ (left). The apparent expansion observed in this plane is a projection effect; stars at positive $z$ are moving toward positive $z$ while stars at negative $z$ are moving towards negative $z$ (see $x-z$ panel). The angular momentum of the galaxy, projected in each plane by a blue arrow, reflects these general rotation patterns.

Combined, all dimensions depict an inclined system that is closer to us on the west and north. Closer areas are moving in our direction, while the furthest are moving even further away. Of course, this is in the COM frame, as the whole galaxy is moving away from the Sun at $\vlos = 142.7 \pm 1.7 \kms$, much larger than the $\sim20\kms$ gradient discussed here.

\subsubsection{galactocentric Projections}

\begin{figure*}
\begin{center}
\includegraphics[width=\linewidth]{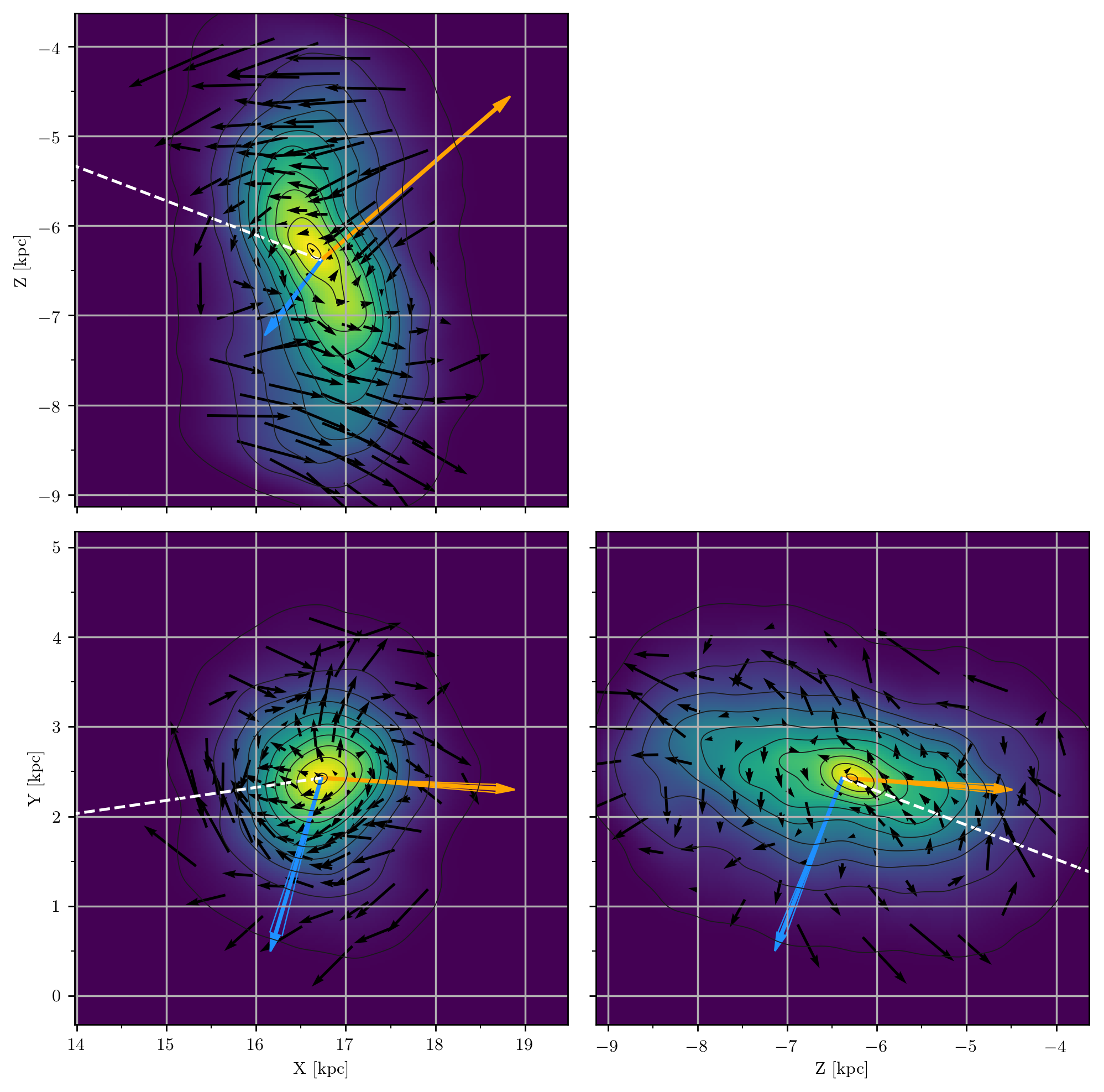}
\caption{Projected stellar density and kinematics of Sgr on galactocentric coordinates. Markers and their scale coincide with those from Figure~\ref{fig:Sky_distro}.}
\label{fig:galactocentric_distro}
\end{center}
\end{figure*}

The most important external factor affecting Sgr morphology and kinematics is the MW potential. Tidal forces stretch the body of the galaxy, accelerating its stars depending on their relative distance to the MW center. Figure~\ref{fig:galactocentric_distro} shows Sgr projected on the galactocentric frame, $(X, Y, Z)$, with the observer located at infinite distance above the MW plane in $(16.73, 2.44, \infty) \kpc$. The $X-Y$ plane is almost aligned to the orbital plane of Sgr.

The largest component of the bulk velocity for Sgr is found in the $X-Z$ plane, which is also the plane where the largest velocity gradient is observed. This suggests that tidal forces are greatly affecting Sgr kinematics, shearing its body and stripping its stars from outer regions. On the other hand, some internal rotation is apparent in the innermost parts of the body, specially in the $X-Z$ and the $Z-Y$ planes.

\subsubsection{Principal Axis Projections}

We can better understand Sgr's inner structure and kinematics by aligning the system with its principal axes of inertia, $(x^\prime, y^\prime, z^\prime)$. We calculated $(x^\prime, y^\prime, z^\prime)$ within a 3D sphere of a given radius $r_{\rm 3D}$, as well as the Euler angles necessary to align Sgr from its orientation in the sky to these axes, $\alpha$, $\beta$ and $\gamma$, defined as $z-y-x$ rotation. 

\begin{figure}
\begin{center}
\includegraphics[width=\linewidth]{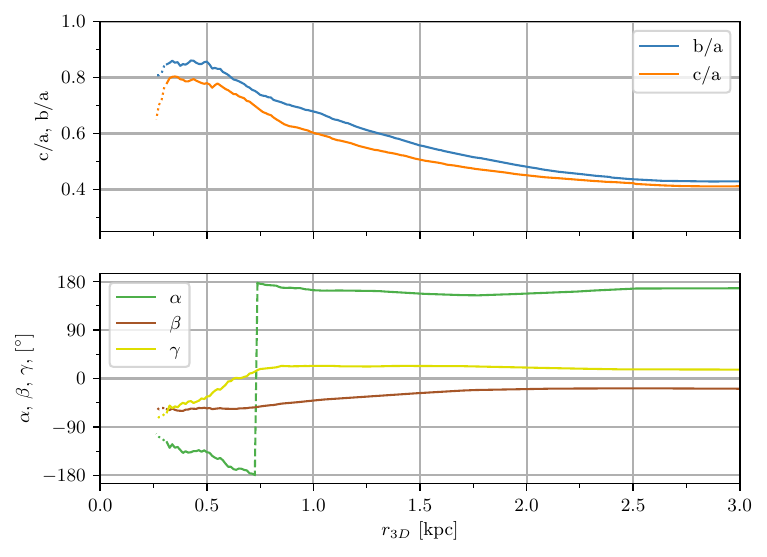}
\caption{Principal axis ratios and their Euler angles with respect to the plane of the sky. These are computed for stars within spheres of radii 3D $r_{\rm 3D} > 0.25 \kpc$.}
\label{fig:axes_ratio}
\end{center}
\end{figure}

The ratio between axes changes depending on the considered sphere radius, as does the orientation. We derived the direction of the principal axes of inertia of Sgr and the ratio between the FWHM of the stellar density along them for stars within spheres of radii $0.25 \kpc \leq r_{\rm 3D} \leq 3 \kpc$ (notice that no star is located farther than $\sim 2.8 \kpc$ in our data). This is shown in Figure~\ref{fig:axes_ratio}.

The Sgr is a triaxial system at all radii considered in our sample. However, the ratio between its intermediate and shortest axes, given by $y^\prime$ and $z^\prime$, is never larger than 1.14, with an average of 1.08, indicating that Sgr is almost prolate within 3 kpc. The inclination of the longest axis, measured by $\beta$, remains almost constant up to distances of $0.75 \kpc \sim r_{\rm 3D}$ as expected, since the bar dominates the stellar distribution at small radii. The angle $\alpha$ measures the inclination of the sky-projected major axis of the galaxy from the $x$-axis (east direction) in the counterclockwise direction, i.e.\ $\alpha = 270\degree - $PA. Its variation within just 1 kpc, $\sim60\degree$, is due to the transition from the core of the galaxy to the tidal tails, that elongate the shape of the galaxy along the stream direction. Here $\alpha$ stabilizes at around $r_{\rm 3D} = 2.5 \kpc$ at $167.5\degree$ (PA = $102.4\degree$). Lastly, $\gamma$ measures the pitch angle, i.e.\ the inclination in the north-south direction after the $\beta$ rotation has been applied. The largest variation in $\gamma$ occurs at $0.5 \kpc \sim r_{\rm 3D} \lesssim 0.75 \kpc$, radii for which the galaxy changes from having its north regions closer to us to the opposite. Given the evident change in morphology for radii larger than 1 $\kpc$, we decided to adopt this value to calculate the nominal orientation of the Sgr principal axes. Using such a radius, the principal axes of Sgr, $(x^\prime, y^\prime, z^\prime)$, form net angles of $(43\degree, 2\degree, 46\degree)$ with respect to the plane of the sky and $(2\degree, 18\degree, 72\degree)$ with respect to the orbital plane of Sgr around the MW. We should emphasize, though, that these angles vary depending on the considered galactocentric radius, $r_{\rm 3D}$.

\begin{figure*}
\begin{center}
\includegraphics[width=\linewidth]{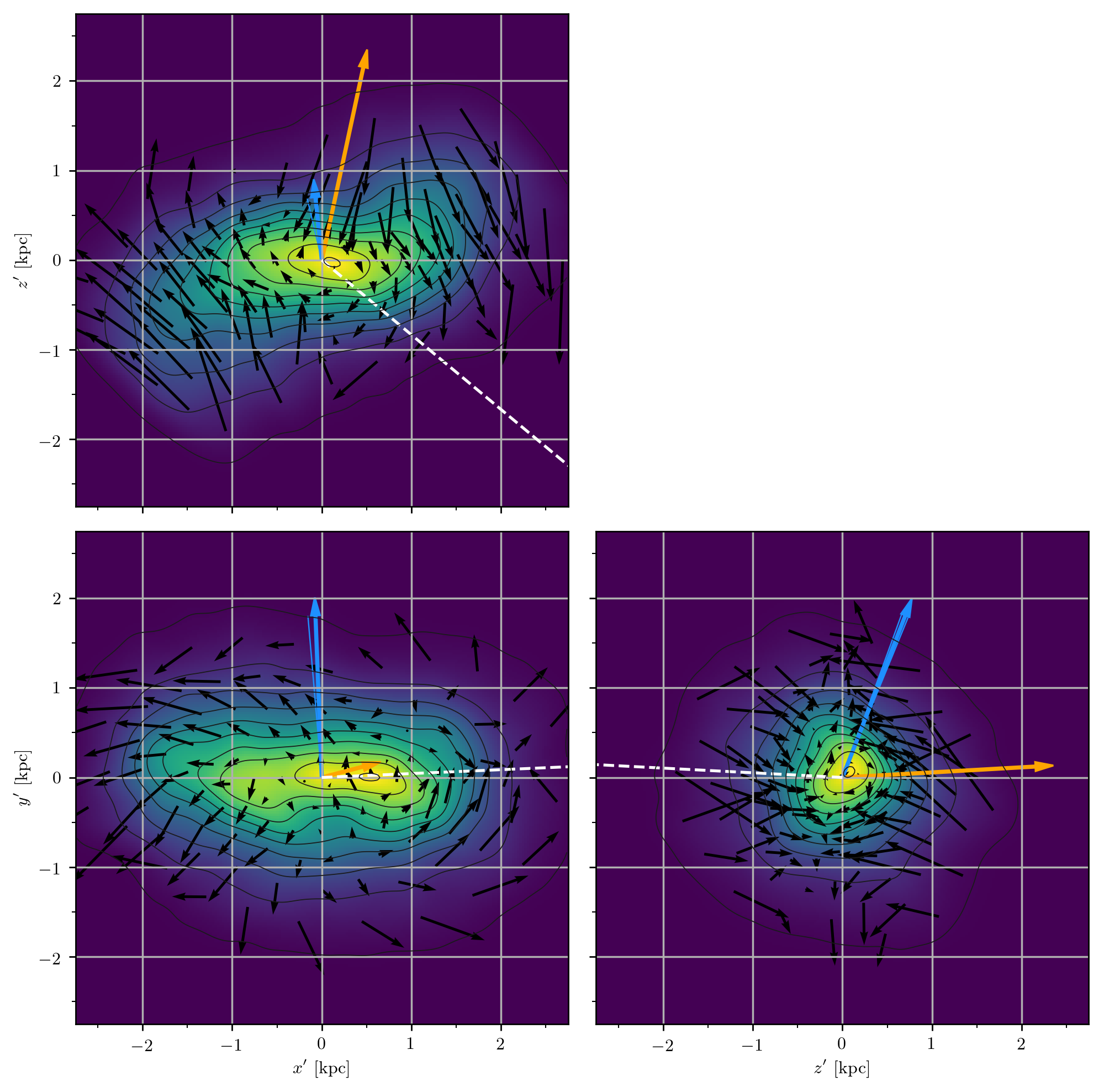}
\caption{Stellar density and kinematics of Sgr projected over the planes defined by principal axes of inertia. The axes  being $x^\prime$, $y^\prime$, and $z^\prime$ the longest, intermediate, and shortest axis, respectively. The principal axes were computed within a sphere of radius $r_{\rm 3D} = 1 \kpc$. Markers coincide with those of Figure~\ref{fig:Sky_distro}.}
\label{fig:Main_distro}
\end{center}
\end{figure*}

Figure~\ref{fig:Main_distro} shows the projected stellar density and kinematics over the principal axes of inertia of Sgr. The Euler angles necessary to align Sgr from its orientation in the sky to its principal axes of inertia are $\alpha = 177 \pm 3 \degree$, $\beta = -40 \pm 2 \degree$ and $\gamma = 16 \pm 2 \degree$, measured at $r_{\rm 3D} = 1 \kpc$. These rotations align Sgr's longest principal axis $x^\prime$ with the $x$-axis, the intermediate $y^\prime$ with $y$-axis, and the shortest axis $z^\prime$ with the $z$-axis with $(x,y,z)$ defined by Equation~\ref{eq:xyz}. 

In this projection the galaxy is aligned with the bar, clearly visible in the $x^\prime$-$y^\prime$ and $x^\prime$-$z^\prime$ planes. The almost spherical view of Sgr from its longest axis, i.e.\ projected over its intermediate and shortest axes shows the little length difference found between the intermediate and the shortest axis ($z^\prime$-$y^\prime$ plane). Indeed, the ratio between the FWHM of the stellar distribution along the three principal axes inside the 1 kpc radius sphere is 1:0.67:0.60, not far from being prolate.

To further investigate the nature and strength of the bar, we measured the bar mode $A_2$ from the positions of the stars projected onto the $x^\prime$-$y^\prime$ plane (along the shortest axis). In general, the amplitude of the $m$th Fourier mode of the discrete distribution of stars in the galaxy is calculated as $A_m = (1/N) \left| \Sigma^{N}_{j=1} \exp (i m \phi_j) \right|$ where $\phi_j$ are the particle phases in cylindrical coordinates and $N$ is the total number of stars \citep{Sellwood1986, Debattista2000}. Values larger than 0.2 of the $m=2$ term of this expansion indicate the presence of a strong bar. With $A_2 = 0.36$, our data suggest that the elongated shape of Sgr is indeed a strong bar. We should also notice that we obtained $A_2 = 0.3$ measured directly on the positions in the sky, also larger than 0.2.

The Sgr rotates mainly around its intermediate principal axis, $y^\prime$. This is clearly visible by the direction of the angular momentum, $L$, almost parallel to $y^\prime$, and by the projected velocity field on the $x^\prime$-$z^\prime$ plane. Counterclockwise rotation can be observed in the $x^\prime$-$y^\prime$, causing the angular momentum, $L$ to point toward positive values of $z^\prime$. The galaxy also shows very little clockwise rotation in its inner regions about its major axis, $x^\prime$, ($z^\prime$-$y^\prime$ plane). The expansion of the outer regions of Sgr is most visible in the $x^\prime$-$y^\prime$ plane, where areas dominated by the arms are being pulled away from the main body of the galaxy.

\begin{figure}[t]
\begin{center}
\includegraphics[width=\linewidth]{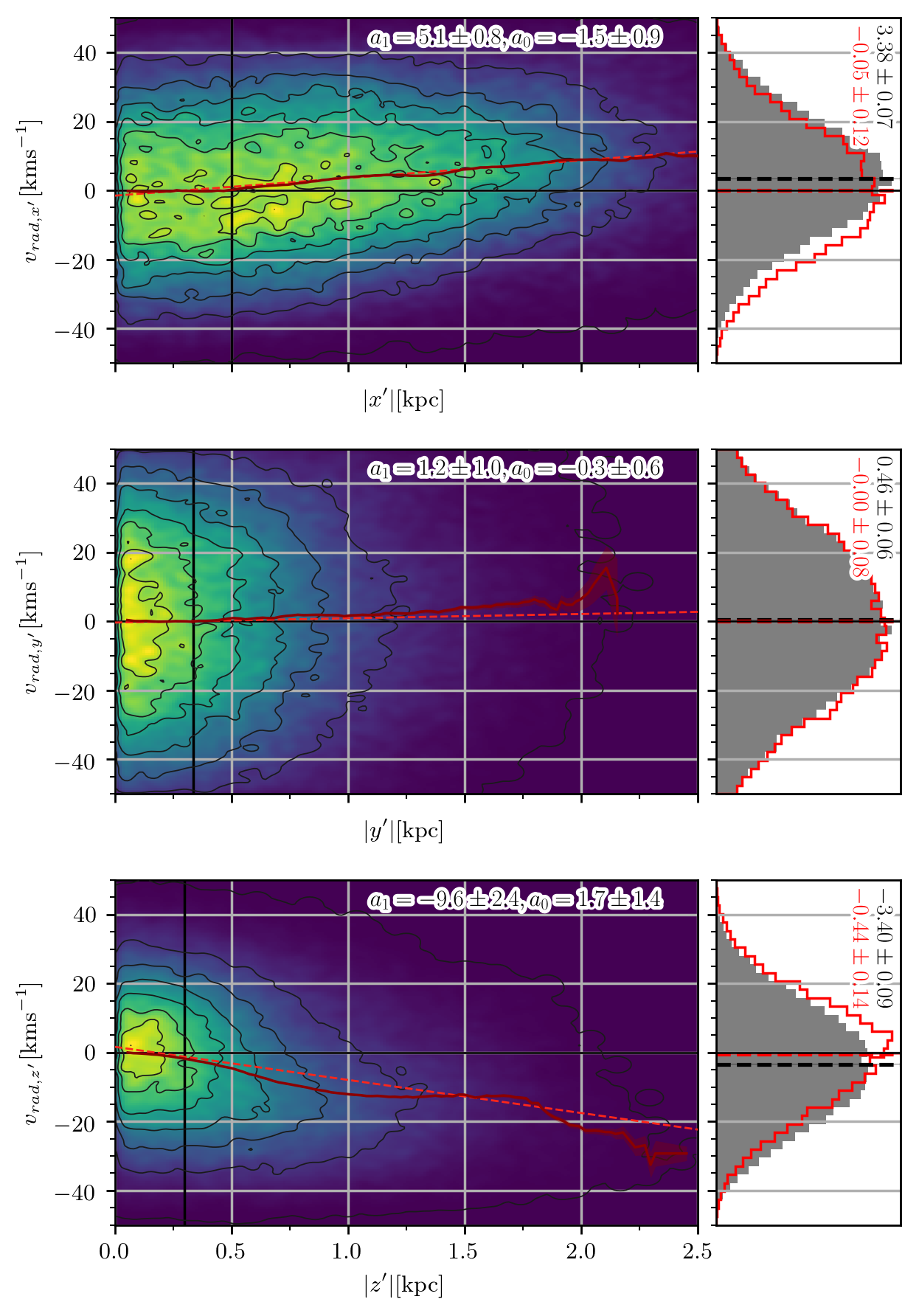}
\caption{Velocity of the stars projected over the principal axes of inertia of Sgr. From top to bottom the panels show velocities along the longest, the intermediate, and the shortest principal axis, respectively. Negative values of distance along each axis and their corresponding velocities are inverted, showing the absolute value along the axis (See text). The color map show the error-weighted distribution of the stars. Black vertical lines mark distances along the axes of 0.5, 0.33, and 0.30 kpc, respectively. An error-weighted linear fit to the data is represented by the red-dashed line, while its coefficients are shown in the top right part of each panel. A dark-red curve shows the box-car error-weighted average of the distribution along each axis with steps of $0.025 \kpc$ and a window of $0.1 \kpc$. The shaded area around the curve represent the error of the average. Right panels show the distribution collapsed along each axis. The gray histogram shows all the stars, whereas the red one shows the stars located within the radius marked by the vertical lines in the left panels measured from the center.}
\label{fig:main_axes_expansion}
\end{center}
\end{figure}

\begin{figure}[ht!]
\begin{center}
\includegraphics[width=\linewidth]{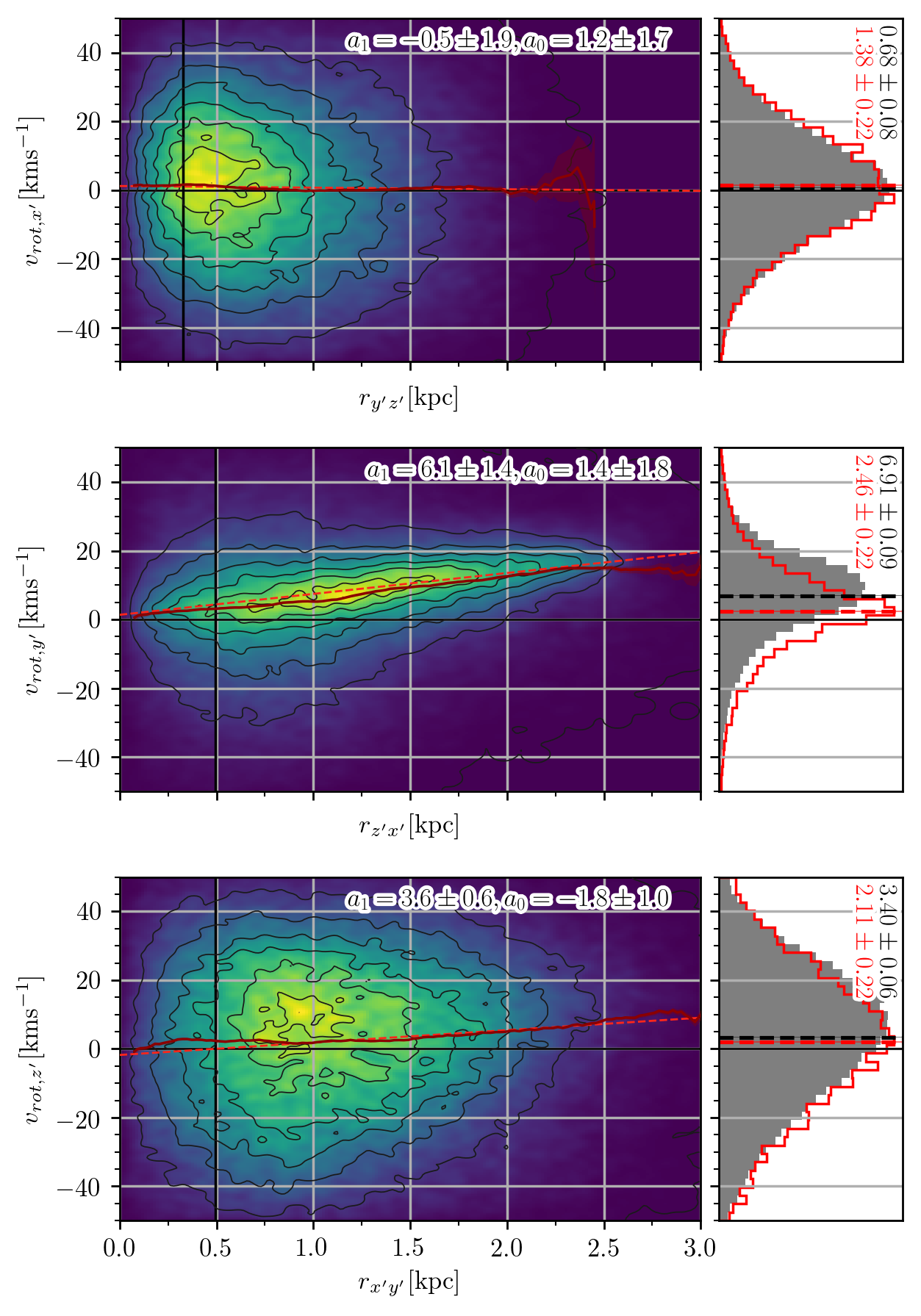}
\caption{Rotation velocity of the stars projected over the planes defined by the principal planes of inertia. From top to bottom the panels show rotation about the longest, the intermediate, and the shortest principal axis, respectively. Black vertical lines mark the projected maximum radial distance of the stars within a 3D ellipsoid with axes of 0.5, 0.33, and 0.30 over the planes defined by $y^\prime-z^\prime$, $z^\prime-x^\prime$ and $x^\prime-y^\prime$, respectively. The rest of the markers coincide with those from Figure~\ref{fig:main_axes_expansion}.}
\label{fig:main_axes_rotation}
\end{center}
\end{figure}

We computed the absolute net velocity along each one of the principal axes of Sgr as

\[
  v_{rad, i} =
  \begin{cases}
                                   v_{rad, i} & \text{if $i\geq 0$} \\
                                   -v_{rad, i} & \text{if $i<0$}
  \end{cases}
\]

with $i$ being each one of the principal axes $x^\prime$, $y^\prime$, or $z^\prime$. These velocities as a function of the absolute value of the radial distance along their respective axis $|i|$ are shown in Figure~\ref{fig:main_axes_expansion}. The rotation velocity around the same axes, computed over the planes defined by tuples of the two other principal axes (as shown in Figure~\ref{fig:Main_distro}) are shown in Figure~\ref{fig:main_axes_rotation}. In this case, the rotation around $|i|$ is rotation measured on the $(j, k)$ plane with ($i \neq j \neq k$), as defined by a right-handed Cartesian coordinate system. The galactocentric radius is defined as $r_{jk} = \sqrt{j^2+k^2}$. In both figures, histograms in the right panels show the total velocity distribution (black), as well as the velocity distribution near the center of the galaxy (red). Stars in the center were selected based on their distance to the COM along each given axis using radii of 0.5, 0.33, and 0.3 kpc for the longest, intermediate, and shortest principal axis, respectively (following the axis ratios of the system).


The outer regions of Sgr are clearly affected by tidal forces. This can be observed as distortions in all of the rotation curves in the outer regions of the galaxy (see, for example, $v_{rot,x^\prime}$ in Figure~\ref{fig:main_axes_rotation}, which shows positive rotation of the core up to distances $r_{y^\prime x^\prime} \sim 0.75 \kpc$). Figure~\ref{fig:main_axes_expansion} shows the expansion of the galaxy along its longest principal axis, $x^\prime$, while the galaxy remains close to stable along $y^\prime$ and shows apparent contraction along $z^\prime$ for distances larger than $\sim0.25 \kpc$. However, this apparent contraction seems to be the projection of the tidal tails' kinematics along $|z^\prime|$ (see plane $x^\prime-z^\prime$ in Figure~\ref{fig:main_distro_3D}), rather than a real contraction of the core of the galaxy. We confirmed these findings, calculating the variation of the radial components of the velocity as a function of distance along each axis through finite differences. Similar results were obtained in all axes with expansion starting at different distances. We also found that the inner core of the galaxy shows almost no net expansion or contraction within an ellipsoid of axes $(0.5, 0.33, 0.3)\kpc$. This ellipsoid is rotating mainly about its intermediate principal axis, $y^\prime$, at an average $v_{rot,y^\prime} 2.46 \pm 0.22 \kms$. Residual rotation can also be observed about the shortest ($v_{rot,z^\prime} 2.11 \pm 0.22 \kms$) and, to a lesser extent, the longest axes ($v_{rot,x^\prime} 1.38 \pm 0.22 \kms$) axis, indicating the presence of precession and nutation due to the MW's tidal torques. We aligned this inner ellipsoid to its angular velocity vector and measured a maximum rotation velocity of $v_{rot} = 4.13 \pm 0.16 \kms$ within this ellipsoid. It is important to notice that the orientation and strength of this vector vary with radius as shearing tidal forces change the shape and kinematics of the galaxy.

Three-dimensional renderings of Sgr aligned to its principal axes of inertia on a galactocentric frame can be found in Appendix~\ref{Apx:3D_projections}. An interactive version of this rendering can be found at \url{https://www.stsci.edu/~marel/hstpromo/Sagittarius_GC.html}.

\section{N-body Counterpart}\label{sec:Nbody_counterparts}

The rotation observed in Sgr seems to be a combination of intrinsic residual rotation and rotation induced by tidal torque from the MW potential well. In an attempt to reproduce the observational results and discern between these two factors, we ran a suite of $N$-body simulations of an interaction between an Sgr-like dwarf and a bigger galaxy with properties similar to the MW. Both galaxies initially contained two components: a spherical dark matter halo and an exponential disk. The structural parameters of the Sgr progenitor were similar to those in \citet{Lokas2010}; its dark matter halo had a Navarro-Frenk-White (NFW; \citealt{Navarro1997}) profile with nominal virial mass $M_{\rm h} = 1.6 \times 10^{10}$ $M_{\odot}$ and concentration $c=15$. In order to mimic the tidal stripping of the halo prior to the simulated period, we introduced a smooth cutoff to its density distribution at a radius of 20 kpc with a width of 10 kpc. The exponential disk had a mass $M_{\rm d} = 3.2 \times 10^{8}$ $M_{\odot}$, scale-length $R_{\rm d} = 2.3$ kpc, and thickness $z_{\rm d} = 0.3 R_{\rm d}=0.69$ kpc. The central values of the radial and vertical velocity dispersions of the disk were chosen to be equal, $\sigma_{z,0}=\sigma_{R,0}=19.5 \kms$. The resulting model was stable against bar formation in isolation. The MW model was the same as in \citet{Lokas2019}; its dark matter halo had an NFW profile with a virial mass $M_{\rm H} = 10^{12}$ $M_{\odot}$ and concentration $c=25$ while the exponential disk had a mass $M_{\rm D} = 4.5 \times 10^{10}$ $M_{\odot}$, scale-length $R_{\rm D} = 3$ kpc, and thickness $z_{\rm D} = 0.42$ kpc.

The $N$-body realizations of the galaxies were initialized using the procedures described in \citet{Widrow2005} and \citet{Widrow2008}, with each component containing $10^6$ particles. The progenitor of Sgr was placed at an apocenter of the orbit with apo- and pericenter distances of 58 and 19 kpc, respectively. The initial configuration was similar to the one shown in Figure 2 of \citet{Lokas2010}, except that the dwarf's disk was rotated a certain angle around the $X$-axis of the simulation box. The evolution of the galaxies was followed for 2 Gyr with the GIZMO code \citep{Hopkins2015}, an extension of the widely used GADGET-2 \citep{Springel2001, Springel2005}, saving outputs every 0.005 Gyr. The adopted softening scales were $\epsilon_{\rm d} = 0.02$ kpc and $\epsilon_{\rm h} = 0.06$ kpc for the disk and halo of Sgr while $\epsilon_{\rm D} = 0.2$ kpc and $\epsilon_{\rm H} = 2$ kpc for the disk and halo of the MW, respectively. Configurations similar to the present state of Sgr are reached soon after the second pericenter passage, around 1.22--1.24 Gyr after the simulation begins.

We completed this set with two pressure-supported models with three pericenter passages instead of two. The \citet{Law-Majewski2010} model (hereafter \citetalias{Law-Majewski2010}) and the \citet{Fardal2019} model (hereafter \citetalias{Fardal2019}), both focused on reproducing the stream's properties.

\begin{figure*}
\begin{center}
\includegraphics[width=\linewidth]{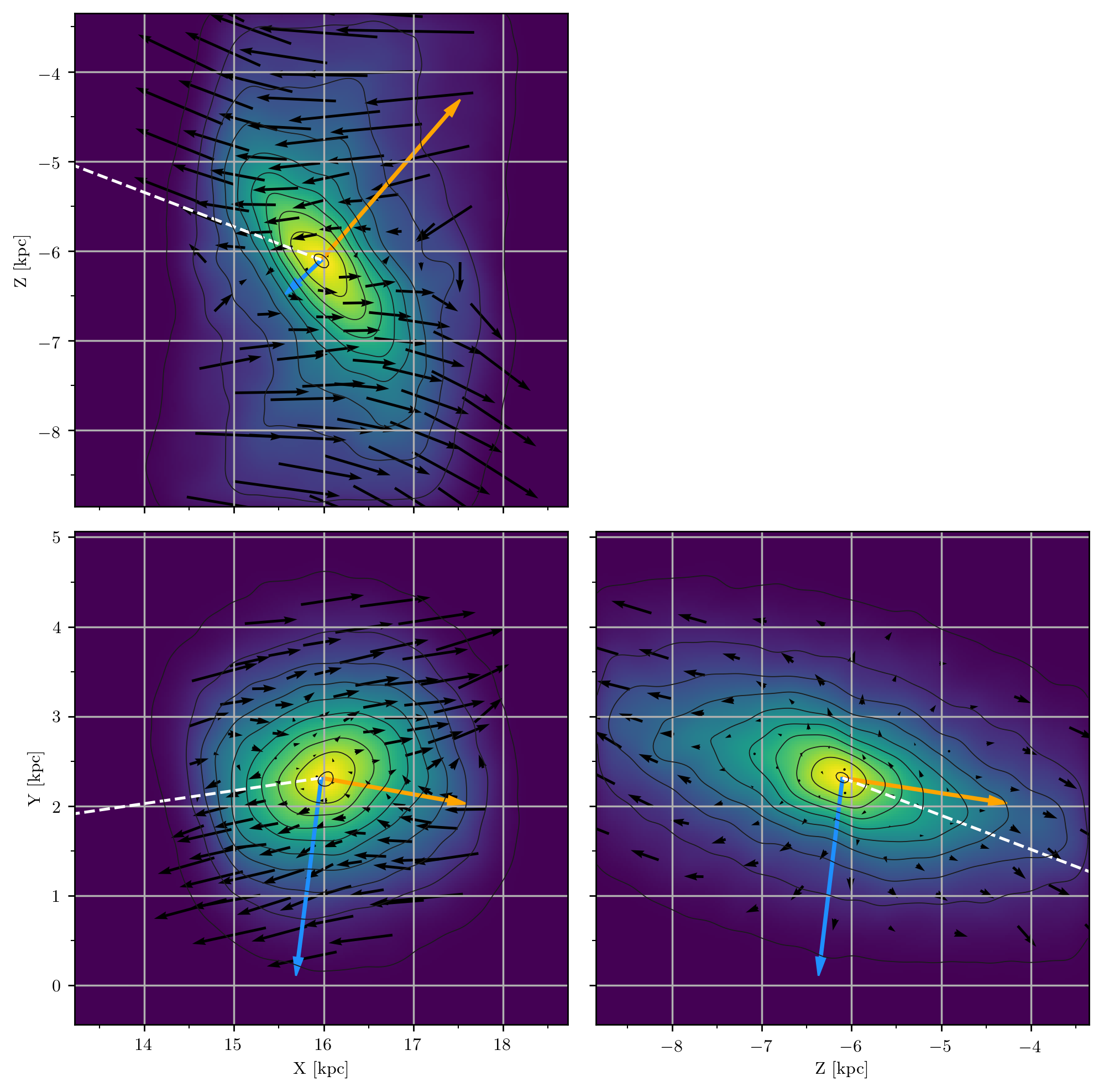}
\caption{Projected stellar density and kinematics of our best $N$-body candidate on galactocentric coordinates (rotating model), which should be compared with Figure~\ref{fig:galactocentric_distro}. Markers and their scale coincide with those from Figure~\ref{fig:Sky_distro}.}
\label{fig:galactocentric_distro_sim}
\end{center}
\end{figure*}

Figure~\ref{fig:galactocentric_distro_sim} shows the best candidate of our originally rotating $N$-body simulations in galactocentric coordinates, which should be compared with Figure~\ref{fig:galactocentric_distro}. This model is was rotated by $10\degree$ in the opposite direction to the one shown in Figure 2 of \citet{Lokas2010}. The similarity with the observations is remarkable. 

\begin{itemize}
    \item The geometry of the $N$-body is almost identical to observations.
    \item The direction of the internal angular momentum, $L$, coincides within the errors.
    \item The bulk motion of the $N$-body around the MW is also remarkably accurate.
\end{itemize}

\begin{figure}
\begin{center}
\includegraphics[width=\linewidth]{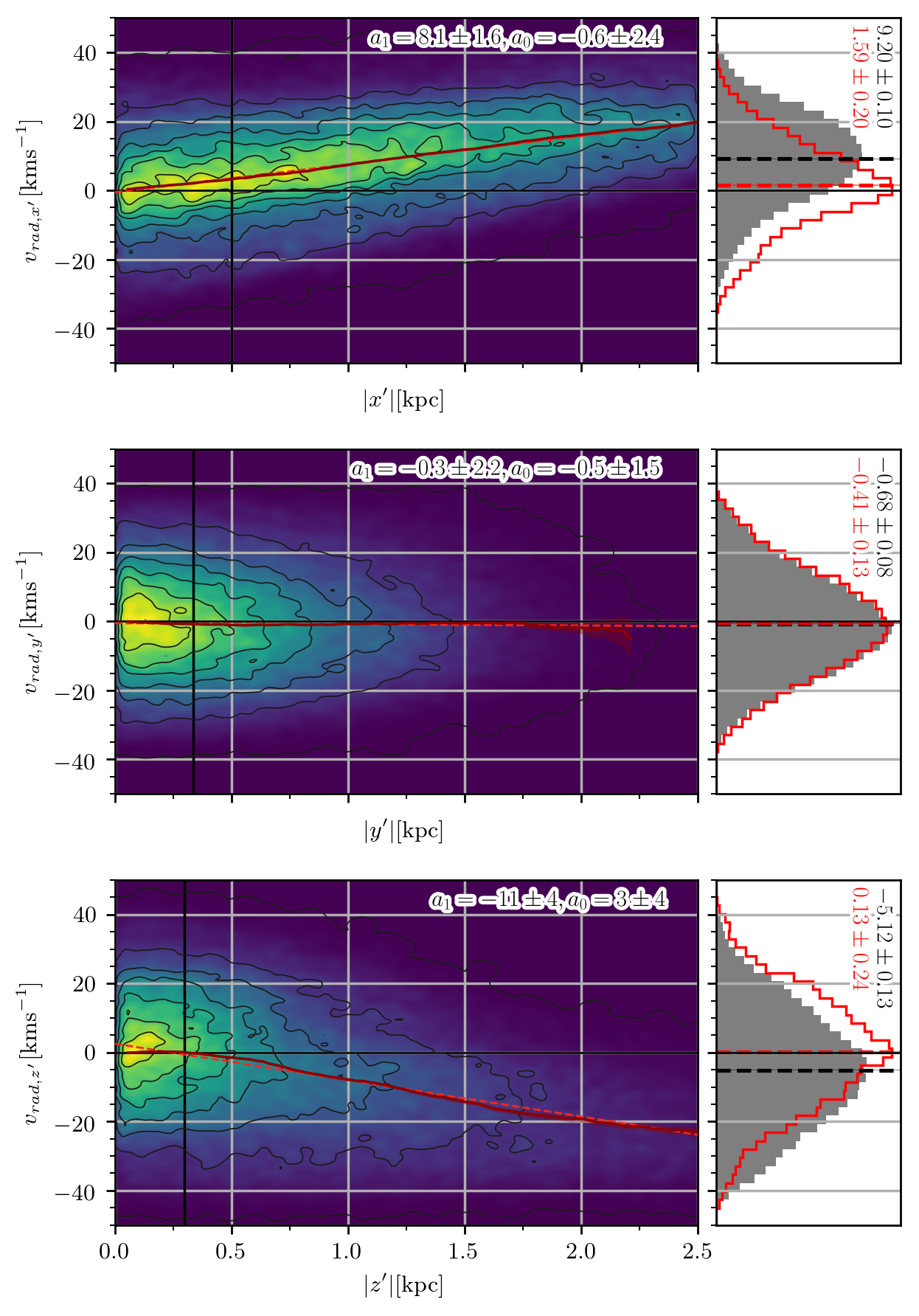}
\caption{Same as Figure~\ref{fig:main_axes_rotation} for the originally rotating $N$-body simulation.}
\label{fig:main_axes_expansion_simL}
\end{center}
\end{figure}

\begin{figure}
\begin{center}
\includegraphics[width=\linewidth]{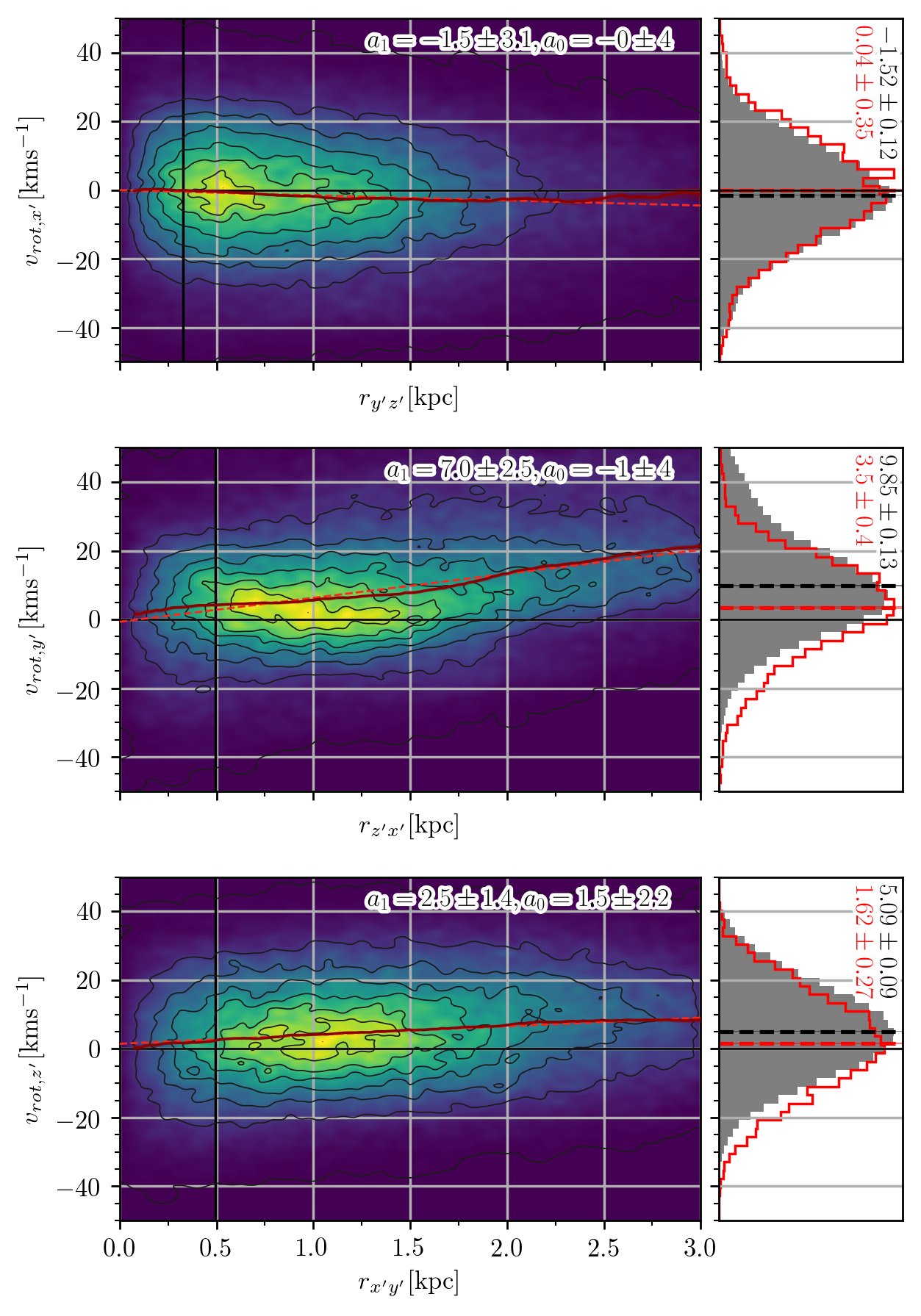}
\caption{Same as Figure~\ref{fig:main_axes_expansion} but for the originally rotating $N$-body simulation.}
\label{fig:main_axes_rotation_simL}
\end{center}
\end{figure}

We computed the principal axes of the simulated dwarf using the same criteria as as for the observations. Figure~\ref{fig:main_axes_expansion_simL} shows the expansion or contraction of the simulated dwarf along its principal axes, while Figure~\ref{fig:main_axes_rotation_simL} shows the rotation measured over the planes defined by such axes. These figures should be compared with Figures~\ref{fig:main_axes_expansion} and~\ref{fig:main_axes_rotation}, respectively. A resemblance between the model and the observations is obvious, with both showing the same general trends as well as similar differences between the inner and more external regions. The two tested spherical models, on the other hand, resulted in flat velocity profiles in the central regions of the galaxy (see Appendix~\ref{Apx:Nbody_Models_properties}).

\begin{figure*}
\begin{center}
\includegraphics[width=\linewidth]{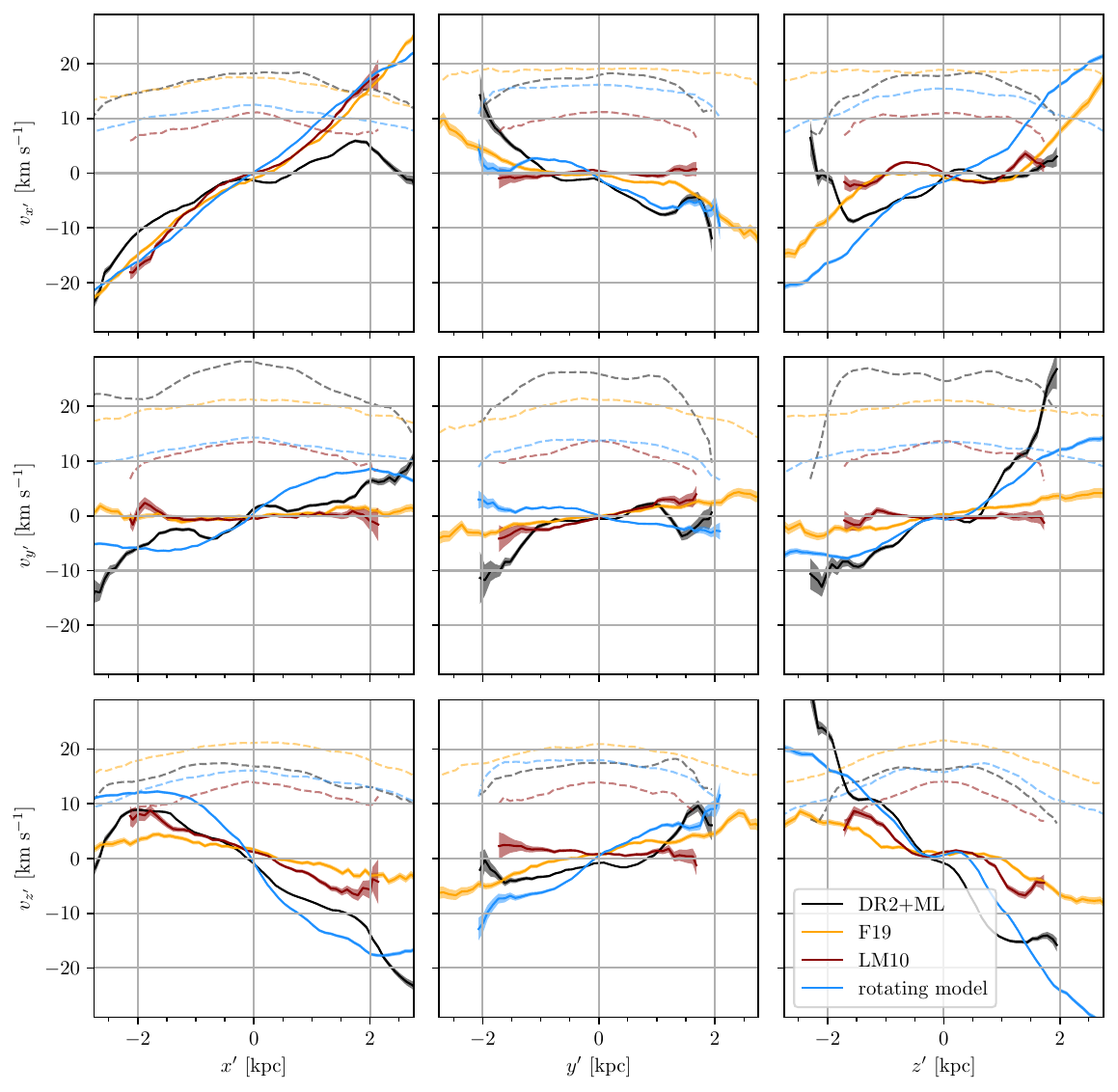}
\caption{Internal kinematics of Sgr and three $N$-body models measured along its longest, intermediate, and shortest principal axes of inertia ($x^\prime$, $y^\prime$, $z^\prime$). Curves show the box-car median value of the velocity along the axis with steps of $0.1 \kpc$ and a window of $0.2 \kpc$. Dashed-line curves show the measured velocity dispersion profile using the same box-car parameters. The black curve shows the observations using \Gaia DR2 and the ML models predictions (DR2+ML), the blue curve shows our best candidate $N$-body model with rotating progenitor, based on \citet{Lokas2010} (rotating model). The dark red curve shows the \citet{Law-Majewski2010} $N$-body model (\citetalias{Law-Majewski2010}), while the orange curve represents the \citet{Fardal2019} $N$-body model (\citetalias{Fardal2019}), both initially spherical and pressure-supported. Errors are shown by the shaded areas around the curves.}
\label{fig:comparison_velocities}
\end{center}
\end{figure*}

Figure~\ref{fig:comparison_velocities} shows a comparison between the internal kinematics of the observed galaxy and three $N$-body models tested in this work along their principal axes.

In general, the rotating model better reproduces the observed velocity profiles. Only the expansion along the intermediate intermediate axis, $y^\prime$, seems to be better reproduced by pressure-supported models (see $v_{y^\prime}$-$y^\prime$ panel). Yet the velocity profiles predicted by the rotating model model are too steep compared to the data ($v_{z^\prime}$-$x^\prime$, $v_{x^\prime}$-$z^\prime$). The seemingly high velocity dispersion observed in $v_{y^\prime}$ and, to a lesser extent, $v_{x^\prime}$ is likely to be caused by the large astrometric errors in the PMs. While their absolute value provides little insight, their shapes give an idea of the velocity anisotropy measured along each axis.

The shape of the core with the observed bar was also better reproduced by the rotating $N$-body model, whereas the others resulted in remnants that are too spherical and do not have a bar. The two pressure-supported models were aimed at reproducing the properties of the stream; therefore, we do not expect them to fit the observed properties of the core. Yet some correlations in the general kinematic trends can be observed, such the expansion of the galaxy or the tidally induced rotation in $v_{z^\prime}$-$x^\prime$. The computed rotation and radial velocity profiles along each axis for these models can be found in Appendix~\ref{Apx:Nbody_Models_properties}.

\section{Discussion}\label{sec:Discussion}

Our knowledge about Sgr is heavily based on theory. Most previous works have tried to reproduce Sgr observables with $N$-body models and studied the full phase space in the latter. New available observational data help to better constrain those models, providing more accurate information about the galaxy physical properties. Yet Sgr has proved to be more complicated than what we can fully understand. Many factors are likely to be affecting its evolution, like the shape of the MW potential well, the influence of the Magellanic System \citep{Erkal2019} or possible close passages with other systems  \citep{Bonaca2020} to give some examples. While we can learn about all these mechanisms from $N$-body models, if properly simulated, the complexity of such simulation forces us to adopt simplifications, normally resulting in an incomplete picture of the evolution of the system.

In this work, we took a different approach. Instead of learning from the full phase space of an $N$-body model that reproduces the observations, we have used ML to predict the full phase space directly on the observations. Using ML has the advantage that no explicit physical priors are imposed, removing the risk of introducing biases due to our lack of knowledge over a system. Our models, with minimal user input, have predicted the 3D positions and 3D velocities of more than $1.2\times10^5$ member stars in the core of the galaxy, which allows us to study the internal dynamics of the galaxy directly over the data and later compare the results to different $N$-body models. The results shown in this paper are fully compatible with those obtained for our training samples, where only one of the variables has been modeled (distance for the $\vlos$ catalog and $\vlos$ for the distance catalog). We tested the robustness of the results against possible statistical and systematic errors, different models, LSR properties, and COM position and velocity.

\subsection{Geometry and Orientation}

Our results allow us to discern between at least two components: a main body with a spheroid shape and the tidal tails.

Previous studies have shown that Sgr has an elongated body with some depth along the line of sight. Using red clump stars, \citet{Ibata1997} found that its core is consistent with a prolate spheroid with ratios 3:1:1 for its major, intermediate, and minor axes. More recently, \citet{Ferguson-Strigari2020} used RR Lyrae stars from \Gaia DR2 and the OGLE collaboration to find that Sgr body is likely to have a triaxial geometry with ratios 1:0.76:0.43. Our results sit somewhere in between these two works, with a triaxial geometry (1:0.67:0.60), but closer to prolate than what \citet{Ferguson-Strigari2020} suggested. In terms of depth, our derived FWHM of $2.48 \kpc$ agrees with that from the OGLE collaboration (Hamanowicz et al. 2016), with an FWHM of $2.42 \kpc$ (see Appendix~\ref{Apx:Error_distance}).

Our derived inclination of the Sgr main body also differs from the \citet{Ferguson-Strigari2020} results. They found the longest principal axis of the galaxy to be almost parallel to the plane of the sky ($-4.9^{+17.5}_{-18.8}\degree$), whereas our results suggest an inclination of $43\degree\pm6\degree$ for the longest and $2\degree\pm7\degree$ for the intermediate axes in the inner regions of the galaxy, $r < 1 \kpc$. It is important to notice that the overall inclination of the system depends on the considered radius. The inclination flattens as we consider larger radii due to the galaxy's elongation along its orbital path but remains as large as $\sim 35\degree$ up to distances of $2.75 \kpc$. This means that the major principal axis is inclined $\sim 30 \degree$ with respect to the MW potential direction (see Figure~\ref{fig:galactocentric_distro}). Stars closer to the MW center are subject to stronger tidal forces and thus get stripped earlier from Sgr body while gaining angular velocity along their orbit. This mechanism is responsible for the observed ``S'' shape, a feature whose orientation is tightly constrained by the position of the galaxy relative to the Galactic center and the direction of motion. While tidal tails also occur in $N$-body models with spherical profiles, only ellipsoids forming certain angle ($30\degree \lesssim \alpha \lesssim 60\degree$) will feature this characteristic ``S'' profile. In turn, models that reproduce well the Sgr stream properties show this shape regardless of the shape and internal kinematics of the progenitor. The fact that both ML and $N$-body models made qualitatively consistent predictions about this feature strengthens the results obtained in the present work.

Another intriguing aspect is the bar observed in the galaxy. Such a bar was also reproduced by our rotating $N$-body models at the first pericenter passage on its orbit around the MW, after $t=0.46$ Gyr from the start of the simulation. As described in detail by \citet{Lokas2014} and \citet{Gajda2017}, the formation of a bar is a characteristic intermediate stage in the process of the transformation of disky dwarfs into dwarf spheroidals. This prolate shape is maintained and again enhanced right after the second pericenter passage, which occurs at 1.2 Gyr. The orientation predicted by the ML models is reached soon after, at around 1.22--1.24 Gyr. Arguably, a third passage would help to better reproduce the observed properties of Sgr, since it could naturally explain some of the stream's farthest features and would reduce the rotation signal of the remnant. However, the present model would result in remnant that is too spherical and almost nonrotating if it survives the third pericenter or, if it is less strongly bound, it will disrupt soon after the second pericenter. A model that survived a third pericenter while keeping some internal rotation would require a new selection of structural parameters of the dwarf, addition of the gas, star formation, etc., which would involve a computationally prohibitive parameter survey. We thus decided to stop the simulation after the second passage. In any case, we emphasize that in order to form a strongly elongated shape similar to the one seen in the \Gaia data, the dwarf should transform into a bar at some stage of its evolution, and for this to occur, it is necessary that the progenitor has the form of a stellar disk.

It is worth mentioning that the 3D shape derived by our ML model (an elongated shape with tidal tails) falls in with those predicted by different $N$-body models whose main goal was to reproduce the properties of the core of the galaxy \citep[see, for example, ][]{Lokas2010, Vasiliev-Belokurov2020}.

\subsection{Internal Kinematics of the Core}

Several works have previously analyzed Sgr's structural and kinematic properties, looking for signs of the ongoing destruction of the galaxy. Aiming to reproduce the stream properties, \citet{Penarrubia2010} proposed a scenario in which Sgr progenitor is a late-type rotating disk galaxy inclined $20 \degree$ with the normal vector of its orbital plane. \citet{Lokas2010} focused on reproducing the properties of the core of the galaxy from a similar starting point. Both works presaged a rotating core remnant for Sgr, which later works were unable to confirm. Of special relevance to this discussion are \citet{Penarrubia2011} and \citet{Frinchaboy2012}. Both used samples of giant stars with $\vlos$, concluding that the Sgr kinematics were well reproduced by those of a pressure-supported system, i.e.\ nonrotating. Interestingly, a preliminary version of the data presented by \citet{Frinchaboy2012} was used in \citet{Lokas2010} as a comparison with the rotating remnant $N$-body galaxy. The \Gaia mission has brought new possibilities of studying Sgr in detail; in the recent contribution of \citet{Vasiliev-Belokurov2020}, the authors did not find any significant rotation signal from their best-fitting $N$-body model. It is worth noting that some residual rotation ($< 4 \kms$) was found by \citet{Frinchaboy2012}, although they concluded that it could be due to other factors rather than intrinsic rotation from Sgr's progenitor.

Many aspects are contributing to the observed velocity field, making this a difficult issue. Only having access to the full phase space information allows us to untangle the most relevant ones.

\begin{itemize}

    \item The contribution of the COM motion (or bulk motion) to the observed velocity field in the sky. Here Sgr is moving along its projected optical major axis toward negative values of $\Lambda$, with a tangential velocity of $254.8\pm-2.8 \kms$ projected on the sky. This tangential motion projects over the line-of-sight as a positive gradient in the same direction of movement, in this case of $4.54 \kms {\rm deg}^{-1}$, assuming a constant distance of $D = 25.97 \kpc$.

    \item The Sgr geometry and inclination with respect to the celestial plane. Sgr is a triaxial system whose longest principal axis is inclined $43 \degree$ with respect to the celestial plane. That translates to an average of $\sim 1.15 \kpc$ distance difference between the areas closer to us and the farthest ones along the surveyed area in this work. This gradient in distance reflects on the projected tangential components of the velocity (PMs), with stars that are farther away apparently moving slower than closer ones. In the particular case of Sgr bulk PMs, the difference in distance projects as $\sim 16.7 \kms$ in tangential velocities.

    \item The internal kinematics of Sgr. This includes both residual kinematics from the Sgr progenitor and perturbations of these from external potentials, such those induced by tidal forces from the MW or the Large Magellanic Cloud.

\end{itemize}

In this paper, we have derived the full phase space for more than $1.2\times10^5$ member star which allows us to simultaneously account for all of these effects and to study the Sgr internal dynamics with unprecedented detail. We have shown that the Sgr core rotates in the plane of the sky at $2.9\pm0.2\kms$ counterclockwise. It does so while also rotating perpendicularly to the plane of the sky, with its east side moving toward the Sun (upon subtraction of the systemic motion of the galaxy).

It is clear that the internal dynamics of Sgr is linked to its orbital history and to its interaction with the MW. All tested $N$-body models show similar velocity gradients in the outer regions of the galaxy, caused by the tidal forces from the MW. However, we found substantial differences in the inner parts of Sgr. Our results show that despite being ripped apart by tidal forces, the galaxy conserves an inner region that is not expanding and is rotating at $v_{rot} = 4.13 \pm 0.16 \kms$. 

The spherical $N$-body models tested in this work failed to reproduce these characteristics, resulting in flat velocity profiles in the central regions of the galaxy (plots of these models are shown in Appendix~\ref{Apx:Nbody_Models_properties}). This was expected, since spherical systems are not subject to torque forces. A flattened system, on the other hand, changes the part of the galaxy that is closer to the MW as it orbits around it, suffering torques that would induce velocity gradients. This results in velocity gradients along axes other than the original axis of rotation, $v_z^\prime$. As the system evolves and stretches along its orbit, the shape and the kinematics of the system change in such way that most of the original rotation is conserved about the intermediate principal axis $v_y^\prime$, while some rotation is also transmitted to the other axes. This can be seen as a negative gradient in $v_{rot,x^\prime}$ along $r_{y^\prime z^\prime}$, or a positive gradient in $v_{rot,z^\prime}$ along $r_{x^\prime y^\prime}$ in Figure~\ref{fig:main_axes_rotation}, that were well reproduced by the rotating model but not by the spherical models.

The flattening of Sgr would imply some initial rotation of the system, which would also contribute to the observed gradients. However, the detected rotation velocity is still  $1 \kms$ below the predicted value of our best rotating $N$-body model candidate ($3.5\pm 0.4 \kms$). As commented in the previous section, a possible explanation for the excess of rotation in the rotating model model would be that the simulated galaxy just suffered its second pericenter passage. During each passage, the progenitor is stirred and perturbed, heating its kinematics and removing rotation. The Sgr is likely to have suffered more than two pericenter passages; only models with three or more passages are able to reproduce the stream's full extension \citep{Dierickx-Loeb2017a}. Moreover, three main star formation events have been observed in the MW's star formation history at the approximate times when such passages occurred \citep{Ruiz-Lara2020}, further strengthening this hypothesis. However, even with just two passages, the rotating model does a good job reproducing the closer parts of the stream \citep[see Fig.5 in ][]{Lokas2010}. Future high-resolution $N$-body models aimed at reproducing both the stream and the core with three or more passages may provide some insight.

\subsection{Orbital and Internal Angular Momentum}

Our data allow us to derive the 3D vector for the internal angular momentum of the galaxy. For stars within $r_{\rm 3D}$ = 1 kpc, this vector forms an angle $\theta =18\degree\pm6\degree$ with respect to the orbital angular momentum of Sgr, i.e. the normal vector to its orbital plane, meaning that Sgr is in an inclined prograde orbit around the MW. The slight inclination of the axis of rotation with respect to the orbital plane could aid in the explanation of the bifurcation of the stream \citep{Belokurov2006, Koposov2012}, as different parts of Sgr get stripped with slightly different angles as the dwarf moves along its orbit. This possibility was investigated by \citet{Penarrubia2010}, who found that a retrograde orbit, $\theta = -20\degree$, was the best inclination that reproduced the observed bifurcation in the south arm. The model was later shown to be unsuccessful in reproducing the internal kinematics of Sgr \citep{Penarrubia2011}. On the other hand, while our rotating model, prograde and initially inclined $10\degree$ with respect to its orbital plane, well reproduces Sgr's internal dynamics, it did not create two clearly distinct arms in the stream but rather a single wider one. Higher initial inclinations were also tested, although they showed a poorer match between the observed and simulated angular momentum vectors.

The 3D positions reveal still more differences between observations and retrograde models, such as the observed bar. Dwarfs in prograde orbits tend to form tidally induced bars after pericenter passages. This is not the case for retrograde systems, whose stellar content remains disky along the entire orbit \citep{Lokas2015}. 

\subsection{Survivability of the Core}

Very recently, \citet{Vasiliev-Belokurov2020} presented an $N$-body model that, despite not having initial rotation, formed an elongated structure that resembles a bar. The authors argued that the lower concentration used in their models allowed tidal forces to perturb the system to form a bar-like structure.
While the morphology could be reproduced in a such way, the authors found that the resulting dwarf would be tidally disrupted in its totality.
Our results, on the other hand, show that Sgr conserves a small inner core that is not expanding, supporting a scenario in which the inner regions of the galaxy might survive this last pericenter passage.
This is also the case for the rotating model, which was the best at reproducing the configuration predicted by the ML models and whose inner core remains tidally bound at least until the next pericenter passage \citep[see Figure 6 of][]{Lokas2010}.
Therefore, we do not confirm the complete disruption of the Sgr after its recent pericenter passage.
 
Differences in the evolution of our $N$-body model from that of \citet{Vasiliev-Belokurov2020} are to be expected given the differences in how models were compared to observations; models in \citet{Vasiliev-Belokurov2020} were fitted directly to Sgr's observables projected on the sky, whereas in this work, a small set of models were compared to 6D predictions from ML.
We also note the qualitative, ``by-eye'' nature of the model selection in both cases, which may weight the various observables differently in each case.
The fact that in our model, Sgr's inner regions remain tidally bound after this pericenter passage might indicate that some adjustments should be made to the \citet{Vasiliev-Belokurov2020} model's central initial conditions.
But, given our own observational errors and model degeneracies, we do not make any strong claim about this particular aspect of Sgr evolution.                                      

\section{Conclusions}\label{sec:Conclusions}

In this work, we have presented the first comprehensive study of Sgr dwarf using ML. We have derived the full phase space, i.e.\ 3D positions and 3D velocities, for more than $1.2\times 10^5$ stars in the core, carefully considering and propagating all known sources of errors through extensive Monte Carlo experiments. We have also tested different data sets, models, and methods, all providing consistent results, and compared these to different $N$-body models. We have presented 3D projections of the galaxy in different projections depicting a complex system. From a first look at the ML predictions, the following can be observed.

\begin{itemize}
    \item The Sgr has a bar $\sim 2.5 \kpc$ long that is inclined $43\pm6 \degree$ with respect to the plane of the sky.
    \item Tidal tails depart from the tips of the bar, conferring Sgr an ``S'' shape along its orbital path.
    \item The Sgr is rotating and expanding along the stream in its outer regions.
\end{itemize}

The Sgr is a highly perturbed system. Tidal forces exerted by the MW are shearing its body and stripping its stars from distances as close as $\sim 1 \kpc$ from the center of the galaxy. However, a closer inspection of the data shows that Sgr conserves a triaxial elliptical core that is not expanding. This region measures approximately $500 \times 330 \times 300$ pc and is rotating at an average velocity of $v_{rot} = 4.13 \pm 0.16 \kms$, mainly about its intermediate principal axis of inertia ($2.46 \pm 0.22\kms$), with some nutation and precession indicated by the presence of residual rotation about its shortest and longest principal axes ($2.11\pm0.22 \kms$ and $1.38\pm0.22\kms$, respectively). The inclination of the inner body and its rotation indicate that Sgr is in an inclined prograde orbit, with its internal angular momentum forming an angle $\theta = 18\degree\pm6\degree$. This slight inclination of the orbit with respect to the internal angular momentum could aid in explaining some of the features observed in the stream.

We compared our results against the predictions of three $N$-body models: a rotating, flattened disk in an inclined prograde orbit with two pericenter passages and two spherical, pressure-supported models with three pericenter passages. Only the rotating model was able to qualitatively reproduce the observed velocity gradients. We notice, though, that the velocity gradients predicted by the rotating model are too steep ($3.5 \pm 0.4 \kms$ along the intermediate axis), which means that the initial structure of the progenitor and/or properties of the orbit require further adjustments.

The two spherical models tested (\citetalias{Law-Majewski2010} and \citetalias{Fardal2019}), show a general dynamical behavior that is similar to the observations in the outskirts of the galaxy but fail to reproduce any of the internal velocity gradients in Sgr. It is important to remark that these models were aimed at reproducing the stream properties, not the core of the galaxy.

Our results suggest that the Sgr progenitor was a flattened system to some extent and that it had some internal rotation. Investigating this scenario in detail will require a large theoretical and computational effort, as many factors are affecting the last 2--3 Gyr of Sgr evolution as it orbits the MW. Other potentials, such as those from other dSphs or the Large Magellanic Cloud, may have also played an important role.

\ \\

\textit{Acknowledgements:} The authors thank the anonymous referee for the comments that have helped to improve this paper. AdP also thanks S. Bertran de Lis for her support and help during the realization of this project. This project benefited from support for Hubble Space Telescope archival proposal 15633 through a grant from STScI, which is operated by AURA, Inc., under NASA contract NAS 5-26555. This work has made use of data from the European Space Agency (ESA) mission \Gaia (\url{https://www.cosmos.esa.int/gaia}), processed by the \Gaia Data Processing and Analysis Consortium (DPAC, \url{https://www.cosmos.esa.int/web/gaia/dpac/consortium}). Funding for the DPAC has been provided by national institutions, in particular the institutions participating in the \Gaia Multilateral Agreement. Topcat \citep{Topcat} was used during the preliminary inspection of the data and astropy aided on the coordinate transformations \citep{astropy1, astropy2}. This project is part of the HSTPROMO (High-resolution Space Telescope PROper MOtion) Collaboration\footnote{http://www.stsci.edu/$\sim$marel/hstpromo.html}, a set of projects aimed at improving our dynamical understanding of stars, clusters and galaxies in the nearby Universe through measurement and interpretation of proper motions from HST, \Gaia, and other space observatories. We thank the collaboration members for the sharing of their ideas and software.

\appendix

\section{Estimating Contaminants}\label{Apx:Contaminants}

We used the Besancon 2016 model of the Galaxy using the GOG model \citep{Luri2014} in order to estimate how many MW contaminants are polluting our sample. To do so, we subjected stars from the model on same region in the sky to the same selection procedure as for observed data described in Section~\ref{sec:DATA}. This requires both samples to be comparable, in star numbers and observational errors. The GOG model does not include all observational effects found on the \Gaia DR2 data. Specifically, the GOG model does not include the global parallax zero-point or any of the systematic errors known to be affecting DR2 data. In addition to that, the GOG model accounts for nominal end-of-mission (EoM) errors, which are estimated to be at least 4 times smaller than the statistical errors present in DR2. These observational effects must be simulated on the GOG model prior to any comparison with real \Gaia DR2 data. We selected the stars and performed such error simulation as follow:

We first download the same region of the sky covered by the real data with wider selection cuts. Systematic errors of $0.66 \masyr$ in the PMs and $0.43\masyr$ in parallax, estimated for length-scales smaller than $1 \degree$, are added in quadrature to our search limits. We also relaxed our error selection criteria to $12\sigma$. Lastly, we shifted our parallax search to compensate for a global zero-point of $\varpi_0 = -0.03 \masyr$ \citep{Lindegren2018}. These criteria result in a wider selection than the one performed in the real data, ensuring that the number of contaminants is not underestimated.

DR2 errors and zero points must be included in the resulting GOG model list prior to any comparison to real data. We simulated errors by randomly shifting the photometric magnitudes, PMs and parallaxes of every synthetic star according to the corresponding errors observed in real stars at the same position in the CMDs. These shifts were introduced over GOG stars without the simulated EoM errors. The new simulated errors are adopted as statistical errors for the GOG stars. For further information about the method, we refer the reader to \citet{delPino2015}. A parallax zero point of $\parallax = -0.055 \pm 0.024 \masyr$, derived from our QSO list, was added to the synthetic stars (see Appendix~\ref{Apx:Systematic_Errors}). After simulating errors and zero-points, DR2 and GOG data should be comparable one to one.

We then proceed to impose the exact same selection and quality cuts as with the real data. First only synthetic stars with errors compatible within 3 sigma with the PMs and parallax of Sgr are selected. The GOG model does not account for poor astrometric measurements due to crowding or more complex observational factors. In the case of real stars, those with bad astrometry are removed using equation C.1 in L18 and the renormalized unit weight error $\text{RUWE} < 1.4$. The fraction of real stars that were screened out by these equations was less than 2\%. We randomly removed the same fraction of stars from the GOG model that had similar colors and magnitudes to the rejected ones in the real data.

Synthetic and real stars are then simultaneously selected in the CMD, PMs and parallaxes following the procedure explained in Section~\ref{sec:DATA}. Only real stars are used to fit the Gaussian mixture model and to define the logarithmic-likelihood condition. The same criteria was applied when dividing the sky in cells. GOG stars that passed our cleaning procedures as well as the rejected ones are shown in Figure~\ref{fig:members_Sky}.

\section{Choosing the Best Selection Parameters}\label{Apx:Choosing_best_sigmas}

The parameter $n$ that we use on Equation~\ref{eq:log_cond} affects our membership selection and may impact our results. The use of a larger value will include more Sgr stars to our sample, yet it will also allow many contaminants to be selected as well. On the other hand, a very conservative value will reject most contaminants, but it will also reject real member stars reducing our signal-to-noise (SN).

A number of factors affect the position of the stars in the PM-parallax space, altering the shape and the concentration of its distribution. In order to account for the most of them, we used the full covariance matrix to whiten the data\footnote{A whitening transformation is a linear transformation that transforms a vector of random variables with a known covariance matrix into a set of new variables whose covariance is the identity matrix, meaning that they are uncorrelated and each have variance 1. It can be decomposed in a decorrelation and a standarization of the data.}, taking into account their statistical errors, possible systematic errors given by equations 16 and 18 from \citet{Lindegren2018}, as well as the correlation present between these quantities. The whitening of the data strongly relies on how well errors and correlations are accounted in the {\tt gaia\_source} table and it will impact our membership selection.

The GOG model allow us to determine the optimal value for $n$ and whether we should whiten the data by maximizing the accuracy of our classification. Being our classification positive for stars selected as members of Sgr, and negative for stars selected as contaminants, we can define stars as:

\begin{itemize}
    \item True-positive (TP) stars: Sgr members and classified as member.
    \item True-negative (TN) stars: not Sgr members and classified as contaminants.
    \item False-positive (FN) stars: Sgr members, but classified as contaminant.
    \item False-positive (FP) stars: not Sgr members, but classified as member.
\end{itemize}

While accurate calculations on the total numbers of stars classified as TP or FN is not feasible from the observations, the GOG model provides an estimation for TN and FP. By assuming $\left\vert \text{TN}\right\vert = \left\vert\text{Rejected GOG} \right\vert$ and $\left\vert \text{FN}\right\vert = \left\vert\text{Selected GOG} \right\vert$, we can estimate $\left\vert \text{TP}\right\vert$ as $ \sim \left\vert\text{Selected DR2} \right\vert - \left\vert\text{Selected GOG} \right\vert$. Hence, the accuracy of our classification can be defined as:

\begin{equation}\label{eq:SN}
\begin{aligned}
\text{accuracy} = \frac{\left\vert \text{TP}\right\vert + \left\vert \text{TN}\right\vert}{\left\vert \text{TP} \right\vert + \left\vert \text{FP} \right\vert + \left\vert \text{TN} \right\vert + \left\vert \text{FN} \right\vert} =
\frac{\left\vert \text{Selected DR2}\right\vert - \left\vert \text{Selected GOG}\right\vert + \left\vert \text{Rejected GOG}\right\vert}{\left\vert \text{All DR2} \right\vert}
\end{aligned}
\end{equation}

We calculated the accuracy for the original and whitened data using $n = [1.5, 2.0, 2.5, 3.0]$ in Equation~\ref{eq:log_cond} for the bulk and the cell selection in the PM-parallax space. Covariance was left as a free parameter when fitting the Gaussian mixture model to the original data. The result that maximized SN was with $n=2$ and no data whitening, although all results provided qualitatively compatible results. These findings indicate that the statistical error values provided in DR2 do not represent well enough the actual errors affecting PMs and parallaxes.

\section{Distances and Line-of-sight Velocities: Models and Caveats}\label{Apx:Models_and_Caveats}

We tested a variety of suitable models to predict distances and $\vlos$. These listed a Epsilon-Support Vector Machine Regression (SVR) with linear and nonlinear kernels, a K-Neighbors Regression (KNR), a Random Forest Regression (RFR), an Extra Trees Regression (ETR), a Gaussian Process, an Elastic Net, a Lasso, and a Multilayer Perceptron Regression (MLP). For each model, $f(x)$, we derived $y_\mathrm{pred}$ through a \revision{nested $k \times l$ cross-validation scheme}, and calculated the residuals as $y_{res} = y_\mathrm{true} - y_\mathrm{pred}$. The experiment was performed methodically, sampling all possible combinations of features, columns, providing astrometric information. Specifically, we tried all the 64 combinations of features listing from just $(x, y)$, to the most complete one with {\tt (x, y, pmx, pmy, parallax, pmra\_pmdec\_corr, parallax\_pmdec\_corr, parallax\_pmra\_corr)}. For each combination of features, we covered a wide range of parameters that tune the fitting in each model. For example, in the case of the MLP, we tried all possible combinations of one, two, or three hidden layers, each one with a number of perceptrons ranging from 5 to 305 in steps of 10, totaling $\sim1.8\times10^6$ combinations. For the SVR we tested a total of $\sim2.3\times10^5$ different combinations, covering $\epsilon$ and $z^\prime$ values over four different kernels: linear, second-degree polynomial, third-degree polynomial, and Gaussian Radial Basis Functions (Rbf). \revision{ The best model architecture was chosen and tested through a nested $k \times l$ cross-validation (NCV). NCVs use an outer loop ($k$-fold) and an inner loop ($l$-fold) to effectively split the data in independent train/validation/test sets. This allows the use of all instances to optimize and test the model. Specifically, we used a leave-one-out configuration for both NCV's loops for the KNR and a $k = 10, l = 5$ for the SR. The performance and generalization error for each model were obtained studying $y_{res}$ derived through the NCV in the training lists.}

Four quantities were used to determine the goodness of the model: the correlation between $y_\mathrm{true}$ and $y_\mathrm{pred}$; the sum of the squared residuals; the dispersion of the residuals; and gradients along the sky, if any, of the residuals. A linear regression was fitted to each output, using as $y_\mathrm{true}$ as independent variable and $y_\mathrm{pred}$ as dependent one, in order to derive the correlation between both. Possible artificial gradients along the sky were studied by fitting a surface to $\left \vert  y_{res} \right \vert $ using $(\left \vert x \right \vert, \left \vert y \right \vert)$ as independent variables, where all quantities had been previously averaged over square cells of $0.5\degree \times 0.5\degree$ in order to avoid over representation in the most dense areas.

Except for the approximately the 10\% worst configurations, all tested models yielded qualitatively compatible results. In fact, no distinction could be made between the $\sim$15\% best configurations of a given model. Only models fitted to at least three features are able to reproduce dispersion of stars along the line-of-sight, this ability being increased, in general, by the use of more features. Models fitted to just $(x, y)$ were used as benchmark for the average distribution of stars along the sky. The different importance of the observational errors made necessary to take a different approach for each training sample.

\subsection{Distances}\label{Apx:Distances}

\subsubsection{Errors in the Training Sample and Estimation of Real FWHM}\label{Apx:Error_distance}

RR Lyrae stars in Sgr have large observational errors due to their relatively faint magnitudes. This affects both, the independent variables and distances inferred from their pulsation modes. We have kept a close control of the independent variables errors by propagating the statistical errors provided in the \Gaia {\tt gaia\_source} table, and through our small systematic errors simulations (see Appendix~\ref{Apx:Systematic_Errors}). 

We have measured the dispersion caused by errors affecting our derived distances, by selecting RR Lyrae stars likely to be members of the globular cluster M54 and measuring the dispersion of their location along the line-of-sight. The selection was based on the position in the sky and on their distance. We first selected all the RR Lyrae stars within M54 tidal radius, $r_t$, and then selected them in distance using a Gaussian mixture model (GMM) whose number of components was chosen through a Dirichlet Bayesian process. We repeated the experiment ranging the initial guess of the number of components in distance from 2 to 8, obtaining always as a result 2 components, or groups. We assumed that the narrow component was formed by M54 stars and obtained a dispersion of 0.946 kpc along the line-of-sight. Assuming M54 sphericity, and a tidal radius of $\sim 73$ pc \citep[using $c=2.04$, $r_c=0.09$][2010 edition]{Harris1996}, we found that 0.944 kpc of artificial dispersion is needed in order to explain the observed dispersion in distances along the line-of-sight.

M54 stars provide only an estimation of the distance error due to \Gaia photometric limitations. Other sources of error, such metallicity spread or reddening correction are not likely to be affecting such an homogeneous stellar population in a very small region in the sky as M54. Therefore, these quantities should be interpreted as a minimum level of error for the entire galaxy of Sgr.

We calculated that a dispersion of 0.2 dex in metallicity \citep{McConnachie2012} would cause a dispersion in distance of $\sim 0.65 \kpc$ in our RR Lyrae list. Errors in the reddening correction do not have critical impact on the results, adding a mere $\sim 0.02 \kpc$ of dispersion when considering a 15\% of error. Thus, assuming that metallicity dispersion and reddening errors caused a dispersion of 0.67 kpc, the total artificial dispersion in the distance amounts up to 1.16 kpc. The subtraction of this value to the FWHM measured in the central $4 \degree$ would yield a real FWHM of 2.48 kpc, rather than the measured 3.69 kpc. This value is close to the one derived by the OGLE team of 2.42 kpc \citep{Hamanowicz2016} and encourages us to adopt 2.48 kpc as the real FWHM of the distribution of RR Lyrae stars along the line-of-sight on its central regions.

We took these results into account when scoring the models used to predict distances on the real data, favoring those models producing distance dispersion of 2.48 kpc in the central regions of the galaxy for HB stars \secrevision{at the instability strip} in the CMD. \secrevision{This ensures that the comparison is made between similar stellar populations.}

\subsubsection{Different Models}\label{Apx:Distance_models}

As shown in Appendix~\ref{Apx:Error_distance}, almost one third of the observed dispersion along the line-of-sight (1.16 kpc) could be caused by observational errors. The ideal model will regularize the fitting to the training list, averaging several stars, avoiding overfitting the errors, and providing the most accurate distance possible. Hence, the dispersion of the residuals from the cross-validation is not indicative of how well the model performs when predicting distances and we decided to use the Spearman correlation coefficient, the dispersion in the predicted distances in the central regions of the galaxy, and the existence of artificial gradients in the predicted distances with respect to the training sample.

SVR with linear or polynomial kernels and KNR models tend to produce better results using less features, while the SVR with Rbf kernel, MLP and the RFR benefit from including extra information, such correlations between features. This behavior is somewhat expected, as SVR and KNR models are simpler and may not obtain useful information from the correlation features, which show a periodic behavior in the sky along the covered area. In all cases, best results were obtained by fitting at least four features (columns).

The best model overall was SVR with the Rbf kernel fitting all the features, although just marginally. This model is computationally expensive, and was very closely followed by the KNR which is the fastest of the 7 tested models. We thus decided to use the KNR until less than 100 stars are rejected during a iteration in our code. Then we shifted to the SVR model with the Rbf kernel until convergence. In our final tests results obtained with this method were the same within errors as if only the KNR model was used, thus we decided to keep the KNR results for its simplicity and faster performance. 

\begin{figure}[ht!]
\begin{center}
\includegraphics[width=\linewidth]{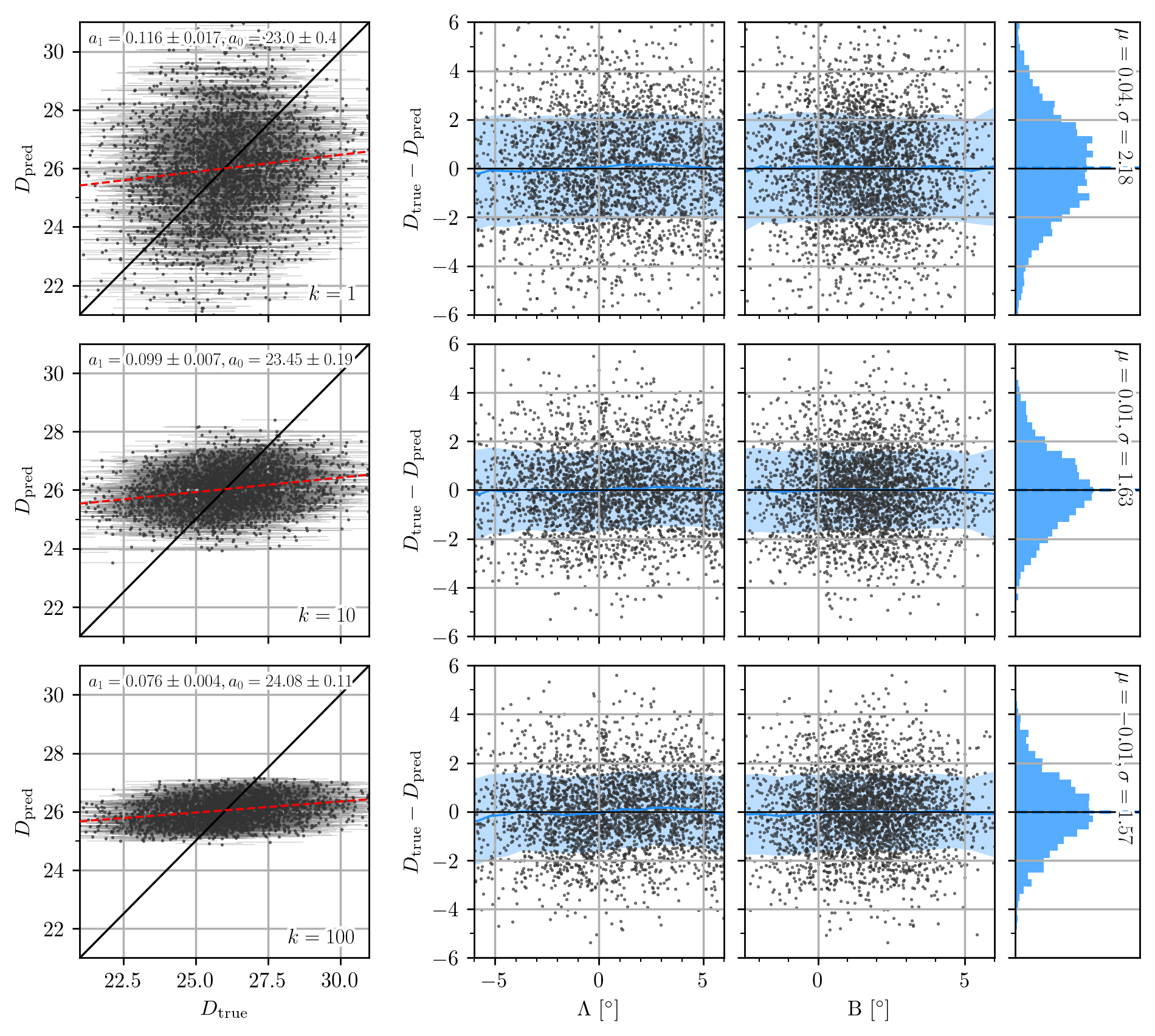}
\caption{\revision{Examples of distance predictions using the KRN model with different number of neighbors, $k = 1$, 10, and 100 from top to bottom, respectively. For each configuration, left panels show predictions, $D_\mathrm{pred}$, compared to the ground truth, $D_\mathrm{true}$. Errors in $D_\mathrm{true}$ are shown by error bars. The red-dashed line shows a linear fit to the data. The parameters of such fit are shown in the top right part of the panel. Second and third panels show the spatial variation of the residuals, $D_\mathrm{true} - D_\mathrm{pred}$, in the $(\Lambda, \Beta)$ coordinate system. Blue curves show the box-car average of the distribution along each axis with steps of $1.5 \degree$ and a window of $0.75 \degree$. Blue-shaded areas show the dispersion of the residuals measured as the standard deviation. Histograms in the right panels show the integrated distribution of the residuals. Their average, $\mu$ and standard deviation, $\sigma$ are also shown.}}
\label{fig:rr_metapar}
\end{center}
\end{figure}

\revision{The number of neighbors, $k$, and the best combination of independent variables, $\mathbf{x}$, was determined on every iteration through the NCV test. Figure~\ref{fig:rr_metapar} shows three examples of results when using $k = 1, 10, 100$ on the final training list. Low values of $k$ produce slightly steeper predictions {\it versus} ground truth (closer to a one-to-one relation), but at the cost of having too high variance. Given our prior knowledge on the training list's errors, we favoured configurations producing results with central distance dispersion around 2.48 kpc measured for HB stars. In other words, we favored bias versus variance in order to contain the artificial dispersion observed in the training list from propagating to our predictions. Notice that, as mentioned before, the error limitations in the training list prevented us from using the one-to-one relation as an indicative of how well the model is performing. In fact, the KNR with the best one-to-one relation between $D_\mathrm{pred}$ and $D_\mathrm{true}$, $k=1$, was the worst model in any other of the performed comparisons.}

\revision{Depending on the iteration of the membership selection, best results were obtained using $k$ between 10 and 30, and $\mathbf{x} = (x, y, \pmx, \pmy, \parallax)$. Larger values of $k$ tend to produce overly flattened results without improving much the variance. The KNR that produced best results in the final iteration was with $k = 15$, although values ranging from $k = 9$ to $k = 32$ provided consistent results. Results removing $\parallax$ or $\pmy$ were also reasonably consistent, indicating that some correlation between these two variables might exist.}

\begin{figure}[ht!]
\begin{center}
\includegraphics[width=\linewidth]{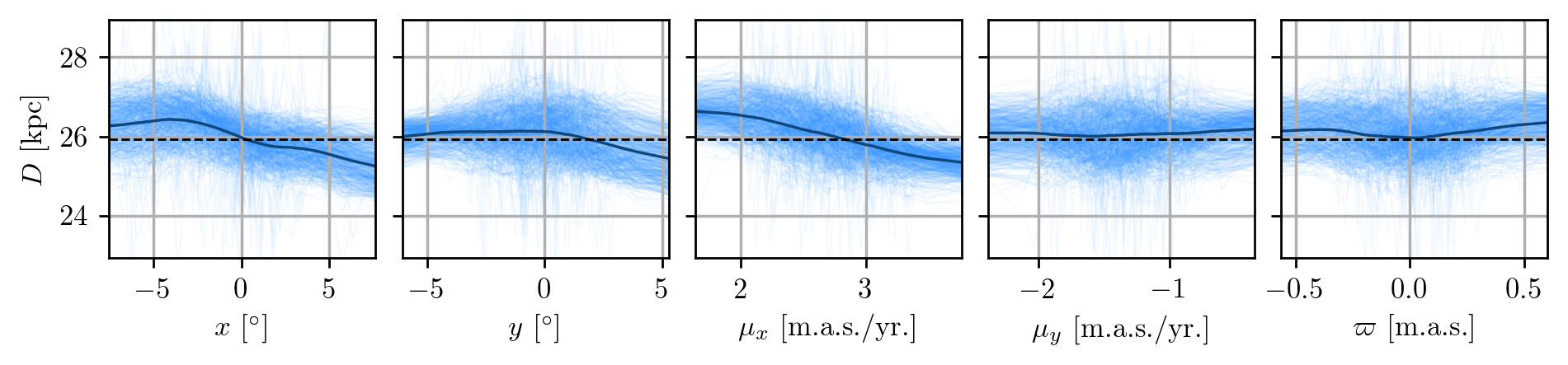}
\caption{\revision{Individual Conditional Expectation (ICE) curves for the KNR model. Thick lines shows the average dependence between the model predictions and each used feature in the training list $\mathbf{x}$. Thin curves show 20\% of all the computed combinations.}}
\label{fig:ICE_distance}
\end{center}
\end{figure}

\revision{In Figure~\ref{fig:ICE_distance}, Individual Conditional Expectation (ICE) curves show the dependence between the KNR predictions and each feature in $\mathbf{x}$. Position in the sky, $(x, y)$, and PM along the West direction, $\pmx$, have the largest impact on the prediction. On average there is also a non-monotonic dependence of the prediction on parallax, $\parallax$, which is not intuitive. This dependence may stem from the complex relationship of the random and systematic errors in Gaia parallaxes and proper motion. In any case, the dependence is weaker than for the first three features, and models not using $\parallax$ or $\pmy$ produced similar results to our standard model.}

\subsection{Line-of-sight Velocities}\label{Apx:Line-of-sight_velocities}

The bright stars in the $\vlos$ training sample show relatively small errors ($< 2 \kms$) compared to the observed dispersion. This mitigates the risk of overfitting, and allow us to use the sum of the squared residuals as indicative of how well the model is performing. On the other hand, the stars in this training sample are sparsely distributed along the sky, which poses a challenge to the models that have to interpolate the gaps in a sensible way.

We tested models with nonlinear kernels such second and third degree polynomial or radial basis functions, with mixed results. While in some cases nonlinear kernels provided better predictions in the training set, they tended to produce overly large or low velocities for extreme values of $x$ in the main sample. This behavior can be controlled by means of regularization, $z^\prime$, but introduces a large dependence of the result on this and other metaparameters of the fitting. When using linear kernels on the other hand, SVR are highly insensitive to the use of larger $z^\prime$ and thus to overfitting. The choice of a linear kernel is thus justified in the outer parts of the galaxy by the simplicity of the model and the readability of its results, although one should be aware of its limitations to reproduce the nonlinear features observed in the training set.

To overcome this, we decided \revision{to combine our best performing linear and nonlinear learners into a Stacked Regressor (SR).} In SR, a model is trained over the prediction of several learners through cross-validation of their results. They generally outperform the best of the individual learners, but are computationally expensive to run. Our SR model used a nonlinear Artificial Neural Network (an MLP) to combine the predictions of six regressors; three nonlinear: Extra Trees, Support Vector Machine with a Radial Basis Function, and Gaussian Process; with three linear: Lasso, Support Vector Machine with a linear kernel, and an Elastic Net. We chose the best configuration for each of the learners by maximizing $R^2 = (1 - \sum (y_\mathrm{true} - y_\mathrm{pred})^2/ \sum((y_\mathrm{true} - \left<y_\mathrm{true}\right>)^2)$ in a grid covering different values of their fitting metaparameters \revision{through a NCV. The experiment showed that the model provides robust predictions against mid to small metaparameters variations. Only shifting all metaparameters towards a specific behaviour, e.g.\ to have more regularization, or large variations on the architecture of the final MLP metaregressor had non-negligible impact on the final predictions. This was somewhat expected, as the metaregressor leverages the input of the six independent models choosing the best possible prediction, hence making it more stable against spurious bad predictions.}

\begin{figure}[ht!]
\begin{center}
\includegraphics[width=\linewidth]{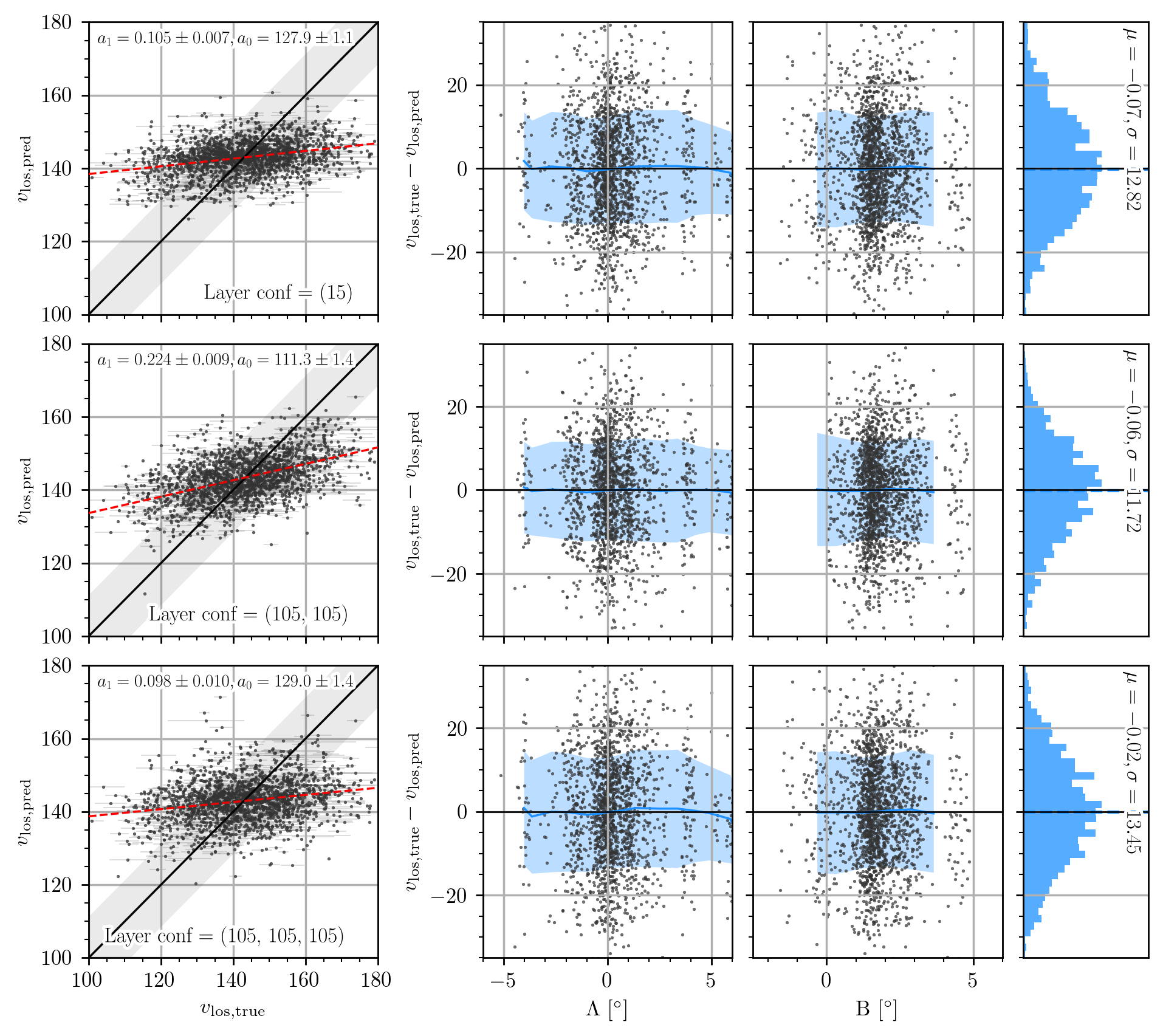}
\caption{\revision{Same as Figure~\ref{fig:rr_metapar} for the SR. Three different configurations are shown with meta parameter architectures of one hidden layer and 15 neurons (15), two hidden layers and 105 neurons per layer (105, 105) and three layers of 105 neurons each (105, 105, 105). The shaded area around the one-to-one line in the left panels shows the internal velocity dispersion of the galaxy, $11.4 \kms$ \citep{McConnachie2012}.}}
\label{fig:vlos_metapar}
\end{center}
\end{figure}

\begin{figure}[ht!]
\begin{center}
\includegraphics[width=\linewidth]{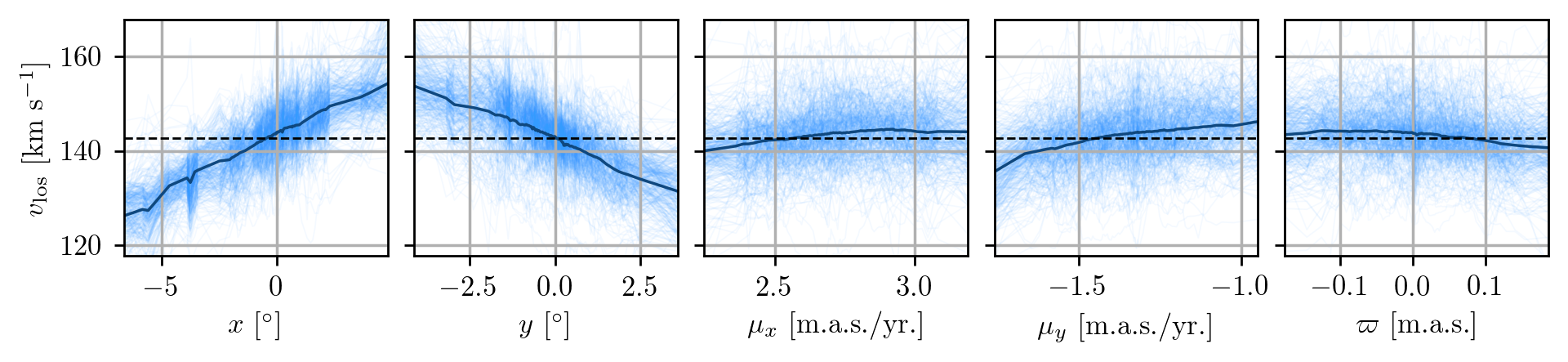}
\caption{\revision{Same as Figure~\ref{fig:ICE_distance} but for the SR model.}}
\label{fig:ICE_radial_velocity}
\end{center}
\end{figure}

\revision{Figure~\ref{fig:vlos_metapar} shows an example for three different MLP architectures in the metaregressor. The use of more neurons and extra hidden layers can help to better predict velocities. However, adding more layers or neurons can result in an overfitted model, producing predictions with needlessly large variances. The selected best architecture for the MLP varies slightly depending on the combination of the input models' metaparameters, and we found that the best 25\% of our models produced indistinguishable results within the errors. Our final model in the last iteration had an architecture of (5, 95, 85, 1), uses a ReLU activation function and a L-BFGS solver. ICE curves showing the dependence of the prediction on each feature in $\mathbf{x}$ are shown in Figure~\ref{fig:ICE_radial_velocity}.}

The resulting model was able to reproduce nonlinear features while showing stability against high-order oscillations on areas where no information is available. However, although the model produced some dispersion in the predicted velocities, cross-validation and final tests showed that the reproduced dispersion is likely to be driven by random differences with the actual velocities rather than an accurate prediction of naturally dispersed values. \revision{This manifests as the inability of the model to reproduce the extreme values of the velocity in the left panels of Figure~\ref{fig:vlos_metapar}, failing to achieve a one-to-one relation.} This was the also the case for all individual regressors, regardless of their degree of complexity and nonlinearity. \revision{These findings were totally expected, as velocity stochasticity can not be reproduced star-by-star by non-physically motivated models such the ones used here. Furthermore, despite having much better errors than the distance training list, the line-of-sight velocity errors in PMs are still high star-by-star ($\sim 0.12 \masyr$), causing random variance in the residuals. This does not pose a problem to our analysis, where our aim is to study the general velocity trends in the galaxy averaging stars in Voronoi cells containing $\sim 1000$ stars. Our NCV tests show that the SR do not produce any undesired spatial bias that could otherwise affect our results.}

\subsection{Caveats}\label{Apx:Caveats}

\revision{We have conscientiously selected, configured and trained our models in order to minimize possible biases and errors in their predictions. The models presented in this work were able to make accurate predictions, corroborated by our extensive cross-validation tests}. The results obtained on the {\tt gaia\_source} table shown in this paper are fully compatible with those obtained for our training samples separately, where only one of the variables has been modelled (distance for the $\vlos$ catalog and $\vlos$ for distance catalog). Furthermore, results obtained imposing a constant $D$ to the $\vlos$ training list, or a constant $\vlos$ to the distance training list also provided compatible similar results, indicating that the ML models used are able to make good predictions based on the training list they used to learn. \revision{Yet ML models are limited by the quality of the target data, $y_\mathrm{pred}$, and the quality of the features used to train the model and make predictions on new data, $\mathbf{x}$. The observational errors affecting these quantities negatively affect the models' prediction abilities. We believe that, in terms of the internal dynamics and structure of Sgr, no meaningful star-by-star prediction can be made with our models given the current limitations of the data. Furthermore, }training data is likely to be affected by systematic errors. A model that reproduced well a feature, $y_\mathrm{true}$, in the training list affected by systematics, would be propagating those systematics to its final predictions, $y_\mathrm{pred}$.

We have performed extensive tests in order to find and analyze systematics in the values of $y_\mathrm{true}$ in our training lists, finding the data to be highly homogeneous and of similar quality. \revision{We have also propagated all sources of errors we are aware of to our final results, and average these in within Voronoi cells of $\sim 1000$ stars allowing statistically meaningful comparisons between cells. Lastly, we have inspected ``by-eye'' the predictions from our best candidates ML models, which all provided qualitatively consistent results between them, with our training lists and previous studies in Sgr. Thus our decision over the plausibility of our solutions are ultimately based on our previous knowledge of Sgr and on comparisons to physically motivated $N$-body models as final check.}

\section{Astrometric zero-points and errors}\label{Apx:Errors}

A proper treatment of errors is crucial to this analysis. Two main sources of error are known to be affecting our measurements: the standard and the systematic errors.

\subsection{Standard errors}\label{Apx:Standard_Errors}

Standard uncertainties, or statistical errors, accompany each magnitude in the \Gaia DR2 catalogue. These measure the precision rather than the accuracy and are known to be underestimated by a factor 1.10 \citep{Lindegren2018, Arenou2018}. We propagated with Monte Carlo all statistical errors through all the coordinates transformations performed. These include errors in the coordinates of the stars and the determination of the COM: $\sigma(\alpha*)$, $\sigma(\delta)$, $\sigma(\alpha*_{0})$ and $\sigma(\delta_{0})$; as well as errors in the PMs: $\sigma(\mu_{\alpha*})$, $\sigma(\mu_{\delta})$, $\sigma(\mu_{\alpha*0})$ and $\sigma(\mu_{\delta 0})$. Errors from external quantities such the distance to the system, $\sigma(D_{0})$, or its $\vlos$, $\sigma(v_{\rm los})$, are also considered and properly propagated.

The procedure for deriving distances and line-of-sight velocities is also performed in a Monte Carlo fashion, considering and propagating the errors from all independent and dependent variables. A total of $10^6$ realizations are performed, randomly sampling each independent and dependent variable from a normal distribution centered in its corresponding nominal value. The predicted values using the nominal values of the independent variables are assumed to be the nominal values of the final result. The quadrature addition of the standard deviation of all realizations plus the typical standard deviation from the cross-validation test is adopted as total error.

\subsection{Systematic errors and zero-points}\label{Apx:Systematic_Errors}

Systematics are known to be affecting DR2 data, artificially shifting the astrometric zero-points. They are yet not well understood, but it appears that their effects can be differentiated in two components. The first one consist in large-scale systematics, with a scale length of $\theta = 20\degree$. With such a large scale length, the main effect these systematics will have in our data is to modify the bulk parallax and PMs of the galaxy. This will, in principle, have negligible effects on the detection of rotation.

We have studied any possible large-scale systematics effects on the bulk PMs and rotation using quasars. We found 278 AllWISE AGN sources \citep{Secrest2015} that cross-match with \Gaia DR2 sources within our field. We cleaned this sample from spurious sources by applying the procedure explained in Section~\ref{sec:DATA} i.e.\ rejecting sources not compatible with the centroid of the distribution up to $3\sigma$. The median of their PMs was added to the PMs of the stars in our sample in Table~\ref{tab:finalpops_sky} and the subsequent calculations.

\begin{figure}[ht!]
\begin{center}
\includegraphics[width=\linewidth]{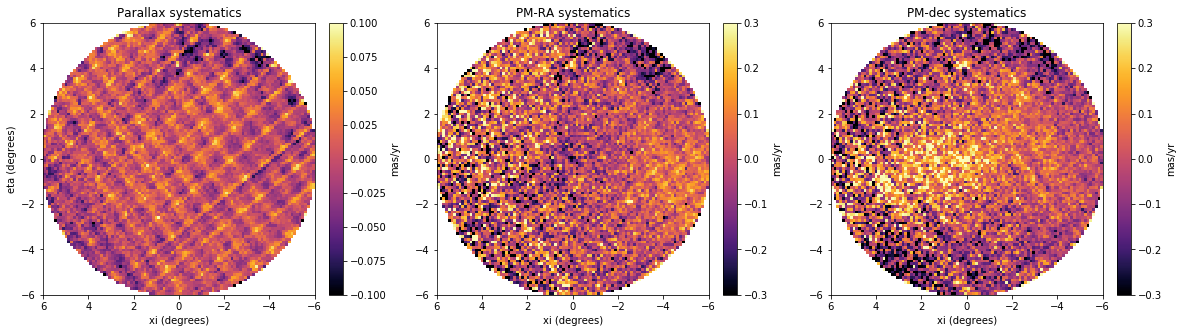}
\caption{Small systematic errors affecting the Parallax and the PMs along the observed region. These manifest as position-dependent variations in the astrometric zero points. Notice that some Sgr stars are polluting these samples, seen as an area of lower PM-R.A. and excess of PM-decl.}
\label{fig:Systematics}
\end{center}
\end{figure}

The small-scale systematics, on the other hand, introduce a variation in the zero-point of parallax and PMs with a regular pattern of scales $\sim 1\degree$ likely related to the \Gaia scanning law. This banding pattern should average out for sufficiently large objects on the sky, such Sgr. We have run tests to quantify how much these small-scale systematics may be affecting our measurements. Figure~\ref{fig:Systematics} shows the observed small variations in the astrometric zero points caused by the small-scale systematics present in \Gaia after removing most of Sgr star. Trying to correct for these systematics could introduce an extra source of error. We therefore decided to calculate and propagate the variance introduced by these patterns in the final solution.

Following same prescription adopted by \citet{Marel2019}, we created mock systematic pattern with an rms of $35\mu$as yr$^{-1}$ \citep[consistent with][]{Helmi2018}. We then shift the origin in sky coordinates ($x, y$) by random amounts between $[-0.5\degree, 0.5\degree)$ and rotate the pattern between $[-\pi, \pi)$. This is done separately for $\mu_{\alpha*}$ and $\mu_{\delta}$ $10^6$ times, re-calculating the COM PMs and the rotation signal about it in each iteration. From all realizations, we obtained a standard deviation of $0.0006 \masyr$ for both $\sigma(\mu_{\alpha\star})$ and $\sigma(\mu_{\delta})$ from the distribution of the COM PMs and $4\times10^{-6} \masyr$ for $\sigma(\mu_{r})$ and $\sigma(\mu_{\rho})$. This suppose a variation of $\sigma(v_W) = 0.8 \kms$ and $\sigma(v_N) = 0.7 \kms$ and $\sigma(v_{3}) = 0.0005 \kms$ and $\sigma(v_{2}) = 0.001 \kms$. These errors are included in all error interval provided along this paper.

\section{Kinematic properties of the spherical $N$-body models}\label{Apx:Nbody_Models_properties}

\begin{figure}[ht!]
\begin{center}
\includegraphics[width=0.48\linewidth]{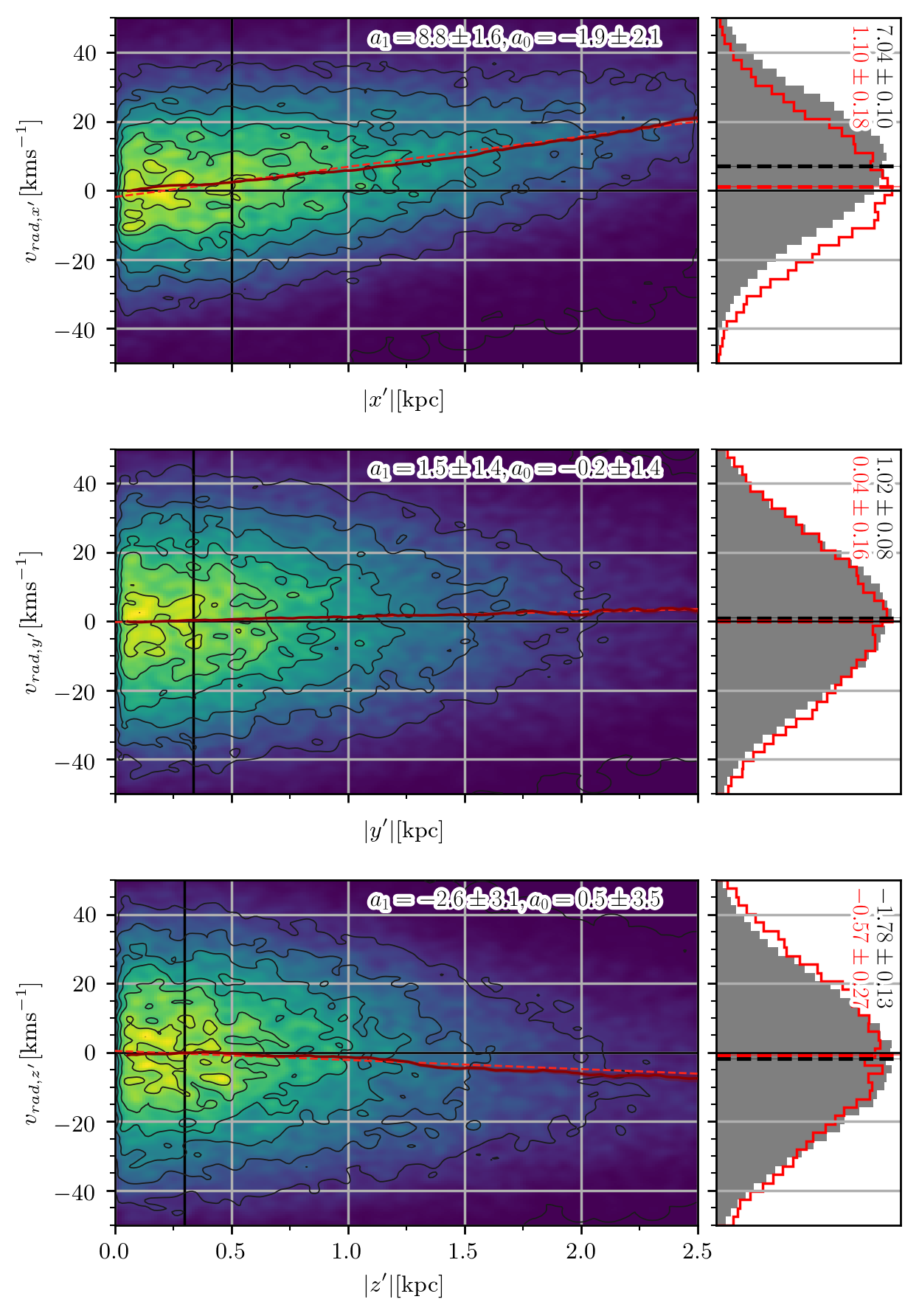}
\includegraphics[width=0.48\linewidth]{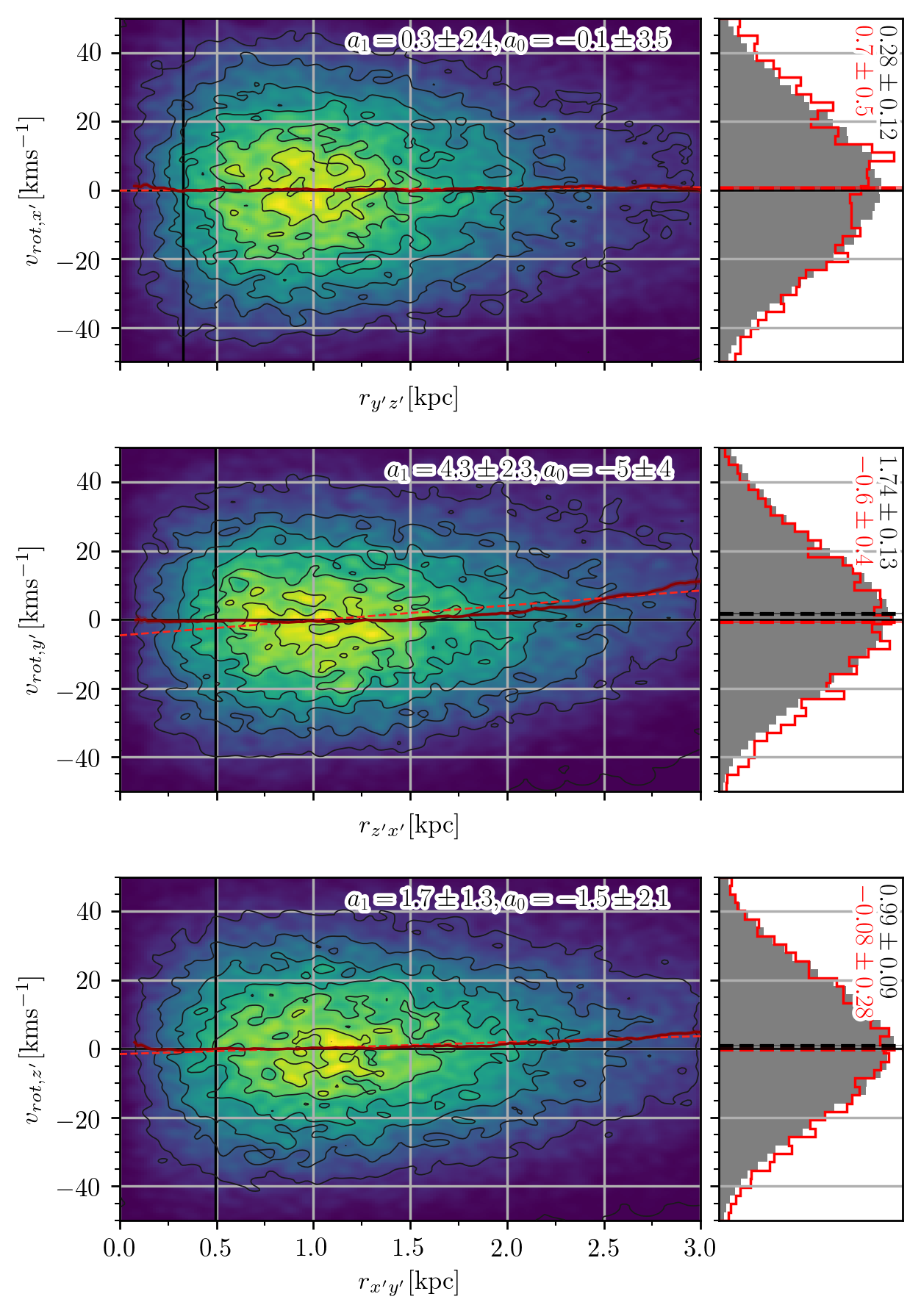}
\caption{Same as Figures~\ref{fig:main_axes_expansion},~\ref{fig:main_axes_rotation}, but for the model in \citet{Fardal2019}. The model has a core that is more extended than the observations, as it can be seen by the kinematic influence of the tidal tails for distances larger than $\sim1.5 \kpc$.}
\label{fig:main_axes_er_simF}
\end{center}
\end{figure}

\begin{figure}[ht!]
\begin{center}
\includegraphics[width=0.48\linewidth]{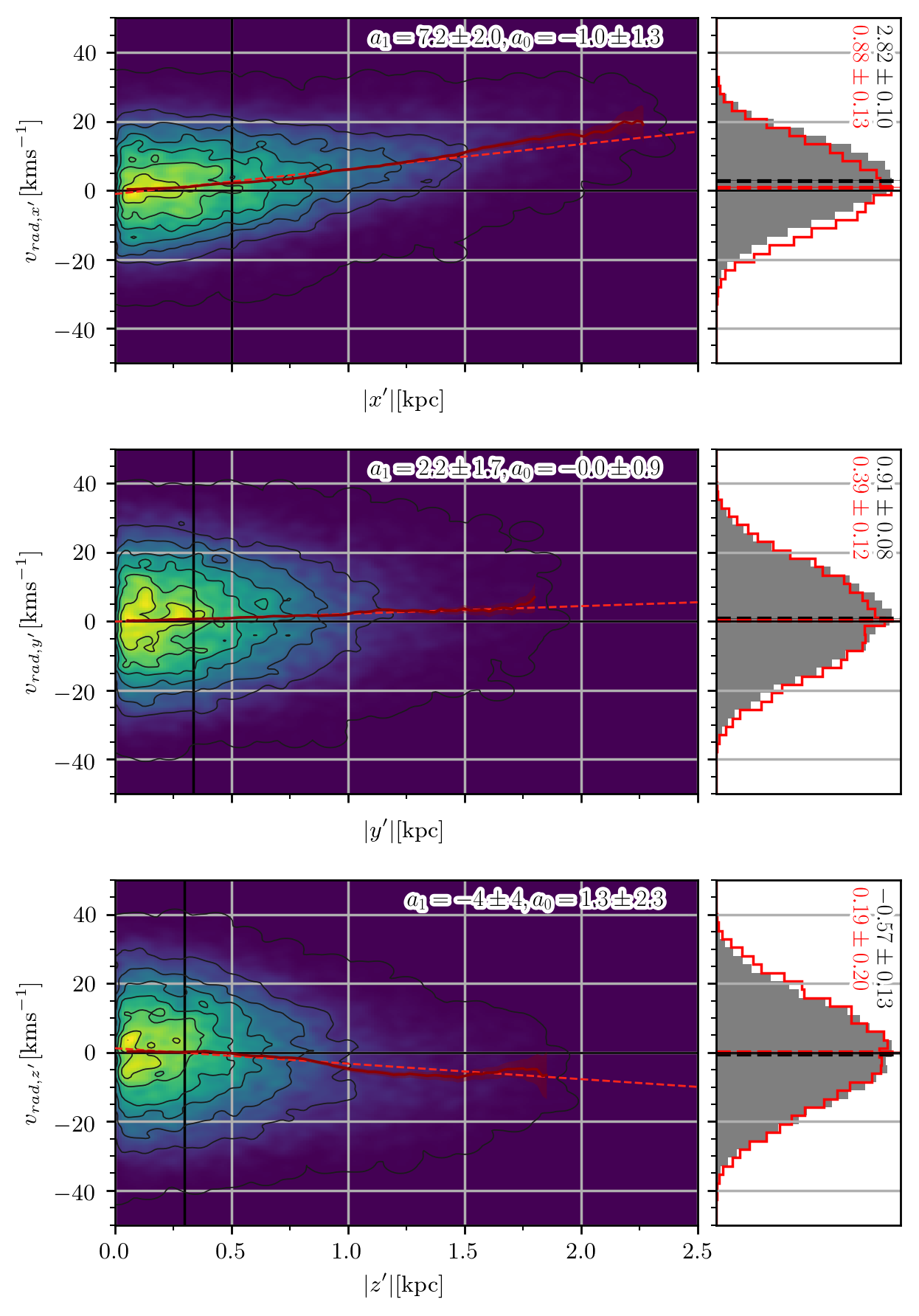}
\includegraphics[width=0.48\linewidth]{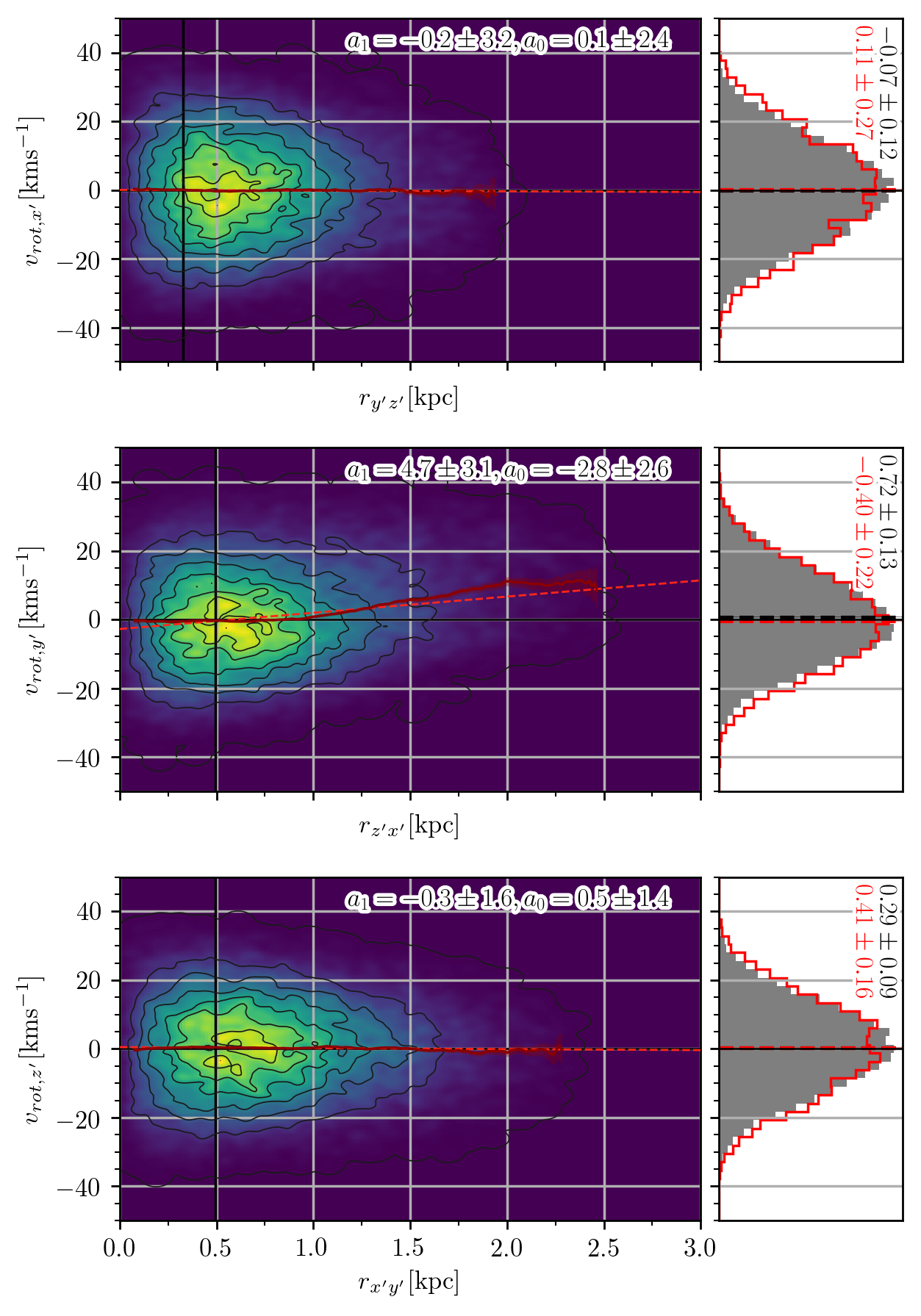}
\caption{Same as Figures~\ref{fig:main_axes_expansion},~\ref{fig:main_axes_rotation}, but for the model in \citet{Law-Majewski2010}. The model has a core that is too spherical, and that transient to the stream too close from the center. This was also observed in \citet{Vasiliev-Belokurov2020}.}
\label{fig:main_axes_er_simLM}
\end{center}
\end{figure}

Spherical models fail to reproduce the internal velocity gradients we observe in the data.  Figure~\ref{fig:main_axes_er_simF} and \ref{fig:main_axes_er_simLM}, show the expansion and rotation signals measured over the principal axes of the simulated Sgr of the \citet{Fardal2019} and the \citet{Law-Majewski2010} models, respectively. These figures should be compared with Figures~\ref{fig:main_axes_rotation} and \ref{fig:main_axes_expansion}.

\section{3D projections of the galaxy}\label{Apx:3D_projections}

\begin{figure}[ht!]
\begin{center}
\includegraphics[trim=0 0 0 0,clip,width=0.32\linewidth]{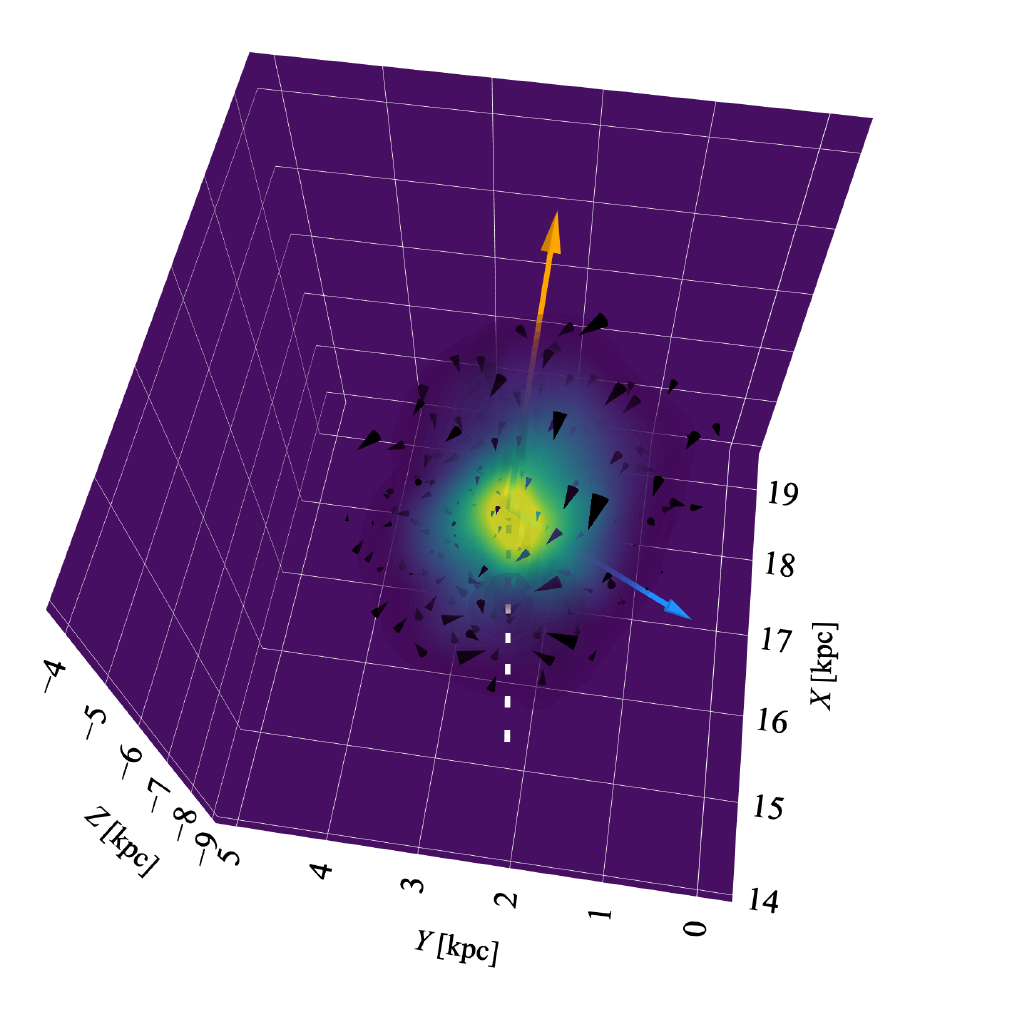}
\includegraphics[trim=0 0 0 0, clip, width=0.32\linewidth]{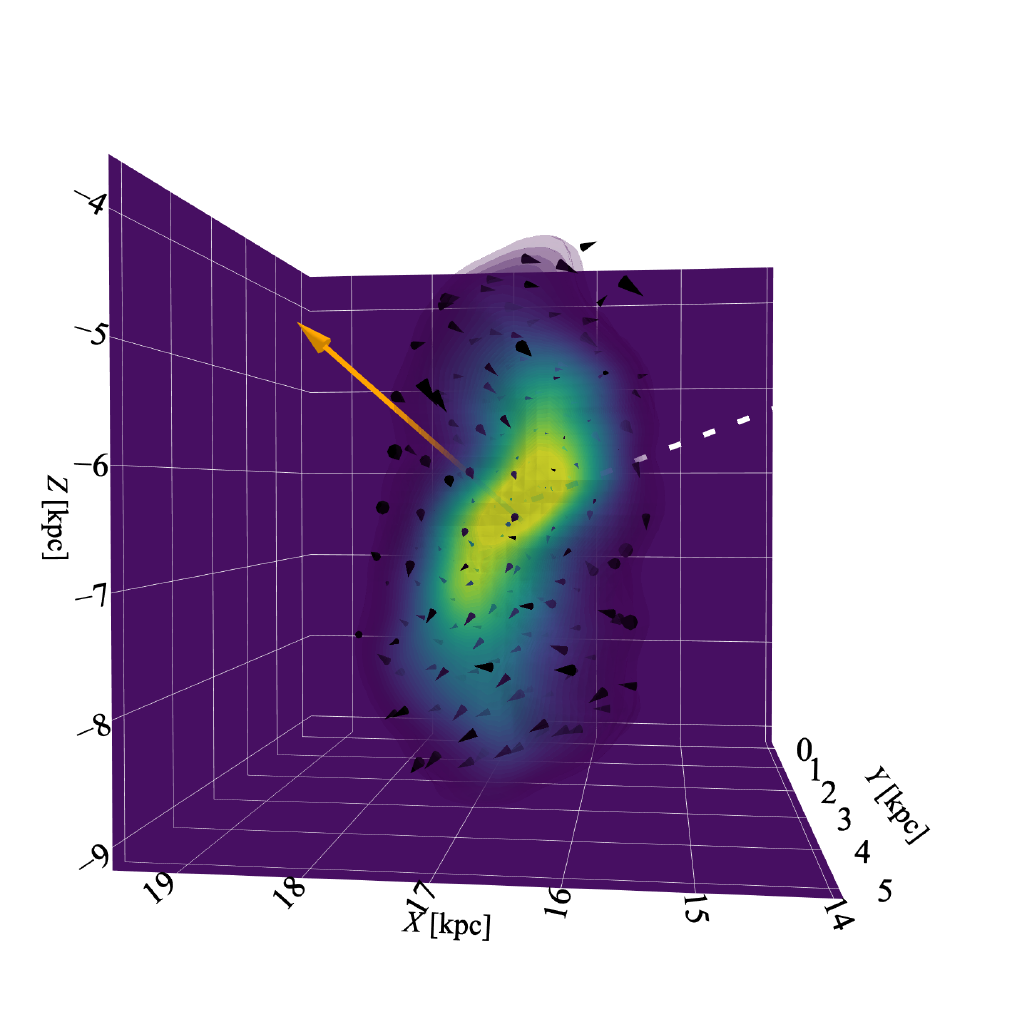}
\includegraphics[trim=0 0 0 0, clip,width=0.32\linewidth]{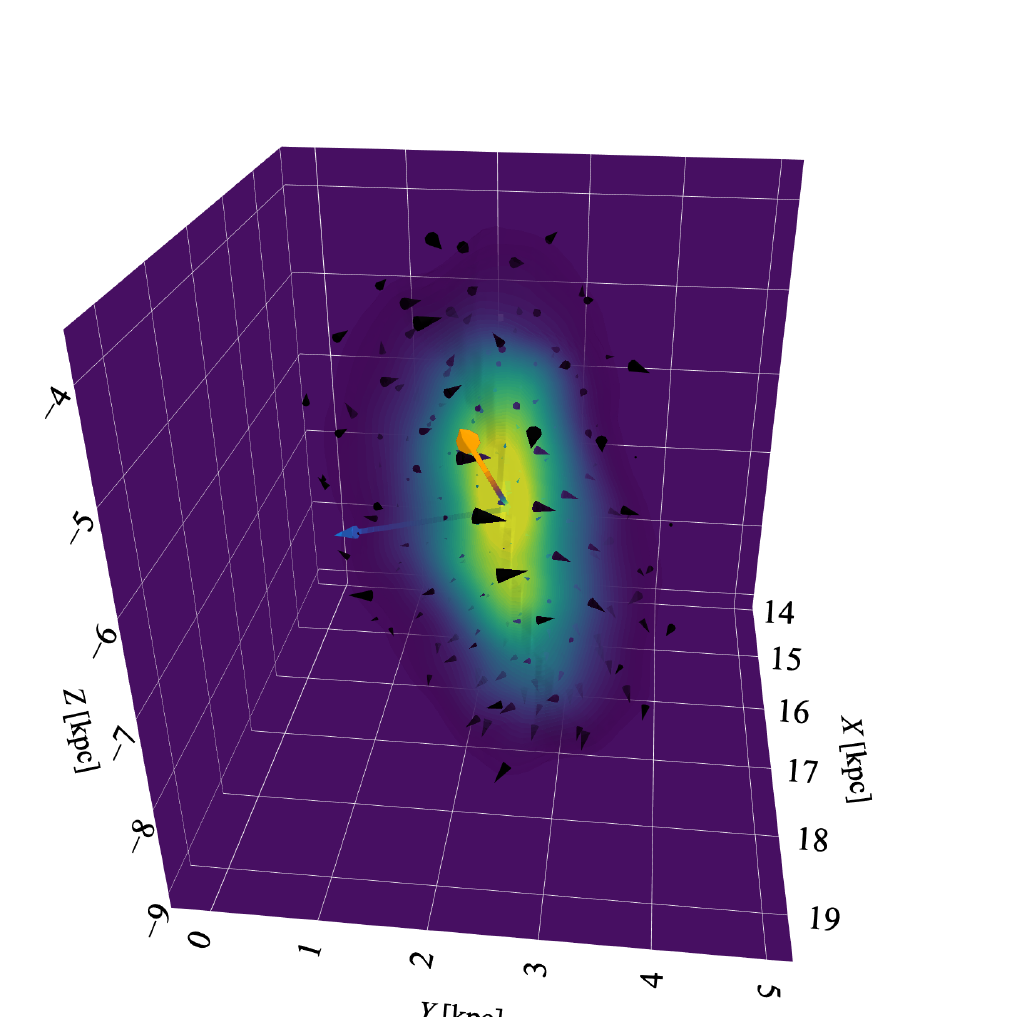}
\caption{3-dimensional renderings of the stellar density and kinematics of Sgr. The Cartesian grid is aligned with the galactocentric coordinates. The points of view are for an observer located at $5.2 \kpc$ from Sgr COM along the longest, the intermediate, and the shortest principal axes of inertia of Sgr, respectively. Black cones mark the error-weighted average motion of the stars within 3-dimensional Voronoi cells of $\sim500$ stars each one. The rest of the markers coincide with those used in Figure~\ref{fig:Sky_distro}. An interactive version of this Figure can be found at \url{https://www.stsci.edu/~marel/hstpromo/Sagittarius_GC.html}.}
\label{fig:main_distro_3D}
\end{center}
\end{figure}

Figure~\ref{fig:main_distro_3D} shows 3-dimensional rendering of the galaxy as seen from each one of its principal axes. In these projections, the axes of the Cartesian grid are aligned to the galactocentric axes. These are the same axes as in Figure~\ref{fig:galactocentric_distro}: the Galactic plane is contained in $X$--$Y$, and the Sun is located at $(X, Y, Z) = (-8.29, 0, 0) \kpc$.

\bibliographystyle{aasjournal}
\bibliography{Sagittarius_core.bib}

\begin{thebibliography}{}
\expandafter\ifx\csname natexlab\endcsname\relax\def\natexlab#1{#1}\fi
\providecommand{\url}[1]{\href{#1}{#1}}
\providecommand{\dodoi}[1]{doi:~\href{http://doi.org/#1}{\nolinkurl{#1}}}
\providecommand{\doeprint}[1]{\href{http://ascl.net/#1}{\nolinkurl{http://ascl.net/#1}}}
\providecommand{\doarXiv}[1]{\href{https://arxiv.org/abs/#1}{\nolinkurl{https://arxiv.org/abs/#1}}}

\bibitem[{{Antoja} {et~al.}(2018){Antoja}, {Helmi}, {Romero-G{\'o}mez}, {Katz},
  {Babusiaux}, {Drimmel}, {Evans}, {Figueras}, {Poggio}, {Reyl{\'e}}, {Robin},
  {Seabroke}, \& {Soubiran}}]{Antoja2018}
{Antoja}, T., {Helmi}, A., {Romero-G{\'o}mez}, M., {et~al.} 2018, \nat, 561,
  360, \dodoi{10.1038/s41586-018-0510-7}

\bibitem[{{Arenou} {et~al.}(2018){Arenou}, {Luri}, {Babusiaux}, {Fabricius},
  {Helmi}, {Muraveva}, {Robin}, {Spoto}, {Vallenari}, {Antoja},
  {Cantat-Gaudin}, {Jordi}, {Leclerc}, {Reyl{\'e}}, {Romero-G{\'o}mez}, {Shih},
  {Soria}, {Barache}, {Bossini}, {Bragaglia}, {Breddels}, {Fabrizio},
  {Lambert}, {Marrese}, {Massari}, {Moitinho}, {Robichon}, {Ruiz-Dern},
  {Sordo}, {Veljanoski}, {Eyer}, {Jasniewicz}, {Pancino}, {Soubiran}, {Spagna},
  {Tanga}, {Turon}, \& {Zurbach}}]{Arenou2018}
{Arenou}, F., {Luri}, X., {Babusiaux}, C., {et~al.} 2018, \aap, 616, A17,
  \dodoi{10.1051/0004-6361/201833234}

\bibitem[{{Astropy Collaboration} {et~al.}(2013){Astropy Collaboration},
  {Robitaille}, {Tollerud}, {Greenfield}, {Droettboom}, {Bray}, {Aldcroft},
  {Davis}, {Ginsburg}, {Price-Whelan}, {Kerzendorf}, {Conley}, {Crighton},
  {Barbary}, {Muna}, {Ferguson}, {Grollier}, {Parikh}, {Nair}, {Unther},
  {Deil}, {Woillez}, {Conseil}, {Kramer}, {Turner}, {Singer}, {Fox}, {Weaver},
  {Zabalza}, {Edwards}, {Azalee Bostroem}, {Burke}, {Casey}, {Crawford},
  {Dencheva}, {Ely}, {Jenness}, {Labrie}, {Lim}, {Pierfederici}, {Pontzen},
  {Ptak}, {Refsdal}, {Servillat}, \& {Streicher}}]{astropy1}
{Astropy Collaboration}, {Robitaille}, T.~P., {Tollerud}, E.~J., {et~al.} 2013,
  \aap, 558, A33, \dodoi{10.1051/0004-6361/201322068}

\bibitem[{{Belokurov} {et~al.}(2006){Belokurov}, {Zucker}, {Evans}, {Gilmore},
  {Vidrih}, {Bramich}, {Newberg}, {Wyse}, {Irwin}, {Fellhauer}, {Hewett},
  {Walton}, {Wilkinson}, {Cole}, {Yanny}, {Rockosi}, {Beers}, {Bell},
  {Brinkmann}, {Ivezi{\'c}}, \& {Lupton}}]{Belokurov2006}
{Belokurov}, V., {Zucker}, D.~B., {Evans}, N.~W., {et~al.} 2006, \apjl, 642,
  L137, \dodoi{10.1086/504797}

\bibitem[{{Bonaca} {et~al.}(2020){Bonaca}, {Conroy}, {Hogg}, {Cargile},
  {Caldwell}, {Naidu}, {Price-Whelan}, {Speagle}, \& {Johnson}}]{Bonaca2020}
{Bonaca}, A., {Conroy}, C., {Hogg}, D.~W., {et~al.} 2020, \apjl, 892, L37,
  \dodoi{10.3847/2041-8213/ab800c}

\bibitem[{Breiman(1996)}]{Breiman1996}
Breiman, L. 1996, Machine Learning, 24, 49, \dodoi{10.1007/bf00117832}

\bibitem[{{Bressan} {et~al.}(2012){Bressan}, {Marigo}, {Girardi}, {Salasnich},
  {Dal Cero}, {Rubele}, \& {Nanni}}]{Bressan2012}
{Bressan}, A., {Marigo}, P., {Girardi}, L., {et~al.} 2012, \mnras, 427, 127,
  \dodoi{10.1111/j.1365-2966.2012.21948.x}

\bibitem[{{Cappellari} \& {Copin}(2003)}]{Cappellari2003}
{Cappellari}, M., \& {Copin}, Y. 2003, \mnras, 342, 345,
  \dodoi{10.1046/j.1365-8711.2003.06541.x}

\bibitem[{{Clementini} {et~al.}(2019){Clementini}, {Ripepi}, {Molinaro},
  {Garofalo}, {Muraveva}, {Rimoldini}, {Guy}, {Jevardat de Fombelle},
  {Nienartowicz}, {Marchal}, {Audard}, {Holl}, {Leccia}, {Marconi}, {Musella},
  {Mowlavi}, {Lecoeur-Taibi}, {Eyer}, {De Ridder}, {Regibo}, {Sarro},
  {Szabados}, {Evans}, \& {Riello}}]{Clementini2019}
{Clementini}, G., {Ripepi}, V., {Molinaro}, R., {et~al.} 2019, \aap, 622, A60,
  \dodoi{10.1051/0004-6361/201833374}

\bibitem[{{Cunningham} {et~al.}(2020){Cunningham}, {Garavito-Camargo},
  {Deason}, {Johnston}, {Erkal}, {Laporte}, {Besla}, {Luger}, \&
  {Sanderson}}]{Cunningham2020}
{Cunningham}, E.~C., {Garavito-Camargo}, N., {Deason}, A.~J., {et~al.} 2020,
  \apj, 898, 4, \dodoi{10.3847/1538-4357/ab9b88}

\bibitem[{{Debattista} \& {Sellwood}(2000)}]{Debattista2000}
{Debattista}, V.~P., \& {Sellwood}, J.~A. 2000, \apj, 543, 704,
  \dodoi{10.1086/317148}

\bibitem[{{del Pino} {et~al.}(2015){del Pino}, {Aparicio}, \&
  {Hidalgo}}]{delPino2015}
{del Pino}, A., {Aparicio}, A., \& {Hidalgo}, S.~L. 2015, \mnras, 454, 3996,
  \dodoi{10.1093/mnras/stv2174}

\bibitem[{{Dierickx} \& {Loeb}(2017)}]{Dierickx-Loeb2017a}
{Dierickx}, M. I.~P., \& {Loeb}, A. 2017, \apj, 836, 92,
  \dodoi{10.3847/1538-4357/836/1/92}

\bibitem[{{Erkal} {et~al.}(2019){Erkal}, {Belokurov}, {Laporte}, {Koposov},
  {Li}, {Grillmair}, {Kallivayalil}, {Price-Whelan}, {Evans}, {Hawkins},
  {Hendel}, {Mateu}, {Navarro}, {del Pino}, {Slater}, {Sohn}, \& {Orphan Aspen
  Treasury Collaboration}}]{Erkal2019}
{Erkal}, D., {Belokurov}, V., {Laporte}, C.~F.~P., {et~al.} 2019, \mnras, 487,
  2685, \dodoi{10.1093/mnras/stz1371}

\bibitem[{{Evans} {et~al.}(2018){Evans}, {Riello}, {De Angeli}, {Carrasco},
  {Montegriffo}, {Fabricius}, {Jordi}, {Palaversa}, {Diener}, {Busso},
  {Cacciari}, {van Leeuwen}, {Burgess}, {Davidson}, {Harrison}, {Hodgkin},
  {Pancino}, {Richards}, {Altavilla}, {Balaguer-N{\'u}{\~n}ez}, {Barstow},
  {Bellazzini}, {Brown}, {Castellani}, {Cocozza}, {De Luise}, {Delgado},
  {Ducourant}, {Galleti}, {Gilmore}, {Giuffrida}, {Holl}, {Kewley}, {Koposov},
  {Marinoni}, {Marrese}, {Osborne}, {Piersimoni}, {Portell}, {Pulone},
  {Ragaini}, {Sanna}, {Terrett}, {Walton}, {Wevers}, \&
  {Wyrzykowski}}]{Evans2018}
{Evans}, D.~W., {Riello}, M., {De Angeli}, F., {et~al.} 2018, \aap, 616, A4,
  \dodoi{10.1051/0004-6361/201832756}

\bibitem[{{Fardal} {et~al.}(2019){Fardal}, {van der Marel}, {Law}, {Sohn},
  {Sesar}, {Hernitschek}, \& {Rix}}]{Fardal2019}
{Fardal}, M.~A., {van der Marel}, R.~P., {Law}, D.~R., {et~al.} 2019, \mnras,
  483, 4724, \dodoi{10.1093/mnras/sty3428}

\bibitem[{{Ferguson} \& {Strigari}(2020)}]{Ferguson-Strigari2020}
{Ferguson}, P.~S., \& {Strigari}, L.~E. 2020, \mnras, 495, 4124,
  \dodoi{10.1093/mnras/staa1404}

\bibitem[{{Frinchaboy} {et~al.}(2012){Frinchaboy}, {Majewski}, {Mu{\~n}oz},
  {Law}, {{\L}okas}, {Kunkel}, {Patterson}, \& {Johnston}}]{Frinchaboy2012}
{Frinchaboy}, P.~M., {Majewski}, S.~R., {Mu{\~n}oz}, R.~R., {et~al.} 2012,
  \apj, 756, 74, \dodoi{10.1088/0004-637X/756/1/74}

\bibitem[{{Gaia Collaboration} {et~al.}(2018{\natexlab{a}}){Gaia
  Collaboration}, {Brown}, {Vallenari}, {Prusti}, {de Bruijne}, {Babusiaux},
  {Bailer-Jones}, {Biermann}, {Evans}, {Eyer}, {Jansen}, {Jordi}, {Klioner},
  {Lammers}, {Lindegren}, {Luri}, {Mignard}, {Panem}, {Pourbaix}, {Randich},
  {Sartoretti}, {Siddiqui}, {Soubiran}, {van Leeuwen}, {Walton}, {Arenou},
  {Bastian}, {Cropper}, {Drimmel}, {Katz}, {Lattanzi}, {Bakker}, {Cacciari},
  {Casta{\~n}eda}, {Chaoul}, {Cheek}, {De Angeli}, {Fabricius}, {Guerra},
  {Holl}, {Masana}, {Messineo}, {Mowlavi}, {Nienartowicz}, {Panuzzo},
  {Portell}, {Riello}, {Seabroke}, {Tanga}, {Th{\'e}venin}, {Gracia-Abril},
  {Comoretto}, {Garcia-Reinaldos}, {Teyssier}, {Altmann}, {Andrae}, {Audard},
  {Bellas-Velidis}, {Benson}, {Berthier}, {Blomme}, {Burgess}, {Busso},
  {Carry}, {Cellino}, {Clementini}, {Clotet}, {Creevey}, {Davidson}, {De
  Ridder}, {Delchambre}, {Dell'Oro}, {Ducourant},
  {Fern{\'a}ndez-Hern{\'a}ndez}, {Fouesneau}, {Fr{\'e}mat}, {Galluccio},
  {Garc{\'\i}a-Torres}, {Gonz{\'a}lez-N{\'u}{\~n}ez}, {Gonz{\'a}lez-Vidal},
  {Gosset}, {Guy}, {Halbwachs}, {Hambly}, {Harrison}, {Hern{\'a}ndez},
  {Hestroffer}, {Hodgkin}, {Hutton}, {Jasniewicz}, {Jean-Antoine-Piccolo},
  {Jordan}, {Korn}, {Krone-Martins}, {Lanzafame}, {Lebzelter}, {L{\"o}ffler},
  {Manteiga}, {Marrese}, {Mart{\'\i}n-Fleitas}, {Moitinho}, {Mora}, {Muinonen},
  {Osinde}, {Pancino}, {Pauwels}, {Petit}, {Recio-Blanco}, {Richards},
  {Rimoldini}, {Robin}, {Sarro}, {Siopis}, {Smith}, {Sozzetti}, {S{\"u}veges},
  {Torra}, {van Reeven}, {Abbas}, {Abreu Aramburu}, {Accart}, {Aerts},
  {Altavilla}, {{\'A}lvarez}, {Alvarez}, {Alves}, {Anderson}, {Andrei},
  {Anglada Varela}, {Antiche}, {Antoja}, {Arcay}, {Astraatmadja}, {Bach},
  {Baker}, {Balaguer-N{\'u}{\~n}ez}, {Balm}, {Barache}, {Barata}, {Barbato},
  {Barblan}, {Barklem}, {Barrado}, {Barros}, {Barstow}, {Bartholom{\'e}
  Mu{\~n}oz}, {Bassilana}, {Becciani}, {Bellazzini}, {Berihuete}, {Bertone},
  {Bianchi}, {Bienaym{\'e}}, {Blanco-Cuaresma}, {Boch}, {Boeche}, {Bombrun},
  {Borrachero}, {Bossini}, {Bouquillon}, {Bourda}, {Bragaglia}, {Bramante},
  {Breddels}, {Bressan}, {Brouillet}, {Br{\"u}semeister}, {Brugaletta},
  {Bucciarelli}, {Burlacu}, {Busonero}, {Butkevich}, {Buzzi}, {Caffau},
  {Cancelliere}, {Cannizzaro}, {Cantat-Gaudin}, {Carballo}, {Carlucci},
  {Carrasco}, {Casamiquela}, {Castellani}, {Castro-Ginard}, {Charlot},
  {Chemin}, {Chiavassa}, {Cocozza}, {Costigan}, {Cowell}, {Crifo}, {Crosta},
  {Crowley}, {Cuypers}, {Dafonte}, {Damerdji}, {Dapergolas}, {David}, {David},
  {de Laverny}, {De Luise}, {De March}, {de Martino}, {de Souza}, {de Torres},
  {Debosscher}, {del Pozo}, {Delbo}, {Delgado}, {Delgado}, {Di Matteo},
  {Diakite}, {Diener}, {Distefano}, {Dolding}, {Drazinos}, {Dur{\'a}n},
  {Edvardsson}, {Enke}, {Eriksson}, {Esquej}, {Eynard Bontemps}, {Fabre},
  {Fabrizio}, {Faigler}, {Falc{\~a}o}, {Farr{\`a}s Casas}, {Federici},
  {Fedorets}, {Fernique}, {Figueras}, {Filippi}, {Findeisen}, {Fonti},
  {Fraile}, {Fraser}, {Fr{\'e}zouls}, {Gai}, {Galleti}, {Garabato},
  {Garc{\'\i}a-Sedano}, {Garofalo}, {Garralda}, {Gavel}, {Gavras}, {Gerssen},
  {Geyer}, {Giacobbe}, {Gilmore}, {Girona}, {Giuffrida}, {Glass}, {Gomes},
  {Granvik}, {Gueguen}, {Guerrier}, {Guiraud}, {Guti{\'e}rrez-S{\'a}nchez},
  {Haigron}, {Hatzidimitriou}, {Hauser}, {Haywood}, {Heiter}, {Helmi}, {Heu},
  {Hilger}, {Hobbs}, {Hofmann}, {Holland}, {Huckle}, {Hypki}, {Icardi},
  {Jan{\ss}en}, {Jevardat de Fombelle}, {Jonker}, {Juh{\'a}sz}, {Julbe},
  {Karampelas}, {Kewley}, {Klar}, {Kochoska}, {Kohley}, {Kolenberg},
  {Kontizas}, {Kontizas}, {Koposov}, {Kordopatis}, {Kostrzewa-Rutkowska},
  {Koubsky}, {Lambert}, {Lanza}, {Lasne}, {Lavigne}, {Le Fustec}, {Le
  Poncin-Lafitte}, {Lebreton}, {Leccia}, {Leclerc}, {Lecoeur-Taibi},
  {Lenhardt}, {Leroux}, {Liao}, {Licata}, {Lindstr{\o}m}, {Lister}, {Livanou},
  {Lobel}, {L{\'o}pez}, {Managau}, {Mann}, {Mantelet}, {Marchal}, {Marchant},
  {Marconi}, {Marinoni}, {Marschalk{\'o}}, {Marshall}, {Martino}, {Marton},
  {Mary}, {Massari}, {Matijevi{\v{c}}}, {Mazeh}, {McMillan}, {Messina},
  {Michalik}, {Millar}, {Molina}, {Molinaro}, {Moln{\'a}r}, {Montegriffo},
  {Mor}, {Morbidelli}, {Morel}, {Morris}, {Mulone}, {Muraveva}, {Musella},
  {Nelemans}, {Nicastro}, {Noval}, {O'Mullane}, {Ord{\'e}novic},
  {Ord{\'o}{\~n}ez-Blanco}, {Osborne}, {Pagani}, {Pagano}, {Pailler},
  {Palacin}, {Palaversa}, {Panahi}, {Pawlak}, {Piersimoni}, {Pineau}, {Plachy},
  {Plum}, {Poggio}, {Poujoulet}, {Pr{\v{s}}a}, {Pulone}, {Racero}, {Ragaini},
  {Rambaux}, {Ramos-Lerate}, {Regibo}, {Reyl{\'e}}, {Riclet}, {Ripepi}, {Riva},
  {Rivard}, {Rixon}, {Roegiers}, {Roelens}, {Romero-G{\'o}mez}, {Rowell},
  {Royer}, {Ruiz-Dern}, {Sadowski}, {Sagrist{\`a} Sell{\'e}s}, {Sahlmann},
  {Salgado}, {Salguero}, {Sanna}, {Santana-Ros}, {Sarasso}, {Savietto},
  {Schultheis}, {Sciacca}, {Segol}, {Segovia}, {S{\'e}gransan}, {Shih},
  {Siltala}, {Silva}, {Smart}, {Smith}, {Solano}, {Solitro}, {Sordo}, {Soria
  Nieto}, {Souchay}, {Spagna}, {Spoto}, {Stampa}, {Steele},
  {Steidelm{\"u}ller}, {Stephenson}, {Stoev}, {Suess}, {Surdej}, {Szabados},
  {Szegedi-Elek}, {Tapiador}, {Taris}, {Tauran}, {Taylor}, {Teixeira},
  {Terrett}, {Teyssand ier}, {Thuillot}, {Titarenko}, {Torra Clotet}, {Turon},
  {Ulla}, {Utrilla}, {Uzzi}, {Vaillant}, {Valentini}, {Valette}, {van Elteren},
  {Van Hemelryck}, {van Leeuwen}, {Vaschetto}, {Vecchiato}, {Veljanoski},
  {Viala}, {Vicente}, {Vogt}, {von Essen}, {Voss}, {Votruba}, {Voutsinas},
  {Walmsley}, {Weiler}, {Wertz}, {Wevers}, {Wyrzykowski}, {Yoldas},
  {{\v{Z}}erjal}, {Ziaeepour}, {Zorec}, {Zschocke}, {Zucker}, {Zurbach}, \&
  {Zwitter}}]{GaiaDR2}
{Gaia Collaboration}, {Brown}, A.~G.~A., {Vallenari}, A., {et~al.}
  2018{\natexlab{a}}, \aap, 616, A1, \dodoi{10.1051/0004-6361/201833051}

\bibitem[{{Gaia Collaboration} {et~al.}(2018{\natexlab{b}}){Gaia
  Collaboration}, {Helmi}, {van Leeuwen}, {McMillan}, {Massari}, {Antoja},
  {Robin}, {Lindegren}, {Bastian}, {Arenou}, {Babusiaux}, {Biermann},
  {Breddels}, {Hobbs}, {Jordi}, {Pancino}, {Reyl{\'e}}, {Veljanoski}, {Brown},
  {Vallenari}, {Prusti}, {de Bruijne}, {Bailer-Jones}, {Evans}, {Eyer},
  {Jansen}, {Klioner}, {Lammers}, {Luri}, {Mignard}, {Panem}, {Pourbaix},
  {Randich}, {Sartoretti}, {Siddiqui}, {Soubiran}, {Walton}, {Cropper},
  {Drimmel}, {Katz}, {Lattanzi}, {Bakker}, {Cacciari}, {Casta{\~n}eda},
  {Chaoul}, {Cheek}, {De Angeli}, {Fabricius}, {Guerra}, {Holl}, {Masana},
  {Messineo}, {Mowlavi}, {Nienartowicz}, {Panuzzo}, {Portell}, {Riello},
  {Seabroke}, {Tanga}, {Th{\'e}venin}, {Gracia-Abril}, {Comoretto},
  {Garcia-Reinaldos}, {Teyssier}, {Altmann}, {Andrae}, {Audard},
  {Bellas-Velidis}, {Benson}, {Berthier}, {Blomme}, {Burgess}, {Busso},
  {Carry}, {Cellino}, {Clementini}, {Clotet}, {Creevey}, {Davidson}, {De
  Ridder}, {Delchambre}, {Dell'Oro}, {Ducourant},
  {Fern{\'a}ndez-Hern{\'a}ndez}, {Fouesneau}, {Fr{\'e}mat}, {Galluccio},
  {Garc{\'\i}a-Torres}, {Gonz{\'a}lez-N{\'u}{\~n}ez}, {Gonz{\'a}lez-Vidal},
  {Gosset}, {Guy}, {Halbwachs}, {Hambly}, {Harrison}, {Hern{\'a}ndez},
  {Hestroffer}, {Hodgkin}, {Hutton}, {Jasniewicz}, {Jean-Antoine-Piccolo},
  {Jordan}, {Korn}, {Krone-Martins}, {Lanzafame}, {Lebzelter}, {L{\"o}ffler},
  {Manteiga}, {Marrese}, {Mart{\'\i}n-Fleitas}, {Moitinho}, {Mora}, {Muinonen},
  {Osinde}, {Pauwels}, {Petit}, {Recio-Blanco}, {Richards}, {Rimoldini},
  {Sarro}, {Siopis}, {Smith}, {Sozzetti}, {S{\"u}veges}, {Torra}, {van Reeven},
  {Abbas}, {Abreu Aramburu}, {Accart}, {Aerts}, {Altavilla}, {{\'A}lvarez},
  {Alvarez}, {Alves}, {Anderson}, {Andrei}, {Anglada Varela}, {Antiche},
  {Arcay}, {Astraatmadja}, {Bach}, {Baker}, {Balaguer-N{\'u}{\~n}ez}, {Balm},
  {Barache}, {Barata}, {Barbato}, {Barblan}, {Barklem}, {Barrado}, {Barros},
  {Barstow}, {Bartholom{\'e} Mu{\~n}oz}, {Bassilana}, {Becciani}, {Bellazzini},
  {Berihuete}, {Bertone}, {Bianchi}, {Bienaym{\'e}}, {Blanco-Cuaresma}, {Boch},
  {Boeche}, {Bombrun}, {Borrachero}, {Bossini}, {Bouquillon}, {Bourda},
  {Bragaglia}, {Bramante}, {Bressan}, {Brouillet}, {Br{\"u}semeister},
  {Brugaletta}, {Bucciarelli}, {Burlacu}, {Busonero}, {Butkevich}, {Buzzi},
  {Caffau}, {Cancelliere}, {Cannizzaro}, {Cantat-Gaudin}, {Carballo},
  {Carlucci}, {Carrasco}, {Casamiquela}, {Castellani}, {Castro-Ginard},
  {Charlot}, {Chemin}, {Chiavassa}, {Cocozza}, {Costigan}, {Cowell}, {Crifo},
  {Crosta}, {Crowley}, {Cuypers}, {Dafonte}, {Damerdji}, {Dapergolas}, {David},
  {David}, {de Laverny}, {De Luise}, {De March}, {de Martino}, {de Souza}, {de
  Torres}, {Debosscher}, {del Pozo}, {Delbo}, {Delgado}, {Delgado}, {Di
  Matteo}, {Diakite}, {Diener}, {Distefano}, {Dolding}, {Drazinos},
  {Dur{\'a}n}, {Edvardsson}, {Enke}, {Eriksson}, {Esquej}, {Eynard Bontemps},
  {Fabre}, {Fabrizio}, {Faigler}, {Falc{\~a}o}, {Farr{\`a}s Casas}, {Federici},
  {Fedorets}, {Fernique}, {Figueras}, {Filippi}, {Findeisen}, {Fonti},
  {Fraile}, {Fraser}, {Fr{\'e}zouls}, {Gai}, {Galleti}, {Garabato},
  {Garc{\'\i}a-Sedano}, {Garofalo}, {Garralda}, {Gavel}, {Gavras}, {Gerssen},
  {Geyer}, {Giacobbe}, {Gilmore}, {Girona}, {Giuffrida}, {Glass}, {Gomes},
  {Granvik}, {Gueguen}, {Guerrier}, {Guiraud}, {Guti{\'e}rrez-S{\'a}nchez},
  {Hofmann}, {Holland}, {Huckle}, {Hypki}, {Icardi}, {Jan{\ss}en}, {Jevardat de
  Fombelle}, {Jonker}, {Juh{\'a}sz}, {Julbe}, {Karampelas}, {Kewley}, {Klar},
  {Kochoska}, {Kohley}, {Kolenberg}, {Kontizas}, {Kontizas}, {Koposov},
  {Kordopatis}, {Kostrzewa-Rutkowska}, {Koubsky}, {Lambert}, {Lanza}, {Lasne},
  {Lavigne}, {Le Fustec}, {Le Poncin-Lafitte}, {Lebreton}, {Leccia}, {Leclerc},
  {Lecoeur-Taibi}, {Lenhardt}, {Leroux}, {Liao}, {Licata}, {Lindstr{\o}m},
  {Lister}, {Livanou}, {Lobel}, {L{\'o}pez}, {Managau}, {Mann}, {Mantelet},
  {Marchal}, {Marchant}, {Marconi}, {Marinoni}, {Marschalk{\'o}}, {Marshall},
  {Martino}, {Marton}, {Mary}, {Matijevi{\v{c}}}, {Mazeh}, {Messina},
  {Michalik}, {Millar}, {Molina}, {Molinaro}, {Moln{\'a}r}, {Montegriffo},
  {Mor}, {Morbidelli}, {Morel}, {Morris}, {Mulone}, {Muraveva}, {Musella},
  {Nelemans}, {Nicastro}, {Noval}, {O'Mullane}, {Ord{\'e}novic},
  {Ord{\'o}{\~n}ez-Blanco}, {Osborne}, {Pagani}, {Pagano}, {Pailler},
  {Palacin}, {Palaversa}, {Panahi}, {Pawlak}, {Piersimoni}, {Pineau}, {Plachy},
  {Plum}, {Poggio}, {Poujoulet}, {Pr{\v{s}}a}, {Pulone}, {Racero}, {Ragaini},
  {Rambaux}, {Ramos-Lerate}, {Regibo}, {Riclet}, {Ripepi}, {Riva}, {Rivard},
  {Rixon}, {Roegiers}, {Roelens}, {Romero-G{\'o}mez}, {Rowell}, {Royer},
  {Ruiz-Dern}, {Sadowski}, {Sagrist{\`a} Sell{\'e}s}, {Sahlmann}, {Salgado},
  {Salguero}, {Sanna}, {Santana-Ros}, {Sarasso}, {Savietto}, {Schultheis},
  {Sciacca}, {Segol}, {Segovia}, {S{\'e}gransan}, {Shih}, {Siltala}, {Silva},
  {Smart}, {Smith}, {Solano}, {Solitro}, {Sordo}, {Soria Nieto}, {Souchay},
  {Spagna}, {Spoto}, {Stampa}, {Steele}, {Steidelm{\"u}ller}, {Stephenson},
  {Stoev}, {Suess}, {Surdej}, {Szabados}, {Szegedi-Elek}, {Tapiador}, {Taris},
  {Tauran}, {Taylor}, {Teixeira}, {Terrett}, {Teyssand ier}, {Thuillot},
  {Titarenko}, {Torra Clotet}, {Turon}, {Ulla}, {Utrilla}, {Uzzi}, {Vaillant},
  {Valentini}, {Valette}, {van Elteren}, {Van Hemelryck}, {van Leeuwen},
  {Vaschetto}, {Vecchiato}, {Viala}, {Vicente}, {Vogt}, {von Essen}, {Voss},
  {Votruba}, {Voutsinas}, {Walmsley}, {Weiler}, {Wertz}, {Wevems},
  {Wyrzykowski}, {Yoldas}, {{\v{Z}}erjal}, {Ziaeepour}, {Zorec}, {Zschocke},
  {Zucker}, {Zurbach}, \& {Zwitter}}]{Helmi2018}
{Gaia Collaboration}, {Helmi}, A., {van Leeuwen}, F., {et~al.}
  2018{\natexlab{b}}, \aap, 616, A12, \dodoi{10.1051/0004-6361/201832698}

\bibitem[{{Gajda} {et~al.}(2017){Gajda}, {{\L}okas}, \&
  {Athanassoula}}]{Gajda2017}
{Gajda}, G., {{\L}okas}, E.~L., \& {Athanassoula}, E. 2017, \apj, 842, 56,
  \dodoi{10.3847/1538-4357/aa74b4}

\bibitem[{{Gibbons} {et~al.}(2014){Gibbons}, {Belokurov}, \&
  {Evans}}]{Gibbons2014}
{Gibbons}, S.~L.~J., {Belokurov}, V., \& {Evans}, N.~W. 2014, \mnras, 445,
  3788, \dodoi{10.1093/mnras/stu1986}

\bibitem[{{Giuffrida} {et~al.}(2010){Giuffrida}, {Sbordone}, {Zaggia},
  {Marconi}, {Bonifacio}, {Izzo}, {Szeifert}, \& {Buonanno}}]{Giuffrida2010}
{Giuffrida}, G., {Sbordone}, L., {Zaggia}, S., {et~al.} 2010, \aap, 513, A62,
  \dodoi{10.1051/0004-6361/200913331}

\bibitem[{{Hamanowicz} {et~al.}(2016){Hamanowicz}, {Pietrukowicz}, {Udalski},
  {Mr{\'o}z}, {Soszy{\'n}ski}, {Szyma{\'n}ski}, {Skowron}, {Poleski},
  {Wyrzykowski}, {Koz{\l}owski}, {Pawlak}, \& {Ulaczyk}}]{Hamanowicz2016}
{Hamanowicz}, A., {Pietrukowicz}, P., {Udalski}, A., {et~al.} 2016, \actaa, 66,
  197.
\newblock \doarXiv{1605.04906}

\bibitem[{{Harris}(1996)}]{Harris1996}
{Harris}, W.~E. 1996, \aj, 112, 1487, \dodoi{10.1086/118116}

\bibitem[{{Holl} {et~al.}(2019){Holl}, {Audard}, \& {Nienartowicz}}]{Holl2019}
{Holl}, B., {Audard}, M., \& {Nienartowicz}, K. 2019, in The Gaia Universe, 14,
  \dodoi{10.5281/zenodo.2636357}

\bibitem[{{Hopkins}(2015)}]{Hopkins2015}
{Hopkins}, P.~F. 2015, \mnras, 450, 53, \dodoi{10.1093/mnras/stv195}

\bibitem[{{Ibata} {et~al.}(1994){Ibata}, {Gilmore}, \& {Irwin}}]{Ibata1994}
{Ibata}, R.~A., {Gilmore}, G., \& {Irwin}, M.~J. 1994, \nat, 370, 194,
  \dodoi{10.1038/370194a0}

\bibitem[{{Ibata} {et~al.}(1997){Ibata}, {Wyse}, {Gilmore}, {Irwin}, \&
  {Suntzeff}}]{Ibata1997}
{Ibata}, R.~A., {Wyse}, R. F.~G., {Gilmore}, G., {Irwin}, M.~J., \& {Suntzeff},
  N.~B. 1997, \aj, 113, 634, \dodoi{10.1086/118283}

\bibitem[{{Koposov} {et~al.}(2012){Koposov}, {Belokurov}, {Evans}, {Gilmore},
  {Gieles}, {Irwin}, {Lewis}, {Niederste-Ostholt}, {Pe{\~n}arrubia}, {Smith},
  {Bizyaev}, {Malanushenko}, {Malanushenko}, {Schneider}, \&
  {Wyse}}]{Koposov2012}
{Koposov}, S.~E., {Belokurov}, V., {Evans}, N.~W., {et~al.} 2012, \apj, 750,
  80, \dodoi{10.1088/0004-637X/750/1/80}

\bibitem[{{Laporte} {et~al.}(2018){Laporte}, {Johnston}, {G{\'o}mez},
  {Garavito-Camargo}, \& {Besla}}]{Laporte2018}
{Laporte}, C. F.~P., {Johnston}, K.~V., {G{\'o}mez}, F.~A., {Garavito-Camargo},
  N., \& {Besla}, G. 2018, \mnras, 481, 286, \dodoi{10.1093/mnras/sty1574}

\bibitem[{{Law} \& {Majewski}(2010)}]{Law-Majewski2010}
{Law}, D.~R., \& {Majewski}, S.~R. 2010, \apj, 714, 229,
  \dodoi{10.1088/0004-637X/714/1/229}

\bibitem[{{Law} \& {Majewski}(2016)}]{Law2016}
---. 2016, Astrophysics and Space Science Library, Vol. 420, {The Sagittarius
  Dwarf Tidal Stream(s)}, ed. H.~J. {Newberg} \& J.~L. {Carlin}, 31,
  \dodoi{10.1007/978-3-319-19336-6_2}

\bibitem[{{Lindegren} {et~al.}(2018){Lindegren}, {Hern{\'a}ndez}, {Bombrun},
  {Klioner}, {Bastian}, {Ramos-Lerate}, {de Torres}, {Steidelm{\"u}ller},
  {Stephenson}, {Hobbs}, {Lammers}, {Biermann}, {Geyer}, {Hilger}, {Michalik},
  {Stampa}, {McMillan}, {Casta{\~n}eda}, {Clotet}, {Comoretto}, {Davidson},
  {Fabricius}, {Gracia}, {Hambly}, {Hutton}, {Mora}, {Portell}, {van Leeuwen},
  {Abbas}, {Abreu}, {Altmann}, {Andrei}, {Anglada}, {Balaguer-N{\'u}{\~n}ez},
  {Barache}, {Becciani}, {Bertone}, {Bianchi}, {Bouquillon}, {Bourda},
  {Br{\"u}semeister}, {Bucciarelli}, {Busonero}, {Buzzi}, {Cancelliere},
  {Carlucci}, {Charlot}, {Cheek}, {Crosta}, {Crowley}, {de Bruijne}, {de
  Felice}, {Drimmel}, {Esquej}, {Fienga}, {Fraile}, {Gai}, {Garralda},
  {Gonz{\'a}lez-Vidal}, {Guerra}, {Hauser}, {Hofmann}, {Holl}, {Jordan},
  {Lattanzi}, {Lenhardt}, {Liao}, {Licata}, {Lister}, {L{\"o}ffler},
  {Marchant}, {Martin-Fleitas}, {Messineo}, {Mignard}, {Morbidelli}, {Poggio},
  {Riva}, {Rowell}, {Salguero}, {Sarasso}, {Sciacca}, {Siddiqui}, {Smart},
  {Spagna}, {Steele}, {Taris}, {Torra}, {van Elteren}, {van Reeven}, \&
  {Vecchiato}}]{Lindegren2018}
{Lindegren}, L., {Hern{\'a}ndez}, J., {Bombrun}, A., {et~al.} 2018, \aap, 616,
  A2, \dodoi{10.1051/0004-6361/201832727}

\bibitem[{{{\L}okas}(2019)}]{Lokas2019}
{{\L}okas}, E.~L. 2019, \aap, 629, A52, \dodoi{10.1051/0004-6361/201936056}

\bibitem[{{{\L}okas} {et~al.}(2014){{\L}okas}, {Athanassoula}, {Debattista},
  {Valluri}, {Pino}, {Semczuk}, {Gajda}, \& {Kowalczyk}}]{Lokas2014}
{{\L}okas}, E.~L., {Athanassoula}, E., {Debattista}, V.~P., {et~al.} 2014,
  \mnras, 445, 1339, \dodoi{10.1093/mnras/stu1846}

\bibitem[{{{\L}okas} {et~al.}(2010){{\L}okas}, {Kazantzidis}, {Majewski},
  {Law}, {Mayer}, \& {Frinchaboy}}]{Lokas2010}
{{\L}okas}, E.~L., {Kazantzidis}, S., {Majewski}, S.~R., {et~al.} 2010, \apj,
  725, 1516, \dodoi{10.1088/0004-637X/725/2/1516}

\bibitem[{{{\L}okas} {et~al.}(2015){{\L}okas}, {Semczuk}, {Gajda}, \&
  {D'Onghia}}]{Lokas2015}
{{\L}okas}, E.~L., {Semczuk}, M., {Gajda}, G., \& {D'Onghia}, E. 2015, \apj,
  810, 100, \dodoi{10.1088/0004-637X/810/2/100}

\bibitem[{{Luri} {et~al.}(2014){Luri}, {Palmer}, {Arenou}, {Masana}, {de
  Bruijne}, {Antiche}, {Babusiaux}, {Borrachero}, {Sartoretti}, {Julbe},
  {Isasi}, {Martinez}, {Robin}, {Reyl{\'e}}, {Jordi}, \& {Carrasco}}]{Luri2014}
{Luri}, X., {Palmer}, M., {Arenou}, F., {et~al.} 2014, \aap, 566, A119,
  \dodoi{10.1051/0004-6361/201423636}

\bibitem[{{Majewski} {et~al.}(2003){Majewski}, {Skrutskie}, {Weinberg}, \&
  {Ostheimer}}]{Majewski2003}
{Majewski}, S.~R., {Skrutskie}, M.~F., {Weinberg}, M.~D., \& {Ostheimer}, J.~C.
  2003, \apj, 599, 1082, \dodoi{10.1086/379504}

\bibitem[{{Mateu} {et~al.}(2020){Mateu}, {Holl}, {De Ridder}, \&
  {Rimoldini}}]{Mateu2020}
{Mateu}, C., {Holl}, B., {De Ridder}, J., \& {Rimoldini}, L. 2020, \mnras, 496,
  3291, \dodoi{10.1093/mnras/staa1676}

\bibitem[{{Mayer}(2010)}]{Mayer2010}
{Mayer}, L. 2010, Advances in Astronomy, 2010, 278434,
  \dodoi{10.1155/2010/278434}

\bibitem[{{McConnachie}(2012)}]{McConnachie2012}
{McConnachie}, A.~W. 2012, \aj, 144, 4, \dodoi{10.1088/0004-6256/144/1/4}

\bibitem[{{McDonald} {et~al.}(2012){McDonald}, {White}, {Zijlstra}, {Guzman
  Ramirez}, {Szyszka}, {van Loon}, {Lagadec}, \& {Jones}}]{McDonald2012}
{McDonald}, I., {White}, J.~R., {Zijlstra}, A.~A., {et~al.} 2012, \mnras, 427,
  2647, \dodoi{10.1111/j.1365-2966.2012.22109.x}

\bibitem[{{McMillan}(2011)}]{McMillan2011}
{McMillan}, P.~J. 2011, \mnras, 414, 2446,
  \dodoi{10.1111/j.1365-2966.2011.18564.x}

\bibitem[{{Navarro} {et~al.}(1997){Navarro}, {Frenk}, \& {White}}]{Navarro1997}
{Navarro}, J.~F., {Frenk}, C.~S., \& {White}, S. D.~M. 1997, \apj, 490, 493,
  \dodoi{10.1086/304888}

\bibitem[{{Niederste-Ostholt} {et~al.}(2010){Niederste-Ostholt}, {Belokurov},
  {Evans}, \& {Pe{\~n}arrubia}}]{Niederste2010}
{Niederste-Ostholt}, M., {Belokurov}, V., {Evans}, N.~W., \& {Pe{\~n}arrubia},
  J. 2010, \apj, 712, 516, \dodoi{10.1088/0004-637X/712/1/516}

\bibitem[{{Pe{\~n}arrubia} {et~al.}(2010){Pe{\~n}arrubia}, {Belokurov},
  {Evans}, {Mart{\'\i}nez-Delgado}, {Gilmore}, {Irwin}, {Niederste-Ostholt}, \&
  {Zucker}}]{Penarrubia2010}
{Pe{\~n}arrubia}, J., {Belokurov}, V., {Evans}, N.~W., {et~al.} 2010, \mnras,
  408, L26, \dodoi{10.1111/j.1745-3933.2010.00921.x}

\bibitem[{{Pe{\~n}arrubia} {et~al.}(2011){Pe{\~n}arrubia}, {Zucker}, {Irwin},
  {Hyde}, {Lane}, {Lewis}, {Gilmore}, {Evans}, \& {Belokurov}}]{Penarrubia2011}
{Pe{\~n}arrubia}, J., {Zucker}, D.~B., {Irwin}, M.~J., {et~al.} 2011, \apjl,
  727, L2, \dodoi{10.1088/2041-8205/727/1/L2}

\bibitem[{{Price-Whelan} {et~al.}(2018){Price-Whelan}, {Sip{\H{o}}cz},
  {G{\"u}nther}, {Lim}, {Crawford}, {Conseil}, {Shupe}, {Craig}, {Dencheva},
  {Ginsburg}, {VanderPlas}, {Bradley}, {P{\'e}rez-Su{\'a}rez}, {de Val-Borro},
  {Paper Contributors}, {Aldcroft}, {Cruz}, {Robitaille}, {Tollerud},
  {Coordination Committee}, {Ardelean}, {Babej}, {Bach}, {Bachetti}, {Bakanov},
  {Bamford}, {Barentsen}, {Barmby}, {Baumbach}, {Berry}, {Biscani}, {Boquien},
  {Bostroem}, {Bouma}, {Brammer}, {Bray}, {Breytenbach}, {Buddelmeijer},
  {Burke}, {Calderone}, {Cano Rodr{\'\i}guez}, {Cara}, {Cardoso}, {Cheedella},
  {Copin}, {Corrales}, {Crichton}, {D{\textquoteright}Avella}, {Deil},
  {Depagne}, {Dietrich}, {Donath}, {Droettboom}, {Earl}, {Erben}, {Fabbro},
  {Ferreira}, {Finethy}, {Fox}, {Garrison}, {Gibbons}, {Goldstein}, {Gommers},
  {Greco}, {Greenfield}, {Groener}, {Grollier}, {Hagen}, {Hirst}, {Homeier},
  {Horton}, {Hosseinzadeh}, {Hu}, {Hunkeler}, {Ivezi{\'c}}, {Jain}, {Jenness},
  {Kanarek}, {Kendrew}, {Kern}, {Kerzendorf}, {Khvalko}, {King}, {Kirkby},
  {Kulkarni}, {Kumar}, {Lee}, {Lenz}, {Littlefair}, {Ma}, {Macleod},
  {Mastropietro}, {McCully}, {Montagnac}, {Morris}, {Mueller}, {Mumford},
  {Muna}, {Murphy}, {Nelson}, {Nguyen}, {Ninan}, {N{\"o}the}, {Ogaz}, {Oh},
  {Parejko}, {Parley}, {Pascual}, {Patil}, {Patil}, {Plunkett}, {Prochaska},
  {Rastogi}, {Reddy Janga}, {Sabater}, {Sakurikar}, {Seifert}, {Sherbert},
  {Sherwood-Taylor}, {Shih}, {Sick}, {Silbiger}, {Singanamalla}, {Singer},
  {Sladen}, {Sooley}, {Sornarajah}, {Streicher}, {Teuben}, {Thomas},
  {Tremblay}, {Turner}, {Terr{\'o}n}, {van Kerkwijk}, {de la Vega}, {Watkins},
  {Weaver}, {Whitmore}, {Woillez}, {Zabalza}, \& {Contributors}}]{astropy2}
{Price-Whelan}, A.~M., {Sip{\H{o}}cz}, B.~M., {G{\"u}nther}, H.~M., {et~al.}
  2018, \aj, 156, 123, \dodoi{10.3847/1538-3881/aabc4f}

\bibitem[{{Purcell} {et~al.}(2011){Purcell}, {Bullock}, {Tollerud}, {Rocha}, \&
  {Chakrabarti}}]{Purcell2011}
{Purcell}, C.~W., {Bullock}, J.~S., {Tollerud}, E.~J., {Rocha}, M., \&
  {Chakrabarti}, S. 2011, \nat, 477, 301, \dodoi{10.1038/nature10417}

\bibitem[{{Ramos} {et~al.}(2020){Ramos}, {Mateu}, {Antoja}, {Helmi},
  {Castro-Ginard}, {Balbinot}, \& {Carrasco}}]{Ramos2020}
{Ramos}, P., {Mateu}, C., {Antoja}, T., {et~al.} 2020, \aap, 638, A104,
  \dodoi{10.1051/0004-6361/202037819}

\bibitem[{{Rimoldini} {et~al.}(2019){Rimoldini}, {Holl}, {Audard}, {Mowlavi},
  {Nienartowicz}, {Evans}, {Guy}, {Lecoeur-Ta{\"\i}bi}, {Jevardat de Fombelle},
  {Marchal}, {Roelens}, {De Ridder}, {Sarro}, {Regibo}, {Lopez}, {Clementini},
  {Ripepi}, {Molinaro}, {Garofalo}, {Moln{\'a}r}, {Plachy}, {Juh{\'a}sz},
  {Szabados}, {Lebzelter}, {Teyssier}, \& {Eyer}}]{Rimoldini2019}
{Rimoldini}, L., {Holl}, B., {Audard}, M., {et~al.} 2019, \aap, 625, A97,
  \dodoi{10.1051/0004-6361/201834616}

\bibitem[{{Ruiz-Lara} {et~al.}(2020){Ruiz-Lara}, {Gallart}, {Bernard}, \&
  {Cassisi}}]{Ruiz-Lara2020}
{Ruiz-Lara}, T., {Gallart}, C., {Bernard}, E.~J., \& {Cassisi}, S. 2020, Nature
  Astronomy, 4, 965, \dodoi{10.1038/s41550-020-1097-0}

\bibitem[{{Schlafly} \& {Finkbeiner}(2011)}]{Schlafly-Finkbeiner2011}
{Schlafly}, E.~F., \& {Finkbeiner}, D.~P. 2011, \apj, 737, 103,
  \dodoi{10.1088/0004-637X/737/2/103}

\bibitem[{{Sch{\"o}nrich} {et~al.}(2010){Sch{\"o}nrich}, {Binney}, \&
  {Dehnen}}]{Schonrich2010}
{Sch{\"o}nrich}, R., {Binney}, J., \& {Dehnen}, W. 2010, \mnras, 403, 1829,
  \dodoi{10.1111/j.1365-2966.2010.16253.x}

\bibitem[{{Secrest} {et~al.}(2015){Secrest}, {Dudik}, {Dorland}, {Zacharias},
  {Makarov}, {Fey}, {Frouard}, \& {Finch}}]{Secrest2015}
{Secrest}, N.~J., {Dudik}, R.~P., {Dorland}, B.~N., {et~al.} 2015, \apjs, 221,
  12, \dodoi{10.1088/0067-0049/221/1/12}

\bibitem[{{Sellwood} \& {Athanassoula}(1986)}]{Sellwood1986}
{Sellwood}, J.~A., \& {Athanassoula}, E. 1986, \mnras, 221, 195,
  \dodoi{10.1093/mnras/221.2.195}

\bibitem[{{Springel}(2005)}]{Springel2005}
{Springel}, V. 2005, \mnras, 364, 1105,
  \dodoi{10.1111/j.1365-2966.2005.09655.x}

\bibitem[{{Springel} {et~al.}(2001){Springel}, {Yoshida}, \&
  {White}}]{Springel2001}
{Springel}, V., {Yoshida}, N., \& {White}, S. D.~M. 2001, \na, 6, 79,
  \dodoi{10.1016/S1384-1076(01)00042-2}

\bibitem[{{Taylor}(2005)}]{Topcat}
{Taylor}, M.~B. 2005, in Astronomical Society of the Pacific Conference Series,
  Vol. 347, Astronomical Data Analysis Software and Systems XIV, ed.
  P.~{Shopbell}, M.~{Britton}, \& R.~{Ebert}, 29

\bibitem[{{van der Marel} {et~al.}(2002){van der Marel}, {Alves}, {Hardy}, \&
  {Suntzeff}}]{Marel2002}
{van der Marel}, R.~P., {Alves}, D.~R., {Hardy}, E., \& {Suntzeff}, N.~B. 2002,
  \aj, 124, 2639, \dodoi{10.1086/343775}

\bibitem[{{van der Marel} \& {Cioni}(2001)}]{Marel-Cioni2001}
{van der Marel}, R.~P., \& {Cioni}, M.-R.~L. 2001, \aj, 122, 1807,
  \dodoi{10.1086/323099}

\bibitem[{{van der Marel} {et~al.}(2019){van der Marel}, {Fardal}, {Sohn},
  {Patel}, {Besla}, {del Pino}, {Sahlmann}, \& {Watkins}}]{Marel2019}
{van der Marel}, R.~P., {Fardal}, M.~A., {Sohn}, S.~T., {et~al.} 2019, \apj,
  872, 24, \dodoi{10.3847/1538-4357/ab001b}

\bibitem[{{Vasiliev} \& {Belokurov}(2020)}]{Vasiliev-Belokurov2020}
{Vasiliev}, E., \& {Belokurov}, V. 2020, \mnras, 497, 4162,
  \dodoi{10.1093/mnras/staa2114}

\bibitem[{{Widrow} \& {Dubinski}(2005)}]{Widrow2005}
{Widrow}, L.~M., \& {Dubinski}, J. 2005, \apj, 631, 838, \dodoi{10.1086/432710}

\bibitem[{{Widrow} {et~al.}(2008){Widrow}, {Pym}, \& {Dubinski}}]{Widrow2008}
{Widrow}, L.~M., {Pym}, B., \& {Dubinski}, J. 2008, \apj, 679, 1239,
  \dodoi{10.1086/587636}

\bibitem[{{Wolpert}(1992)}]{Wolpert1992}
{Wolpert}, D.~H. 1992, Neural Networks, 5, 241 ,
  \dodoi{https://doi.org/10.1016/S0893-6080(05)80023-1}

\end{thebibliography}

\end{document}